\documentclass[11pt]{article}

\usepackage{jcappub}

\usepackage[english]{babel}
\usepackage{amsmath,amssymb,amsfonts}
\usepackage{color}
\usepackage{booktabs}
\usepackage{comment}
\usepackage{graphicx}
\usepackage{hyperref}

\def\YU{\mathbf{Y}_u}
\def\YD{\mathbf{Y}_d}
\def\YE{\mathbf{Y}_e}
\def\YN{\mathbf{Y}_\nu}
\def\YUL{\YU^\dagger\YU}
\def\YUR{\YU\YU^\dagger}
\def\YDL{\YD^\dagger\YD}
\def\YDR{\YD\YD^\dagger}
\def\YEL{\YE^\dagger\YE}
\def\YER{\YE\YE^\dagger}
\def\YNL{\YN^\dagger\YN}
\def\YNR{\YN\YN^\dagger}
\def\TR{\mathrm{Tr}}
\def\MNU{\mathbf{M}_\nu}

\def\AU{\mathbf{A}_u}
\def\AD{\mathbf{A}_d}
\def\AE{\mathbf{A}_e}
\def\AN{\mathbf{A}_\nu}

\def\hZ{h^0}
\def\HZ{H^0}
\def\AZ{A^{0}}
\def\HP{H^{+}}
\def\HM{H^{-}}
\def\HPM{H^{\pm}}

\def\DT{\tfrac{d}{dt}}

\def\M{\mathbf{m}}
\def\ONE{\mathbf{1}}
\def\SU{\mathrm{SU}}
\def\SO{\mathrm{SO}}
\def\Re{\mathrm{Re}}

\def\MGUT{M_{\text{GUT}}}
\def\atMGUT{\big|_{M_{\text{GUT}}}}
\def\MSUSY{M_{\text{SUSY}}}
\def\atMSUSY{\big|_{M_{\text{SUSY}}}}

\begin{document}

\begin{titlepage}

\vspace*{-15mm}
\vspace*{0.7cm}

\begin{center}

{\Large {\bf
Comparatively Light Extra Higgs States as Signature \\[1mm] of SUSY $\mathrm{SO}(10)$ GUTs
with 3rd Family Yukawa Unification
}
}\\[8mm]
Stefan Antusch$^{\star}$\footnote{Email: \texttt{stefan.antusch@unibas.ch}},
Christian Hohl$^{\star}$\footnote{Email: \texttt{ch.hohl@unibas.ch}},
and Vasja Susi\v{c}$^{\star}$\footnote{Email: \texttt{vasja.susic@unibas.ch}}
\end{center}

\vspace*{0.20cm}

\centerline{$^{\star}$
\it Department of Physics, University of Basel,}
\centerline{\it
Klingelbergstr.~82, CH-4056 Basel, Switzerland}

\vspace*{0.4cm}

\vspace*{1.2cm}

\begin{abstract}
\noindent
We study $3$rd family Yukawa unification in the context of supersymmetric (SUSY) $\mathrm{SO}(10)$ GUTs and $\mathrm{SO}(10)$-motivated boundary conditions for the SUSY-breaking soft terms. We consider $\mu<0$ such that the SUSY loop-threshold effects enable a good fit to all third family masses of the charged Standard Model (SM) fermions. We find that fitting the third family masses together with the mass of the SM-like Higgs particle, the scenario predicts the masses of the superpartner particles and of the extra Higgs states of the MSSM: while the sparticles are predicted to be comparatively heavy (above the present LHC bound but within reach of future colliders), the spectrum has the characteristic feature that the lightest new particles are the extra MSSM Higgses. We show that this effect is rather robust with respect to many deformations of the GUT boundary conditions, but turns out to be sensitive to the exactness of top-bottom Yukawa unification. Nevertheless, with moderate deviations of a few percent from exact top-bottom Yukawa unification (stemming e.g.\ from GUT-threshold corrections or higher-dimensional operators), the scenario still predicts extra MSSM Higgs particles with masses not much above $1.5\,\mathrm{TeV}$, which could be tested e.g.\ by future LHC searches for ditau decays $\HZ/\AZ\to\tau\tau$. Finding the extra MSSM Higges before the other new MSSM particles could thus be a smoking gun for a Yukawa unified $\mathrm{SO}(10)$ GUT.
\end{abstract}
\vspace{1cm}
Keywords: t-b-tau unification, Yukawa unification, SO(10) SUSY GUT, MSSM Higgs masses, radiative electroweak symmetry breaking

\end{titlepage}

\newpage

\tableofcontents

\section{Introduction\label{section:introduction}}

Grand Unified Theories (GUTs) \cite{Georgi:1974sy,Fritzsch:1974nn,Georgi:1974my} 
present an attractive setup for Physics Beyond the Standard Model (BSM). While gauge coupling unification in GUT is necessary for consistency, the unification of Yukawa couplings is optional, depending on the GUT operators generating the Yukawa interactions. Conversely, barring a numerical accident, Yukawa unification at high energies might indicate a bigger gauge symmetry. 

The most convenient setup for Yukawa unification are supersymmetric (SUSY) GUT models; while supersymmetry helps with gauge coupling unification by modifying the renormalization group (RG) slopes, it can also help with Yukawa unification indirectly via loop-threshold corrections at the SUSY scale $\MSUSY$ \cite{Hempfling:1993kv,Hall:1993gn,Carena:1994bv,Blazek:1995nv}.

The simplest example of some Yukawa couplings unifying would be $b$-$\tau$ unification in the 3rd family within the context of $\mathrm{SU}(5)$ GUTs~\cite{Georgi:1979df}. An even more restrictive and predictive setup is that of $t$-$b$-$\tau$(-$\nu$) unification, which is most straightforwardly achieved in $\SO(10)$, where all SM fermions of one family, with an addition of a right-handed neutrino, constitute a single irreducible representation $\mathbf{16}$ of $\SO(10)$. In such a setup, the neutrino $3$rd family coupling also has the same value as the top, bottom and tau Yukawa coupling, coming from the operator $\mathbf{16}_3\cdot \mathbf{16}_3\cdot\mathbf{10}$, where $\mathbf{16}_3$ contains the entire Standard Model (SM) $3$rd family and the Minimal Supersymmetric SM (MSSM) Higgs doublets are contained in the representation $\mathbf{10}$. Henceforth, we shall refer to this scenario simply as $t$-$b$-$\tau$ unification and omit the $\nu$, despite its coupling also unifying.  

In this work we study $t$-$b$-$\tau$ unification and assume its origin to be in a SUSY $\SO(10)$ GUT. Below the GUT scale, we take the effective theory to be a softly broken MSSM. In such a framework, GUT symmetry would impose relations between the soft breaking terms of the MSSM at the GUT scale. The attractive phenomenological feature of such a setup is that Yukawa unification with GUT-like boundary conditions for the soft terms results potentially in a predictive sparticle spectrum.

In the most direct ``vanilla'' approach, $\SO(10)$ symmetry would result in all the sfermion mass parameters to unify in a single value $m_{16}$, the soft Higgs masses to unify in $m_{10}$, universal gaugino masses $M_{1/2}$, and a universal factor $a_0$ for the proportionality between the Yukawa and $A$-matrices. The only other SUSY parameters in the theory would then be the ratio of the MSSM Higgs vacuum expectation values (VEVs) $\tan\beta$, and the sign of the coupling $\mu$ of the term $H_u\cdot H_d$ in the superpotential. It is known that for $t$-$b$-$\tau$ unification $\tan\beta$ has to be large ($\sim 50$) due to the top-bottom mass hierarchy $m_t\gg m_b$. Recall that with no SUSY threshold corrections, the Yukawa coupling ratio $y_\tau/y_b$ tends to run via renormalization group equations (RGEs) 
to a GUT value of $1.3$ (see e.g.~\cite{Antusch:2013jca}), and $\mu<0$ gives the correct sign in the threshold correction of $y_b$ to help lower this ratio to $1$, see e.g.~\cite{Antusch:2008tf,Antusch:2009gu}.
For this reason we consider $\mu < 0$ to be the better motivated setup for $t$-$b$-$\tau$ unification. Interestingly enough, fits to low energy data within this specific setup, at least to our knowledge, have not really been attempted, mostly due to the region being disfavored by RGE estimates showing no electroweak symmetry breaking (EWSB), to be discussed later. In this paper, we investigate this ``vanilla'' region and find it viable from the point of view of EWSB. Furthermore, we obtain good fits to the low energy Yukawa data and the SM Higgs mass, resulting in a predictive sparticle spectrum. The most striking feature of the entire setup is the prediction of a typically $\sim\mathrm{TeV}$ mass for the additional neutral and charged Higgses in the MSSM, a prediction which is now being tested by the LHC. The extra Higgs prediction is very sensitive especially to top-bottom unification, and is very hard to observe with a bottom-up approach, especially if the mass $m_{\AZ}$ is assumed \textit{a priori} as in some studies, e.g.~\cite{Elor:2012ig}.

To be more specific in what our setup achieves, and to put our results in context, it is necessary to survey the existing extensive literature on the topic of $t$-$b$-$\tau$ unification. Many early studies \cite{Kubo:1994xa,Rattazzi:1994bm,Allanach:1995ji,Kubo:1995zg,Murayama:1995fn,Rattazzi:1995gk,Bagger:1996ei,Tobe:2003bc,Carena:1995ud} predate the Higgs mass measurement in 2012, or even the top quark mass measurement in 1995. Beside considering the viability of Yukawa unification, they also had to contend  with predicting the top quark or Higgs mass, e.g.~\cite{Carena:1994bv,Polonsky:1994sv,Carena:1995ud}, or were considering naturalness based criteria~\cite{Baer:2012jp}.

In the literature, a number of important issues have been identified:
\begin{enumerate}
\item \textbf{The $\mu$ term: $\mu>0$ or $\mu<0$?}
\par
The Higgs connecting coupling $\mu$ from the superpotential is present in the potential $V$ of the Lagrangian only via $|\mu|$. Assuming no additional CP violation, $\mu\in\mathbb{R}$, so the choice of the sign of $\mu$ is free. 
\par
Historically, the case with $\mu>0$ was investigated far more in-depth, see \cite{Baer:2012jp,Baer:2008xc,Altmannshofer:2008vr,Baer:2012cp,Blazek:2001sb,Baer:2001yy,Blazek:2002ta,Anandakrishnan:2012tj,Ajaib:2013zha,Anandakrishnan:2014nea,Joshipura:2012sr,Baer:2008jn,Karagiannakis:2012sv,Shafi:2015lfa}.  
The $\mu<0$ case was studied in e.g.~\cite{Altin:2017sxx,Gogoladze:2010fu,Anandakrishnan:2013cwa,Gogoladze:2011ce}, while both cases of $\mathrm{Sign}(\mu)$ were considered in \cite{Baer:1999mc,Baer:2000jj,Auto:2003ys,Gogoladze:2011aa}.
\par
The main preference for $\mu>0$ in the literature stems from considerations of the anomalous magnetic moment of the muon $g_\mu-2$, see e.g.~motivation in \cite{Ajaib:2013zha}. This was measured to be above what the Standard Model predicts (see e.g.~PDG~\cite{Tanabashi:2018oca}), and $\mu>0$ would provide a SUSY contribution in the positive direction, potentially explaining the discrepancy. Despite this there are indications that a fit of $g_\mu-2$ for $\mu>0$ with universal gaugino masses is difficult to achieve in $\SO(10)$~\cite{Anandakrishnan:2012tj}.
\par 
The study of $\mu>0$ scenarios, typically within parametrization as close as possible to the constrained MSSM (CMSSM, a.k.a.~mSUGRA) with universal gaugino masses, furthermore showed that there is a preferred ``funnel'' region for the soft MSSM parameters \cite{Baer:2001yy}, and that the universal gaugino mass parameter should be quite small: $M_{1/2}\lesssim 500\,\mathrm{GeV}$~\cite{Blazek:2001sb}. Consequently, these scenarios prefer a light gluino $m_{\tilde{g}}\lesssim 450\,\mathrm{TeV}$ \cite{Baer:2008jn} and suggest an upper bound on the attainable gluino mass of around $m_{\tilde{g}}<2\,\mathrm{TeV}$~\cite{Anandakrishnan:2012tj}, a constraint coming from fitting the SM Higgs mass. Due to the non-observation of such low gluino mass scenarios at the LHC, the possibility of increasing its mass was investigated in subsequent works: it was found in~\cite{Baer:2012cp} that the gluino mass can be raised to $\mathrm{2}$-$3\,\mathrm{TeV}$ by relaxing the Yukawa unification to be approximate at a few $\%$ level, or to introduce a split in the squark mass parameters~\cite{Joshipura:2012sr}. Note that all these results are specific to the preferred soft parameter region for $\mu >0$.
\par
From the point of view of a fit to the data, however, it was already realized a long time ago that $\mu<0$ is preferred, see e.g.~\cite{Baer:2000jj,Gogoladze:2010fu}, since it gives the correct sign to the threshold corrections to the $y_b$ Yukawa coupling. Since the sign of the contribution to $g_{\mu}-2$ depends on $\mathrm{Sign}(\mu M_{2})$, see e.g.~\cite{Stockinger:2006zn}, this prompted a consideration of non-universal gaugino masses, see \cite{Gogoladze:2010fu,Anandakrishnan:2013cwa,Altin:2017sxx}, with $M_{2}<0$. Such boundary conditions can most conveniently be achieved by considering Yukawa unification within the context of the Pati-Salam symmetry instead of fully unified $\SO(10)$, see \cite{Kubo:1994xa,Gomez:2002tj,Gogoladze:2010fu,Karagiannakis:2012sv,Shafi:2015lfa,Altin:2017sxx} for various Pati-Salam setups and studies of Yukawa unification. Another possible approach to $g_{\mu}-2$ with $\mu<0$ is to 
only demand that the $g_\mu-2$ prediction is no worse than in the Standard Model, see~\cite{Gogoladze:2011ce}. This last case still considered non-universal gaugino masses due to EWSB considerations, see next point. 
\item \textbf{EWSB and the split between $m^{2}_{H_d}$ and $m^{2}_{H_u}$ at $\MGUT$}
\par
Another issue in Yukawa unification models important for their consistency turns out to be electroweak symmetry breaking. In a softly broken MSSM, a necessary condition for EWSB is to obtain $m^{2}_{H_u} < 0$ at the SUSY scale. This is typically automatically achieved by RGE running from $\MGUT$, where this parameter value is positive; the scenario where RG running triggers EWSB is referred to as \textit{radiative EWSB} (REWSB). Another necessary non-tachyonicity condition, however, also requires $m^{2}_{H_d}>m^{2}_{H_u}$ at the SUSY scale. Assuming the equality $m^{2}_{H_d}=m^{2}_{H_u}$ at the GUT scale, the $m^{2}_{H_d}$ is driven down faster than the  $m^{2}_{H_u}$ essentially due to the former having positive contributions to its beta function from both $y_b$ and $y_\tau$, while the latter has only contributions from $y_t$ (and potentially from $y_\nu$), cf.~\cite{Gogoladze:2011aa}.
\par
For this reason, most models in the literature introduce a split $m^{2}_{H_d}>m^{2}_{H_d}$ already at the GUT scale~\cite{
Baer:1999mc,
Baer:2000jj,
Baer:2001yy,
Blazek:2001sb,
Blazek:2002ta,
Baer:2009ie,
Gogoladze:2010fu,
Gogoladze:2011ce,
Anandakrishnan:2012tj,
Joshipura:2012sr,
Anandakrishnan:2013cwa,
Ajaib:2013zha,
Anandakrishnan:2014nea,
Altin:2017sxx}. 
The simplest way to achieve this is by imposing the split \textit{ad hoc}, which is called ``just so'' Higgs splitting and assumes $m^{2}_{H_{d,u}}=m_0^2\pm\Delta$ at the GUT scale, e.g.~\cite{Blazek:2001sb,Blazek:2002ta}, with the relative split amounting to $\sim 13\,\%$. An alternative mechanism to generate this split is by $D$-term splitting \cite{Murayama:1995fn,Baer:1999mc,Baer:2000jj,Baer:2001yy}, which also splits up the other soft scalar masses in a particular way due to $D$-term contributions to the masses. Attempts to avoid $m^{2}_{H_d}$ slipping below the value of $m^{2}_{H_u}$ have also been studied in the context of adding right-handed neutrinos or introducing a first/third scalar mass split in the GUT boundary conditions, see~\cite{Baer:2009ie}, both options essentially modifying the RGE beta functions for $m^{2}_{H_d}$ and $m^{2}_{H_u}$.
\par
The well known issue regarding REWSB with $m^{2}_{H_d}=m^{2}_{H_u}$ at the GUT scale has been studied in \cite{Carena:1994bv,Matalliotakis:1994ft,Olechowski:1994gm,Murayama:1995fn}, and reiterated later in e.g.~\cite{Ajaib:2013zha} based on an approximate expression for $m^{2}_{H_d}-m^{2}_{H_u}$ at low scales taken from \cite{Hempfling:1994sa}. It should be noted, however, that these papers use semi-analytic formulas for RGEs running from the GUT scale to the SUSY scale, which hold only approximately. In the context of the GUT boundary condition $m^{2}_{H_d}=m^{2}_{H_u}$, successful REWSB was achieved for the case of non-universal gaugino masses~\cite{Gogoladze:2011aa,Gogoladze:2014hca}, while the old arguments for the universal gaugino mass case are reiterated. On the other hand, successful REWSB was found for the case of CMSSM with $\mu<0$ in \cite{Auto:2003ys}, albeit with only approximate Yukawa unification due to their bottom-up approach of running Yukawa parameters.
\par
In contrast to most considerations in past works presented above, we find that exact Yukawa unification with universal gaugino mass terms and $m^{2}_{H_u}=m^{2}_{H_d}$ is in fact possible. We show this explicitly by performing the RGE running numerically; although we use 2-loop RGEs for the MSSM + soft terms for (most) results, the 1-loop RGE solutions already confirm this qualitative picture. While we agree with prior analyses that RG running just below the GUT scale causes $m^{2}_{H_u}>m^{2}_{H_d}$ in the running parameters, this relation reverses later by RG running a few orders of magnitude above the SUSY scale, thus achieving successful REWSB. This holds true at least in a large part of the soft parameter space. Crucially, however, the running value of $m^{2}_{H_d}-m^{2}_{H_u}$ is typically below $(1\,\mathrm{TeV})^2$ at the SUSY scale, causing the extra MSSM Higgs bosons to be the lightest part of the sparticle spectrum.
\vskip 1cm 
\item \textbf{Experimental constraints and considerations}\par
The most obvious type of prediction studied in Yukawa unification models is the MSSM spectroscopy, see \cite{Baer:1999mc,Baer:2008xc,Poh:2015wta,Gogoladze:2015tfa,Shafi:2015lfa} for studies which focus on this.
\par
Constraints on the masses and mixing of the SUSY partners come e.g.\ from  FCNC processes induced via SUSY loop effects~\cite{Altmannshofer:2008vr}. An important process studied in this regard is $b\to s\gamma$, see e.g.~\cite{Borzumati:1994te,Baer:2000jj,Chattopadhyay:2001mj,Auto:2003ys,Tobe:2003bc,Anandakrishnan:2012tj,Ajaib:2013zha,Anandakrishnan:2014nea,Anandakrishnan:2013cwa}, usually considered in the context of $B$ meson decays such as $B\to X_{s}\gamma$. Typically the most stringent constraint, however, comes from the meson decay $B_s\to\mu\mu$~\cite{Dermisek:2003vn,Karagiannakis:2012sv,Anandakrishnan:2012tj, Ajaib:2013zha,Anandakrishnan:2013cwa,Anandakrishnan:2014nea}.
\par
Two more observables that are not directly measured in accelerators have also received attention: the $g_\mu-2$ of the muon, see e.g.~\cite{Auto:2003ys,Altin:2017sxx} and \cite{Chattopadhyay:2001mj} in the $b$-$\tau$ context, and the relic abundance of the neutralino dark matter (DM), see \cite{Borzumati:1994te,Baer:2000jj,Dermisek:2003vn,Baer:2008jn,Dar:2011sj,Karagiannakis:2012sv,Ajaib:2013zha,Shafi:2015lfa}.
\par
Studies which fit GUT models to the experimental data usually consider some or all of these constraints. It was found in many specific realizations of Yukawa unification, however, that potential experimental tensions can usually be relieved by relaxing the demand for exact Yukawa unification and impose it only at a level of some $\%$. This essentially works due to relaxing constraints on the superpartner masses. Such scenarios have been dubbed ``quasi-unification'', see e.g.~\cite{Gomez:2002tj,Altmannshofer:2008vr,Dar:2011sj,
Karagiannakis:2011pb,Karagiannakis:2012sv,Gogoladze:2014hca,Shafi:2015lfa,Altin:2017sxx}. Alternative setups to improve fits have also been tried, such as splitting the $A$-terms \cite{Guadagnoli:2009ze}, considering $4$ Higgs doublets instead of $2$~\cite{Dutta:2018yos}, introducing certain extra vector-like fermions motivated by an $\mathrm{E}_6$ GUT context~\cite{Hebbar:2016gab}, or introducing an entire vector-like family of SM fermions~\cite{Dermisek:2018hxq}.
\end{enumerate}
In this paper, as motivated earlier, we consider $\mu<0$ and numerically find a good solution for REWSB despite the relation $m^2_{H_d}=m^{2}_{H_u}$ and universal gaugino masses. In the literature, as far as we are aware, the only case directly comparable with ours is in \cite{Auto:2003ys}, with the limitation that the SM Higgs mass was not yet measured at the time. One of the scenarios they consider successful (including EWSB) is the CMSSM (implying universal gaugino masses and no GUT split between $m^{2}_{H_d}$ and $m^{2}_{H_u}$) with $\mu<0$. They use, however, a bottom-up approach for Yukawa RGE, and therefore consider only the quasi-unification scenario with a parameters scan. They consequently do not find the low MSSM Higgs mass effect, since it is very sensitive to exact unification, as we show in this paper.

Given the effect of the low extra Higgses we study in this paper, the most acute experimental constraints would come from two possible sources. The first is the $B_s\to\mu^+ \mu^{-}$ decay, with the extra Higgs contribution estimated as, see e.g.~\cite{Rosiek:2012kx}, 
\begin{align}
		\mathcal{B}(B^{0}_{s}\to\mu^{+}\mu^{-})&\approx 5\cdot 10^{-7}\,\left(\frac{\tan\beta}{50}\right)^6\;\left(\frac{300\,\mathrm{GeV}}{M_A}\right)^4,
		\end{align}
compared to the PDG measured value of $(3.2\pm 0.7)\cdot 10^{-9}$~\cite{Tanabashi:2018oca}. The second constraint is the increasingly competitive LHC searches for ditau decays $\HZ/\AZ\to \tau^{+}\tau^{-}$ of the neutral MSSM Higgses, see~\cite{Aaboud:2017sjh,Sirunyan:2018zut}, with current bounds implying $m_{A}\gtrsim 1.5\,\mathrm{TeV}$ (for $\tan\beta=50$). Given this most recent estimate and future trends of bounds, we find the ditau search to be comparable or more stringent than the $B_s\to\mu^{+}\mu^{-}$ process; we thus focus only on the ditau decay in this paper for simplicity. The other parts of the SUSY spectrum in our setup are heavy, larger than $4\,\mathrm{TeV}$ for gluinos and squarks, far above the present ATLAS and CMS bounds but within reach of future colliders such as the FCC-hh or SppC. 

The organization of the paper is as follows: in Section~\ref{section:MSSM-Higgs-masses} we introduce our notation and conventions, and analyze the salient points regarding EWSB and the masses of the extra Higgs bosons in the MSSM. In Section~\ref{section:RGE-analysis}, we perform an RGE analysis of the quantity $m^{2}_{H_d}-m^{2}_{H_u}$ relevant for both those aspects and perform a sensitivity analysis to deformations of various parameter relations around an example point. In Section~\ref{section:mass-scale}, we perform a more general investigation of the CMSSM parameter space and show that the masses of the extra Higgses are predicted to be low in general. Finally, in Section~\ref{section:experimental-constraints}, we analyze how constraints from the LHC challenge exact Yukawa unification and how a quasi-unification scenario helps in this regard. Then we conclude. For completeness, we also include two appendices. In Appendix~\ref{appendix:general_rge} the general $1$-loop RGEs for a softly broken MSSM with right-handed neutrinos are presented. In Appendix~\ref{appendix:rge-simplified} a simplified version of the RGEs neglecting the Yukawa couplings of the first 2 families is given.

\section{MSSM, EWSB and the Higgs masses --- Conventions \label{section:MSSM-Higgs-masses}}
In this section we briefly summarize the situation with EWSB and Higgs masses in the MSSM, which facilitates a more detailed analysis in later sections. Throughout the paper we use the right-left (RL) convention for the Yukawa matrices as in REAP~\cite{Antusch:2005gp} and SusyTC~\cite{Antusch:2015nwi}. A short note on the relation to other conventions can be found in Appendix~\ref{appendix:general_rge}.

We consider the MSSM extended by right-handed neutrinos as the effective theory below the GUT scale. The matter content consists of chiral multiplets of the group $G_{321}\equiv \mathrm{SU}(3)\times\mathrm{SU}(2)\times\mathrm{U}(1)$. The ``fermionic'' sector consists of the chiral multiplets
\begin{align}
	Q_i&\sim (3,2,+\tfrac{1}{6}),&
	L_i&\sim (1,2,-\tfrac{1}{2}),\nonumber\\
	U_i^c&\sim (\bar{3},1,-\tfrac{2}{3}),&
	E_i^c&\sim (1,1,+1),\\
	D_i^c&\sim (\bar{3},1,+\tfrac{1}{3}),&
	N_i^c&\sim (1,1,\phantom{+}0),\nonumber
\end{align}
\noindent
where the family index $i$ goes from $1$ to $3$. The Higgs sector consists of
\begin{align}
	H_{u}&\sim (1,2,+\tfrac{1}{2}),&
	H_{d}&\sim (1,2,-\tfrac{1}{2}).
\end{align}
As mentioned above, we use the RL convention for the Yukawa matrices $\YU$, $\YD$, $\YE$, $\YN$ in the superpotential $W$ for the MSSM:
\begin{align}
\begin{split}
	W_{\text{MSSM}}&=-(\YU)_{ij}\,U_i^c\;H_u\cdot Q_j+(\YD)_{ij}\,D^{c}_{i}\;H_d\cdot Q_j\\
	&\quad + (\YN)_{ij}\, N^c_i\;H_u\cdot L_j + (\YE)_{ij}\,E^c_i\;H_d\cdot L_j +\tfrac{1}{2}(\mathbf{M}_\nu)_{ij}\,N^c_i\,N^c_j\\
	&\quad + \mu\,H_u\cdot H_d.\\
\end{split} \label{eq:MSSM-superpotential}
\end{align} 
The indices $i$ and $j$ are family indices, the $\mathrm{SU}(2)$ contractions between doublets are denoted by a dot and defined by $\Phi\cdot\Psi\equiv \epsilon_{ab}\Phi^a \Psi^b$ with $\epsilon_{12}=-\epsilon_{21}=1$, while the $\mathrm{SU}(3)$ indices are suppressed. Also note that a left-chiral superfield $\Phi^c$ contains the charge conjugated fermion field $\psi^\dagger$, as well as the conjugated complex scalar field $\tilde\phi_R^*$.

The soft-breaking terms consist of gaugino mass terms, the scalar trilinear $A$-terms, the scalar soft-mass terms, and the $b$-term:  
\begin{align}
	-\mathcal{L}_{\text{soft}}&= -\frac{1}{2} \left(\sum_{a=1}^{8} M_3 \lambda_3^a\lambda_3^a+\sum_{b=1}^{3} M_2 \lambda_2^b\lambda_2^b+M_1 \lambda_1 \lambda_1\right) + h.c. \nonumber \\
&+ (\M^2_{Q})_{ij}\;\tilde Q_i^\dagger \tilde Q_j + (\M^2_{L})_{ij}\;\tilde L_i^\dagger \tilde L_j + (\M^2_u)_{ij}\;\tilde u_{R_i}^* \tilde u_{R_j} + (\M^2_d)_{ij}\; \tilde d_{R_i}^* \tilde d_{R_j} + (\M^2_e)_{ij}\;\tilde e_{R_i}^* \tilde e_{R_j} + (\M^2_{\nu})_{ij}\;\tilde\nu_{R_i}^* \tilde \nu_{R_j} \nonumber \\
& + (\AE)_{ij}\; \tilde e_{R_i}^* H_d \cdot \tilde L_j + (\AN)_{ij}\; \tilde\nu_{R_i}^* H_u \cdot \tilde L_j +  (\AD)_{ij}\; \tilde d_{R_i}^* H_d \cdot \tilde Q_j - (\AU)_{ij}\; \tilde u_{R_i}^* H_u \cdot \tilde Q_j + h.c. \nonumber \\
& + m^2_{H_u} |H_u|^2 + m^2_{H_d} |H_d|^2 + (b\;H_u \cdot H_d + h.c.)\;.\label{eq:MSSM-soft-terms}
\end{align}
We labeled the $\SU(3)_C$, $\SU(2)_L$ and $\mathrm{U}(1)_Y$ gauginos by $\lambda_3^a$, $\lambda_2^b$ and $\lambda_1$, respectively. The tildes above the fields indicate the scalar component of the superfield, with the exception of $H_u$ and $H_d$, which also indicate scalar parts.

The neutral components of $H_u$ and $H_d$ each acquire an EW breaking VEV:
\begin{align}
	v_u&:=\langle H_u^0\rangle,&v_d&:=\langle H_d^0\rangle,\label{eq:EW-VEVs}
\end{align}
which --- motivated by EW symmetry breaking in the SM --- are parametrized by
\begin{align}
	\tan\beta&\equiv v_u/v_d,&v_u^2+v_d^2&\equiv v^2=(174\,\mathrm{GeV})^2.
\end{align}
This leaves $\tan\beta$ as the only free parameter, and $v_u,v_d\in\mathbb{R}$.

Minimization of the potential 
\begin{align}
	V&=\sum_{\phi}\left|\frac{\partial W}{\partial \phi}\right|-\mathcal{L}_{\text{soft}}
\end{align}
with respect to the electrically neutral components $H_u^0$ and $H_d^0$ of $\mathrm{SU}(2)$ doublets leads to a (tree-level) vaccum solution
\begin{align}
    2|\mu|^2_{\text{tree}}&=\frac{m^2_{H_d}-m^2_{H_u}}{\sqrt{1-\sin^2(2\beta)}}-m^2_{H_u}-m^2_{H_d}-m^2_Z,\label{eq:vacuum-mu}\\
	b_{\text{tree}}&=\tfrac{1}{2}\sin(2\beta)\left(m^2_{H_u}+m^2_{H_d}+2|\mu|^2_\text{tree}\right).\label{eq:vacuum-b}
\end{align}
Note that we have solved the vacuum equations for the superpotential parameter $|\mu|^2$ and the soft parameter $b$, while treating the unknown VEVs $v_u$ and $v_d$ as independent variables, appearing implicitly via $v_u/v_d=\tan\beta$. In the large $\tan\beta$ regime, we can make the approximation
\begin{align}
|\mu|^2_{\text{tree}}&\approx -m^{2}_{H_u}-\tfrac{1}{2}\,m^{2}_Z,\label{eq:approx-mu}
\end{align}
implying that a solution to EWSB (at tree level) is possible only if the soft mass parameter is negative at the energy scale of computation, i.e.~$m^{2}_{H_u}<0$ at the SUSY scale.

After EW symmetry breaking, $3$ real scalar degrees of freedom in $H_u$ and $H_d$ become part of the longitudinal components of the massive gauge bosons $W^{\pm}$ and $Z^{0}$ via the Higgs mechanism, leaving $5$ real degrees of freedom to be physical. We label them in the standard way by $\hZ$, $\HZ$, $\AZ$, $\HP$ and $\HM$, where their superscripts denote their EM charge. The low mass Higgs at $125\,\mathrm{GeV}$ is denoted by $\hZ$, while $\HZ$ and $\AZ$ denote heavier neutral scalars with even and odd parity $P$, respectively. We get the following well-known expressions for their tree-level masses:
\begin{align}
    m^2_{\AZ}&=2|\mu|^2+m^2_{H_u}+m^2_{H_d},\label{eq:mass-A0}\\
    m^2_{\hZ}&=\tfrac{1}{2}\,m^2_{\AZ}+\tfrac{1}{2}\,m_Z^2-\tfrac{1}{2}\,\sqrt{(m^2_{\AZ}-m^2_{Z})^2+4m^2_Z\,m^2_{\AZ}\,\sin^2(2\beta)},\\
    m^2_{\HZ}&=\tfrac{1}{2}\,m^2_{\AZ}+\tfrac{1}{2}\,m_Z^2+\tfrac{1}{2}\,\sqrt{(m^2_{\AZ}-m^2_{Z})^2+4m^2_Z\,m^2_{\AZ}\,\sin^2(2\beta)},\\
    m^2_{\HPM}&=m^2_{\AZ}+m^2_{W}.\label{eq:mass-HPM}
\end{align}
Considering the regime $m_{\AZ}^2\gg m_Z^2,m_W^2$ leads in leading order to
\begin{align}
m^2_{\HZ}&\approx m^2_{\AZ}\;(1+\sin^2(2\beta)\;m^2_Z/m^2_{\AZ}),
\end{align}
showing that all extra Higgs particles $\HZ$, $\AZ$ and $\HPM$ are near the scale $m^2_{\AZ}$. The scale of $m^2_{\AZ}$ in turn depends on the vacuum solution for $|\mu^2|$; combining Eq.~\eqref{eq:mass-A0} and \eqref{eq:vacuum-mu} gives the tree level value

\begin{align}
m^2_{\AZ,\text{tree}}&=\frac{m^2_{H_d}-m^2_{H_u}}{\sqrt{1-\sin^2(2\beta)}}-m^2_Z\nonumber\\
&=(m^2_{H_d}-m^2_{H_u})\frac{\tan^2\beta +1}{\tan^2\beta-1}-m^2_{Z}.\label{eq:mass-A0-tree}
\end{align}
We see that, crucially, the scale $m^2_{\AZ}$ depends on the difference $m^2_{H_d}-m^2_{H_u}$ of the mass-square soft parameters. In the large $\tan\beta$ regime, this approximates to
\begin{align}
m^2_{\AZ,\text{tree}}&\approx m^2_{H_d}-m^2_{H_u}-m^2_{Z}, \label{eq:approx-ma}
\end{align}
so that a non-tachyonic tree-level mass for $\AZ$ requires $m^{2}_{H_d}-m^{2}_{H_u}>m^{2}_Z$, implying also $m^{2}_{H_d} > m^{2}_{H_u}$ as a necessary condition.

We now briefly turn to a discussion of the scale of masses at $1$-loop level.
The vacuum solutions at $1$-loop become (see \cite{Pierce:1996zz,Antusch:2015nwi})

\begin{align}
|\mu|^2_{\text{1-loop}} &= \frac{1}{2} \left(\tan(2 \beta) \left(\hat{m}^2_{H_u} \tan\beta
      - \hat{m}^2_{H_d}\cot\beta \right)- \hat{m}_Z^2 \right)\;,\label{eq:vacuum-mu-1loop}\\
b_{\text{1-loop}}&=\frac{1}{2} \left(\tan(2 \beta) \left(\hat{m}^2_{H_u} - \hat{m}^2_{H_d} \right)- \hat{m}_Z^2 \sin(2 \beta)\right).\label{eq:vacuum-b-1loop}
\end{align}
 
The hatted quantities, including $\hat{m}_W^2$ for later convenience, are defined by 
\begin{align}
\begin{split}
\hat{m}^2_{H_u}&:= m^2_{H_u} - t_u,\\
\hat{m}^2_{H_d}&:= m^2_{H_d} - t_d,\\
\hat{m}_Z^2&:= m_Z^2 + \Re\left[\Pi_{ZZ}^T(m_Z^2)\right],\\
\hat{m}_W^2&:= m_W^2 + \Re\left[\Pi_{WW}^T(m_W^2)\right],\\
\end{split}\label{eq:definitions-1loop}
\end{align}
where $t_u$ and $t_d$ are $1$-loop tadpole expressions, and $\Pi_{ZZ}^T$ and $\Pi_{WW}^T$ are the transverse $Z$ and $W$-boson $1$-loop self-energies. The hatted masses $\hat{m}^2_{Z}$ and $\hat{m}^2_{W}$ are the 1-loop masses computed in the $\overline{\mathrm{DR}}$ renormalization scheme. Their explicit expressions can be found in \cite{Antusch:2015nwi} and will not be reproduced here. For a consistent loop calculation, the quantities in the expressions for $1$-loop corrections can be taken to be the parameters at tree-level.

When the quantities in the superpotential of Eq.~\eqref{eq:MSSM-superpotential} are complex, the neutral states $\hZ$, $\HZ$ and $\AZ$ mix: with the $1$-loop correction, the masses may no longer be CP eigenstates. We shall not be considering complex phases in the SUSY parameters, so this complication need not be considered.

Due to the breaking of CP symmetry at next to leading order in the general case, rather than the mass $m^2_{\AZ,\text{tree}}$ from \eqref{eq:mass-A0-tree}, a more convenient quantity to consider is the mass of the charged Higgses $\HPM$, since the charged Higgses $\HPM$ have no other states to mix with. The expression at $1$-loop order 
for the mass of $\HP$ is known to be
\begin{align}
m^2_{\HP,\text{1-loop}}  &= \frac{\left(\hat{m}^2_{H_d}-\hat{m}^2_{H_u}\right)}{-\cos(2 \beta)} -\hat{m}_Z^2 +\hat{m}_W^2 + t_d\sin^2\beta +t_u\cos^2\beta - \Re\left[\Pi_{\HP\HM}\left(m_{\HP}\right)\right],\label{eq:extra-Higgs-1loop}
\end{align}
with $\Pi_{\HP\HM}$ denoting the self-energy of $\HPM$, see \cite{Antusch:2015nwi}.
Since all the $1$-loop corrections have $1/16\pi^2$ suppression factors, the dominant contribution determining the overall scale should come from the term $m^2_{H_d}-m^2_{H_u}$, unless this quantity is unexpectedly small. Note that the prefactor $-1/\cos(2\beta)\to 1$ as $\tan\beta\to\infty$.

We conclude this section by collecting together the stated reasons for the importance of the quantity $m^2_{H_d}-m^2_{H_u}$. First, EWSB requires $m^2_{H_d}-m^{2}_{H_u} > 0$ alongside $m^{2}_{H_u}<0$ to work at tree level. Second, the expression $m^2_{H_d}-m^2_{H_u}$ is a good proxy for the mass scale of the extra Higgs states, at least when $m^{2}_{H_d}-m^{2}_{H_u}\gg m^{2}_Z$ and $\tan\beta$ is large. In Eq.~\eqref{eq:extra-Higgs-1loop}, if the expression $m^2_{H_d}-m^2_{H_u}$ is roughly of the same scale as the soft parameters, the 1-loop contributions are expected to be subdominant due to the $1/16\pi^2$ suppression factor; if $m^2_{H_d}-m^2_{H_u}$ is unexpectedly small, loop contributions might be of comparable size or even dominate. 

\section{RGE analysis of $m^2_{H_d}-m^2_{H_u}$ in $t$-$b$-$\tau$ unification\label{section:RGE-analysis}}
As a first step in assessing models with Yukawa unification and $\SO(10)$ boundary conditions for soft parameters, we study the RG running of the quantity $m^2_{H_d}-m^2_{H_u}$. This quantity must be positive at the SUSY scale, a feature crucial for EWSB, and its magnitude sets the mass scale of the extra MSSM Higgs states $\HZ$, $\AZ$ and $\HPM$, as was discussed in Section~\ref{section:MSSM-Higgs-masses}. An often cited requirement in the literature for REWSB to occur is a split in the GUT scale boundary conditions for $m^{2}_{H_d}$ and $m^{2}_{H_u}$, see Section~\ref{section:introduction} and references therein. We show here, however, that such a split is not necessary, since we obtain $m^2_{H_d}-m^2_{H_u}>0$ at the SUSY scale regardless. The value of this difference, however, is small compared to the magnitude of each term, implying low lying extra Higgs states in the MSSM, an effect that we show to be especially sensitive to $t$-$b$ unification.

To facilitate the RGE analysis, we make use of simplified RGEs at 1-loop and CMSSM boundary conditions, as explained in separate subsections below. Note that these simplifications are specific to this section of the paper and do not change the general conclusions, confirmed by comprehensive analyses in later sections by use of 2-loop RGEs and $\SO(10)$ motivated boundary conditions. The analysis of the simplified case nevertheless gives valuable insights into EWSB and the low spectrum of the extra MSSM Higgses, confirming that this striking feature can be understood as an RGE effect, and is seen already at 1-loop order.

\subsection{The simplified boundary conditions --- CMSSM}
In this section we make a slight simplification and consider the CMSSM boundary conditions (see e.g.~\cite{Martin:1997ns}) as the default scenario, instead of the $\SO(10)$ motivated split in the sfermion and Higgs soft masses to be studied later. We also study how RG running changes under various deformations of the default CMSSM boundary conditions, obtaining a number of important conclusions applicable to the more general scenario beyond CMSSM.

More explicitly, we assume the following for the RGE analysis in this section:
\begin{itemize}
	\item The boundary conditions are set at a high energy: $\MGUT=2\cdot 10^{16}\,\mathrm{GeV}$.
	\item The MSSM is extended by right-handed neutrinos at a scale $M_R$, with $M_R\leq\MGUT$, below which they are integrated out.
	\item The boundary conditions of the soft parameters are those of CMSSM:
		\begin{align}
		m^2_{H_x}\atMGUT&=m_0^2,\quad x\in\{u,d\};\label{eq:RGE-boundary-begin}\\
		\M^2_{x}\atMGUT&=m_0^2 \;\mathbf{1},\quad x\in\{Q,L,u,d,e,\nu\};\\
		M_i\atMGUT&=M_{1/2},\quad i\in\{1,2,3\};\\
		\mathbf{A}_x\atMGUT&=a_0\;\mathbf{Y}_x\atMGUT,\quad x\in\{u,d,e,\nu\}.\label{eq:RGE-boundary-end}
		\end{align}
		The RGE boundary conditions for the soft parameters are thus parametrized by the $3$ CMSSM parameters $m_0^2$, $M_{1/2}$ and $a_0$. 
	\item Unification of 3rd family Yukawa couplings at the scale $M_{\text{GUT}}$:
		\begin{align}
		y_\tau\atMGUT=y_b\atMGUT=y_\tau\atMGUT=y_\nu\atMGUT.	
		\end{align}
\end{itemize}  

The above assumptions are a simplified version of the ``$\SO(10)$ boundary conditions'' with only one soft scalar mass parameter $m_0$ and with universal sfermion soft matrices (typical leading order pattern in ``flavored GUTs'' with family symmetry): the constraints are implied in the unification of all fermion sectors, and $t$-$b$-$\tau$ unification arises in the simple case when the Yukawa contribution to the $3$rd family of $16_F$ comes from the $16_{F3}\cdot 16_{F3}\cdot 10_H$ operator in $\SO(10)$. 
We note that although the stated class of $\SO(10)$ models gives rise to the MSSM setup described below $\MGUT$, we do not necessarily commit to a particular $\SO(10)$ UV completion. In this context, we would also like to remark that the exact Yukawa unification will be subject to model-dependent corrections such as e.g.\ GUT threshold corrections, which however depend on the details of the UV completion. We will study the effects of such perturbations of the scenario later in the paper.

\subsection{The simplified $1$-loop RGE}
The complete set of RGEs for the neutrino-extended and softly-broken MSSM are given in Appendix~\ref{appendix:general_rge} (also cf.~\cite{Antusch:2015nwi}). The full RGEs can be simplified by eliminating some  degrees of freedom which are either numerically irrelevant or unnecessary for our considerations. In the quark sector, for example, there is little mixing, and the Yukawa matrices in both quark sectors as well as the charged lepton sector have hierarchical masses. A good approximation is therefore to consider only the $3$rd family of fermions. Also, we assume family universality in all sfermion mass matrices at the GUT scale. 

To simplify the RGE, we consider the minimal amount of variables consistent with the above assumptions. It turns out that the following $28$ variables in the RGEs are required:
\begin{itemize}
\item The $3$ gauge couplings $g_1$, $g_2$ and $g_3$.
\item The $3$ gaugino mass parameters $M_1$, $M_2$, $M_3$.
\item $4$ Yukawa couplings of the 3rd family $y_t$, $y_b$, $y_\tau$, $y_{\nu}$.
\item The $4$ $A$-term factors $a_u$, $a_d$, $a_e$, $a_{\nu}$, so that $\mathbf{A}_x=a_x\,\mathbf{Y}_x$ with $x\in\{u,d,e,\nu\}$.
\item The $6\times 2+2$ soft mass parameters: $m^2_{x_i}$, where $x\in\{Q,L,u,d,e,\nu\}$ and $i\in\{1,3\}$ are independent, and the Higgs mass parameters $m^2_{H_d}$ and $m^2_{H_u}$. The case $i=2$ does not have to be studied separately since, in our setup, the $i=2$ quantities have exactly the same running and boundary conditions as those for $i=1$.
\end{itemize}

The resulting simplified 1-loop RGE are presented in Appendix~\ref{appendix:rge-simplified}, which contains also more details on the above variables, cf.~Eq.~\eqref{eq:RGE-ansatz-begin}--\eqref{eq:RGE-ansatz-end}. Making use of the RGEs from Appendix~\ref{appendix:rge-simplified}, the running of the expression $m^2_{H_d}-m^2_{H_u}$ is then determined to be
\begin{align}
    \begin{split}
        c_1\,\DT(m^2_{H_d}-m^2_{H_u})&=6 |y_b|^2 \left(|a_d|^2+m^2_{H_d}+m^2_{Q_3}+m^2_{d_3}\right)-6 |y_t|^2\left(|a_u|^2+m^2_{H_u}+m^2_{Q_3}+m^2_{u_3}\right)+\\
        &+2 |y_\tau|^2 \left(|a_e|^2+m^2_{H_d}+m^2_{L_3}+m^2_{e_3}\right)-2 |y_\nu|^2\left(|a_\nu|^2+m^2_{H_u}+m^2_{L_3}+m^2_{\nu_3}\right)-\\
& -\tfrac{6}{5} g_1^2 S,
    \end{split}\label{eq:RGE-UD}
\end{align}
where $c_1$ is the loop factor and $S$ is a linear combination of soft masses:
\begin{align}
	c_1&:=16\pi^2,\label{eq:loop-factor}\\
	S&:=m^2_{H_u}-m^2_{H_d}+2m^2_{Q_1}+m^2_{Q_3}-2m^2_{L_1}-			m^2_{L_3}-4m^2_{u_1}-2m^2_{u_3}+2m^2_{d_1}+m^2_{d_3}+2m^2_{e_1}+m^2_{e_3}.
\end{align}
We see that the first $4$ terms of the result in Eq.~\eqref{eq:RGE-UD} are analogous to each other, the quantities in the terms correspond respectively to the particles $b$, $t$, $\tau$ and $\nu_\tau$ (and their superpartners). Each term contains the modulus-squared of its Yukawa coupling, and the factor next to it contains a modulus-squared of the appropriate $A$-term factor, as well $3$ more terms with the soft masses of particles present in the corresponding superpotential Yukawa term. The $b$ and $t$ terms have an additional numerical factor $3$ compared to $\tau$ and $\nu$ due to the $3$ possible $\mathrm{SU}(3)$ colors they can take. Crucially, the terms also come into the RG beta function with different signs, so it may happen that they cancel. Below the right-handed neutrino mass scale $M_R$, the $\nu$ term  vanishes. The boundary conditions imply that at exactly $\MGUT$, the last term vanishes due to $S=0$, and the $b$ and $t$ terms cancel each other, and as well as the $\tau$ and $\nu$ terms, such that we have 
\begin{align}
(m^2_{H_d}-m^2_{H_u})\atMGUT=\tfrac{d}{dt}(m^2_{H_d}-m^2_{H_u})\atMGUT&=0.
\end{align}
As already stated, the scale of the masses of the extra MSSM Higgs bosons will be determined by 
\begin{align}
(m^2_{H_d}-m^2_{H_u})\atMSUSY.
\end{align}
This same quantity must be positive at low energies also for successful EWSB. It is computed numerically by solving the RGE differential equations of Appendix~\ref{appendix:rge-simplified}. We shall often allude to Eq.~\eqref{eq:RGE-UD} for a better understanding of the numerical results, which we now consider.

\subsection{Numerical RGE results}

We now investigate the RGE properties of the system numerically. To do this as explicitly as possible, we take an example parameter point, whose neighborhood we study. We stress that the conclusions of the RGE behavior in this section nevertheless hold generally, i.e.~different example points of Yukawa unification at high energies and consistent with experimental data at low energies yield the same qualitative conclusions, which we checked explicitly by considering different parameter points. Furthermore, we identify the underlying reasons for certain RG behaviors throughout this section, and the generality (where applicable) is also confirmed by results in later sections.

We take the following boundary values for the parameters at the scale $\MGUT=2.0\cdot 10^{16}\,\mathrm{GeV}$:

\begin{align}
g_1(\MGUT)&=0.7044,\label{eq:example-point-begin}\\
g_2(\MGUT)&=0.6965,\\
g_3(\MGUT)&=0.6980.\label{eq:example-point-gauge-end}
\end{align}
\begin{align}
\tan\beta&=51,\\
\mathrm{sign}(\mu)&=-1,\\
m_0^2(\MGUT)&=(2400\,\mathrm{GeV})^2,\\
M_{1/2}(\MGUT)&=3700\,\mathrm{GeV},\\
a_0(\MGUT)&=-3200\,\mathrm{GeV}.
\end{align}
\begin{align}
y_0:=y_t(\MGUT)=y_b(\MGUT)=y_\tau(\MGUT)=y_{\nu}(\MGUT)=0.483.
\end{align}
\begin{align}
M_R(\MGUT)&=2.0\cdot 10^{16}\,\mathrm{GeV}.\label{eq:example-point-end}
\end{align}
The gauge coupling $g_1$ is given in the GUT normalization, and $M_R$ is the mass of the added right-handed neutrino. The above values are to be understood as boundary conditions for the RGE in Appendix~\ref{appendix:rge-simplified}. At the scale $M_R$, the right-handed neutrino is integrated out; below this threshold, the RGE are corrected by removing all terms containing $y_{\nu}$. For the example point under consideration, we have taken $M_R=\MGUT$ so that by default no effects arise due to the right-handed neutrinos, since the $y_{\nu}$ term with the large 3rd family neutrino Yukawa coupling is removed already at the GUT scale; its effect is studied separately below.

The values of the gauge couplings at the GUT scale are taken from the high-energy data provided by~\cite{Antusch:2013jca}, which uses $2$-loop RGEs and takes the SUSY scale at $3\,\mathrm{TeV}$; note that their values are consistent with a typical unified gauge coupling value of $\approx 0.7$.

The overall scale of the soft parameters $m_0$, $M_{1/2}$ and $a_0$ has been taken at the order of a few $\mathrm{TeV}$, which tends to be the preferred scale for the fits to low energy data, as will be seen in the next sections. Also, the main effect we are after in this paper is that the extra MSSM Higgs particles are unexpectedly light compared to the SUSY scale, for example $\lesssim 1\,\mathrm{TeV}$; this effect will be obscured if the SUSY scale is also taken to be lighter than $1\,\mathrm{TeV}$, as used to be popular in past SUSY studies. The few $\mathrm{TeV}$ scale for sparticles is compatible with (as of yet) non-observation of SUSY particles at the LHC.

Note that the chosen point is such that it gives the correct 3rd generation Yukawa couplings $y_t$, $y_b$ and $y_\tau$ at the scale $M_Z$ in the $\overline{\text{MS}}$ scheme,
\begin{align}
y_t(M_Z)&=0.9861,\\
y_b(M_Z)&=1.63\cdot 10^{-2},\\
y_\tau(M_Z)&=1.003\cdot 10^{-2},
\end{align}
based on the data from~\cite{Antusch:2013jca}. An intuitive qualitative description of how the GUT scale parameters control the fit of the 3rd family Yukawa parameters is the following: 
\begin{itemize}
\item The value $y_0$ controls the overall scale of the $3$ Yukawa couplings, and needs to have the value $y_0\approx 0.5$. 
\item The effect of the soft parameters $m_0^2$, $M_{1/2}$ and $a_0$ is to control the SUSY spectrum, through which SUSY threshold effects give the correct ratio $y_\tau/y_b$. 
\item The quantity $\tan\beta$ controls for the ratio $y_t/y_b$ (alongside SUSY threshold corrections). Low energy data demands a large value of $\tan\beta\approx 50$, a well-known feature of MSSM based $t$-$b$-$\tau$ unification models.  
\end{itemize}

We plot the running under $1$-loop RGE from Appendix~\ref{appendix:rge-simplified} for the various quantities of the MSSM, with the boundary conditions at $\MGUT$ given by the example parameter point in Eq.~\eqref{eq:example-point-begin}-\eqref{eq:example-point-end}. We shall also investigate the effect of changing one feature of the boundary conditions at a time, understanding its impact; note that we do not evaluate the worsening of the fit to low energy data under such a deformation, since we are for now interested only in the (numerical) effect on the RGE running. We plot quantities in the range $[\MSUSY,\MGUT]$; note that the lower scale is the SUSY scale, since that is the scale where the sparticle spectrum is computed. This scale is also where a match between the SM and MSSM theories is performed, and it is taken to be the geometric mean of the masses of the two stops (computed for our example point using SusyTC~\cite{Antusch:2015nwi} to be $\MSUSY=5901\,\mathrm{GeV}$). While we used a custom computer code for RGE running based on Appendix~\ref{appendix:rge-simplified} for greater control, the results were compared and confirmed with SusyTC when applicable.

The RGE running of the system,  based on the results of the example point, turns out to have the following properties:

\begin{enumerate}
	
	\item \textbf{Running of gauge and Yukawa couplings, gaugino masses and the $A$-terms}\par
	The RGE running of the gauge couplings, Yukawa couplings, gaugino mass parameters, as well as the the $A$-term factors $a_x$ from Eq.~\eqref{eq:a-term-factor} is shown in Figure~\ref{Figure:RGE-gMYA}. 
	\par
	As always in the MSSM, each of the gauge couplings evolves independently from other quantities (at $1$-loop level); the couplings approximately meet at  $\sim 0.7$, and their running values are determined; when the renormalization scale $\mu_r$ decreases to low energies, $g_3$ runs upwards and $g_1$ and $g_2$ run downwards, see Eq.~\eqref{eq:RGE-simple-g}, due to the signs of MSSM beta coefficients $\beta_3<0$ and $\beta_1,\beta_2 >0$ from Eq.~\eqref{eq:RGE-beta-coefficients}.
	\par	
	The running of gaugino mass parameters, according to Eq.~\eqref{eq:RGE-simple-M}, is influenced by the gauge couplings. It is the differences in gauge couplings which drive the gaugino mass-parameter differences from a common boundary point $M_{1/2}$ at $\MGUT$. This explains why the gluino mass parameter $M_3$ increases when approaching $\MSUSY$, while $M_1$ and $M_2$ decrease, but all are at a scale of $2\,\mathrm{TeV}$ or higher.
	\par 
	The RGEs of the Yukawas have two competing contributions to the beta functions, cf.~\eqref{eq:RGE-simple-t}--\eqref{eq:RGE-simple-tau}: a positive contribution from the Yukawas themselves, and a negative contribution from gauge bosons (terms proportional to $g_i^2$). The Yukawa couplings can then rise or fall with smaller $\mu_r$, depending on whether the gauge or Yukawa contributions to the beta function are dominant, respectively. 
	\par 
	The 3rd family Yukawa parameters $y_t$ and $y_b$ rise with lower scale $\mu_r$ essentially due to the relatively large negative $g_3^2$ term from the gluons, while $y_\tau$ stays mostly flat, since realistic unified values of the gauge couplings of $\approx 0.7$ and Yukawa couplings of $\approx 0.5$ give the Yukawa and gauge contributions approximately equal. The difference between the top and bottom Yukawa, on the other hand, is small and is 
	essentially driven by the $|y_\tau|^2$ term in $\beta({y_b})$ and the difference in the $g_1^2$ terms in $\beta(y_t)$ and $\beta(y_b)$, see Eq.~\eqref{eq:RGE-simple-t} and \eqref{eq:RGE-simple-b}. This ensures a small relative difference $y_t-y_b$, with $y_t>y_b$ at all energies; the very different values of $y_t$ and $y_b$ at $M_Z$, as implied by the different masses of the $t$ and $b$ quarks, 
must thus come from the MSSM to SM matching at $\MSUSY$, implying a large $\tan\beta$ of around $50$. 
	\par
	The RGEs for the $A$-term factors are given in Eq.~\eqref{eq:RGE-simple-au}--\eqref{eq:RGE-simple-ae}. We can see that the difference between $a_u$ and $a_d$ is essentially driven by the difference between $y_t$ and $y_b$, as well as the $|y_\tau|^2$ and $g_1^2$ terms, which essentially already drive the $y_t$ and $y_b$ difference, as discussed earlier. For this reason, there is again only a small deviation between $a_u$ and $a_d$. The slope of $a_e$ in absolute terms is smaller due to no gluino related terms, and because of smaller numerical factors in front of the Yukawa terms.
	\begin{figure}[htb]
		\begin{center}
		\includegraphics[width=7.8cm]{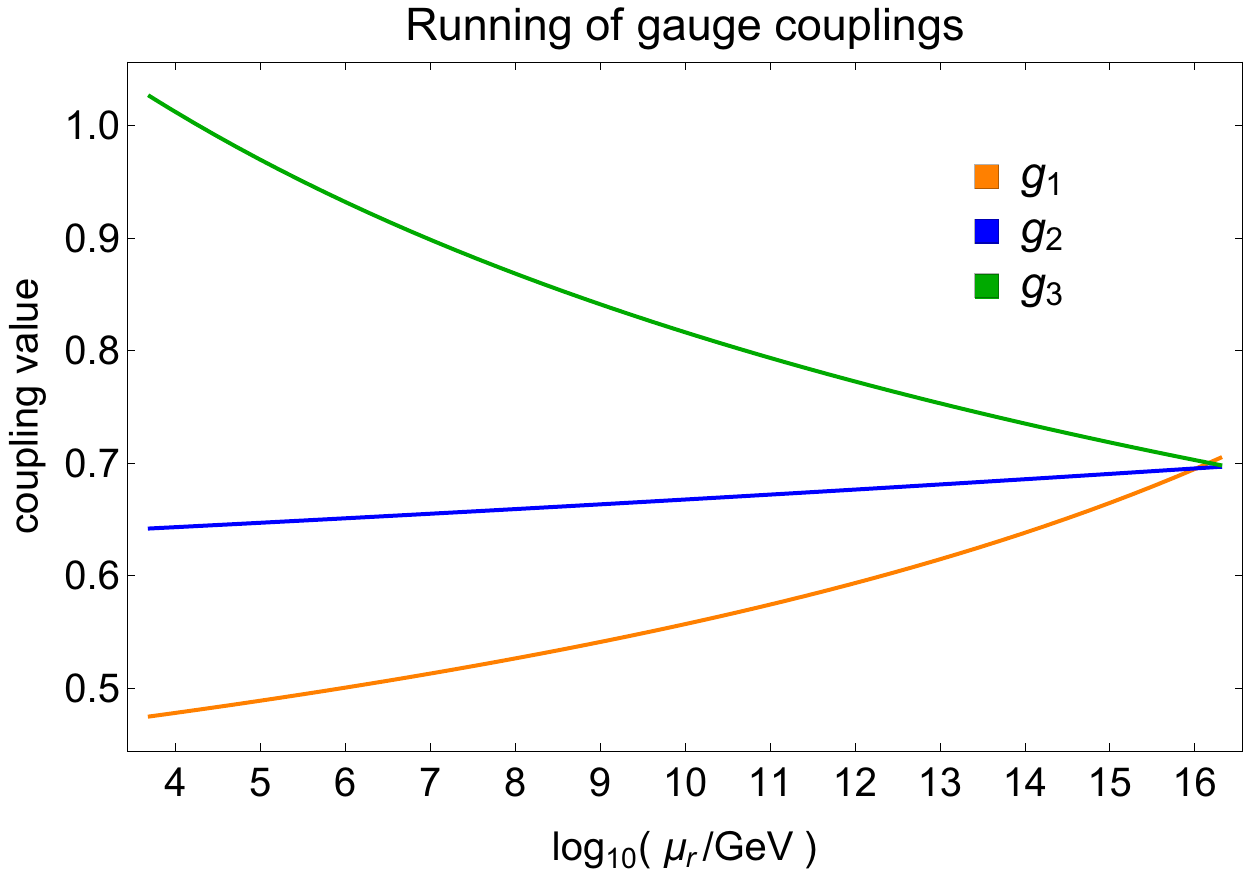}\hspace{0.5cm}
		\includegraphics[width=7.8cm]{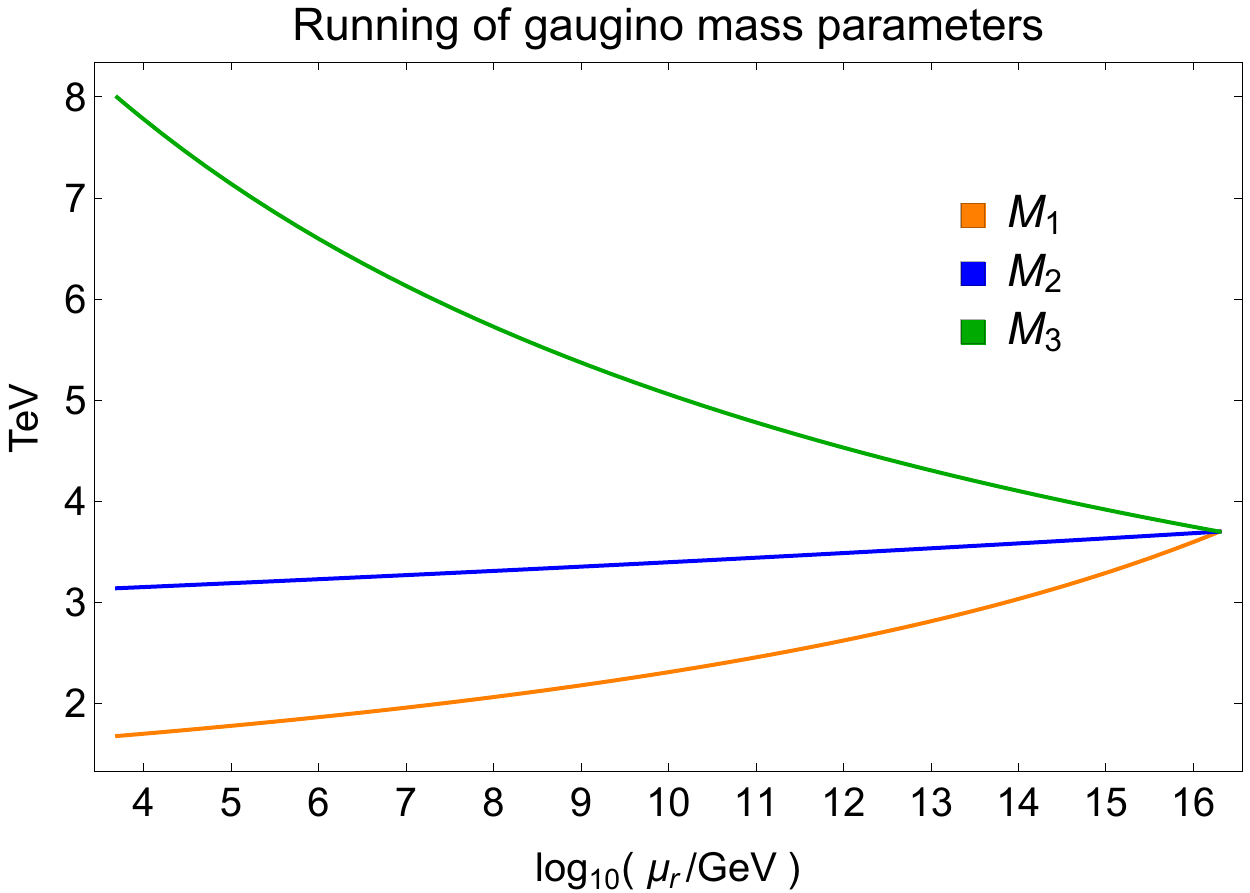}
		\vskip 0.5cm
		\includegraphics[width=7.8cm]{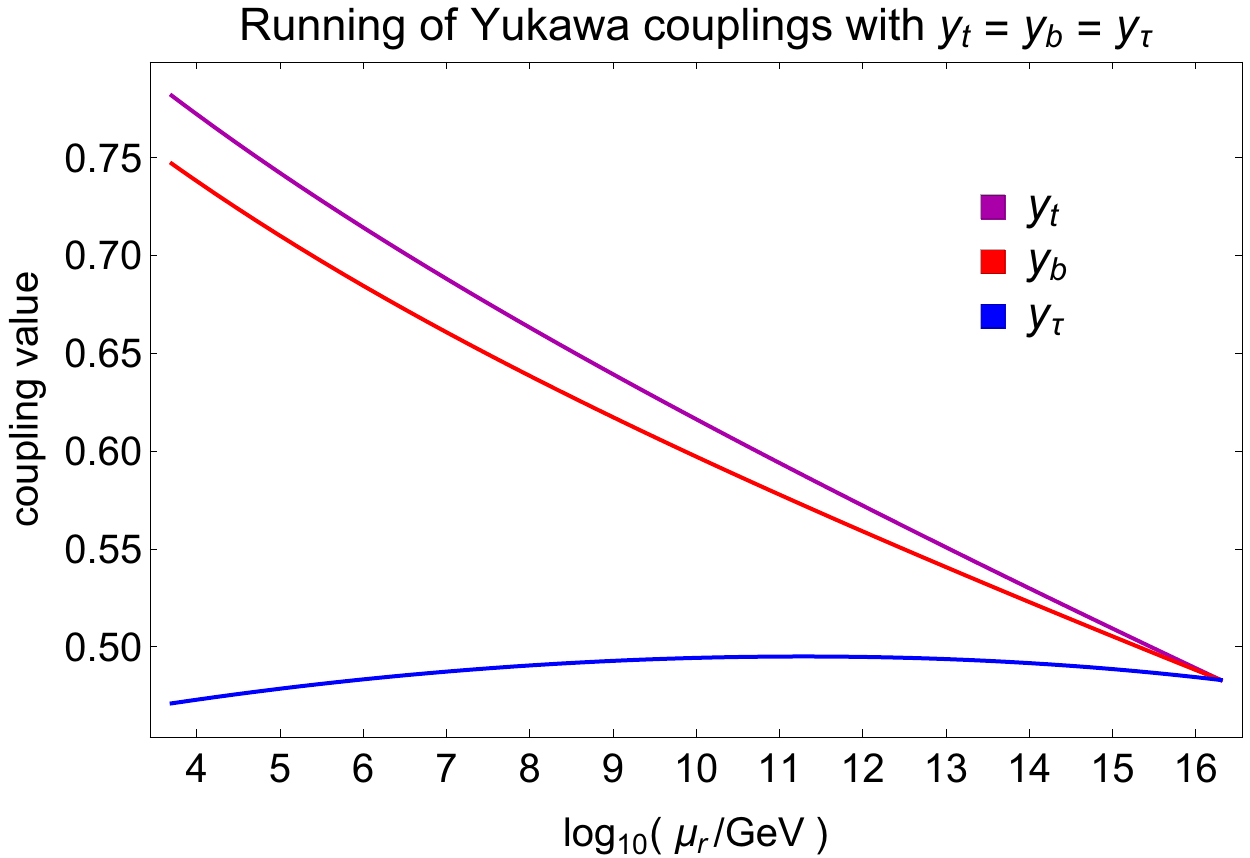}\hspace{0.5cm}
		\includegraphics[width=7.8cm]{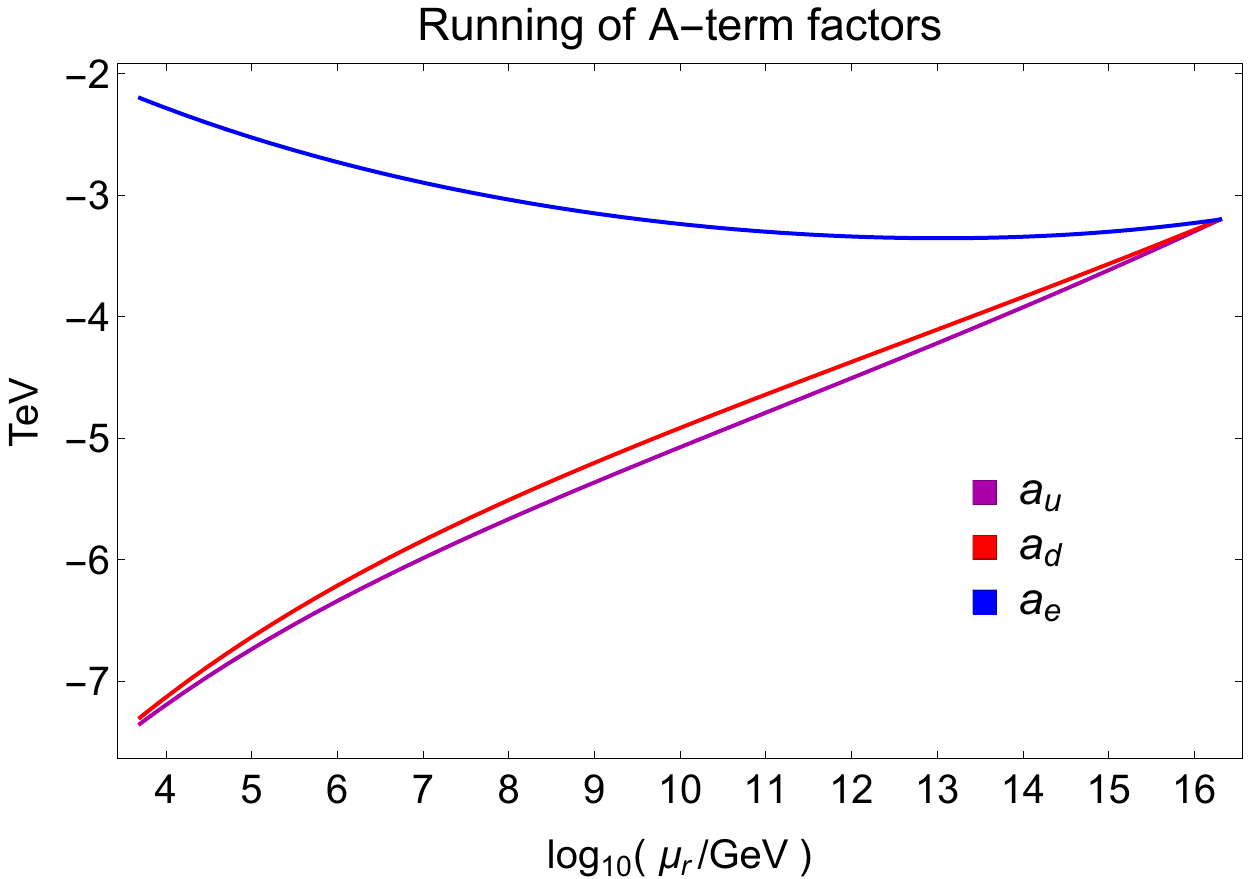}
		\end{center}
		\caption{The RGE running between $\MSUSY$ and $\MGUT$ for the example point in Eq.~\eqref{eq:example-point-begin}-\eqref{eq:example-point-end} of the the gauge couplings (top left), gaugino mass parameters (top right),the 3rd family Yukawa couplings (bottom left), and the $A$-term proportionality factors (bottom right). \label{Figure:RGE-gMYA}
		}
	\end{figure}

\item \textbf{Running of soft masses}\par
	The RGE running of all the soft mass parameters for the example point and a nearby point, where only the top Yukawa $y_t$ is changed to the value $y_t\equiv 1.1\,y_0$ while $y_b=y_\tau\equiv y_0$, are shown in Figure~\ref{Figure:RGE-soft-masses}. The relevant RGEs for these quantities are in Eq.~\eqref{eq:RGE-simple-mHu}--\eqref{eq:RGE-simple-me3}. The patterns are easy to understand; we comment on some of them below. 
	\par 
	For $m^{2}_{H_u}$ and $m^2_{H_d}$, the positive Yukawa term contributions to the $\beta$ functions dominate, leading to a positive slope and thus the parameters becoming smaller and eventually negative with smaller $\mu_r$. The drive to $m^{2}_{H_u}<0$ at low $\mu_r$ confirms that the EWSB is radiative. Crucially, the necessary condition for EWSB $m^{2}_{H_d}>m^{2}_{H_u}$ is also satisfied at low scales, as will be discussed in more detail later.  
	\par 
	The soft mass parameters related to the squarks grow fast with smaller $\mu_r$ due to the large negative contribution of the gluino related terms $g_3^2 |M_3|^2$.	 These terms are not present in the $\beta$ function for soft-mass parameters of leptons, so the slepton masses stay almost flat.  
	\par 
	Another general feature of the soft-mass parameter running is that the masses of the 1st and 2nd family of squarks and sleptons (index 1) become larger than those of the 3rd family (index 3); we are comparing here the soft-mass parameters of particles of the same flavor, but from different families. The simple reason is the additional positive terms proportional to squares of Yukawa couplings, which appear only for 3rd family squarks and sleptons (since the 1st and 2nd family Yukawa coupling are negligible compared to the 3rd family, and they are set to zero in our simple scenario). We thus have the usual \textit{inverted hierarchy} in the squark and slepton masses.
	\par We now discuss how the scenario of $t$-$b$ unification and $y_t=1.1\,y_b$ compare. We see that there is little qualitative difference for the values of any one soft parameter taken on its own. Visually though, major quantitative changes in relative terms can be spotted when comparing the quantity $m^2_{H_d}-m^2_{H_u}$ in the two scenarios, as well as changes in the quantity $m^2_{u_3}-m^2_{d_3}$. These changes might be deemed to have an insignificant effect on the 
low energy observables. But as shown in the previous section, the difference $m^2_{H_d}-m^2_{H_u}$ turns out to determine the mass scale of the extra MSSM Higgs bosons. That means that the exactness of $t$-$b$ unification at the GUT scale, as demonstrated by the two scenarios in Figure~\ref{Figure:RGE-soft-masses}, has a big impact on the sparticle spectrum, i.e.~on the extra Higgs sector to be precise. This is the major effect that this paper investigates.  
	\begin{figure}[htb]
	\begin{center}
	\includegraphics[width=12cm]{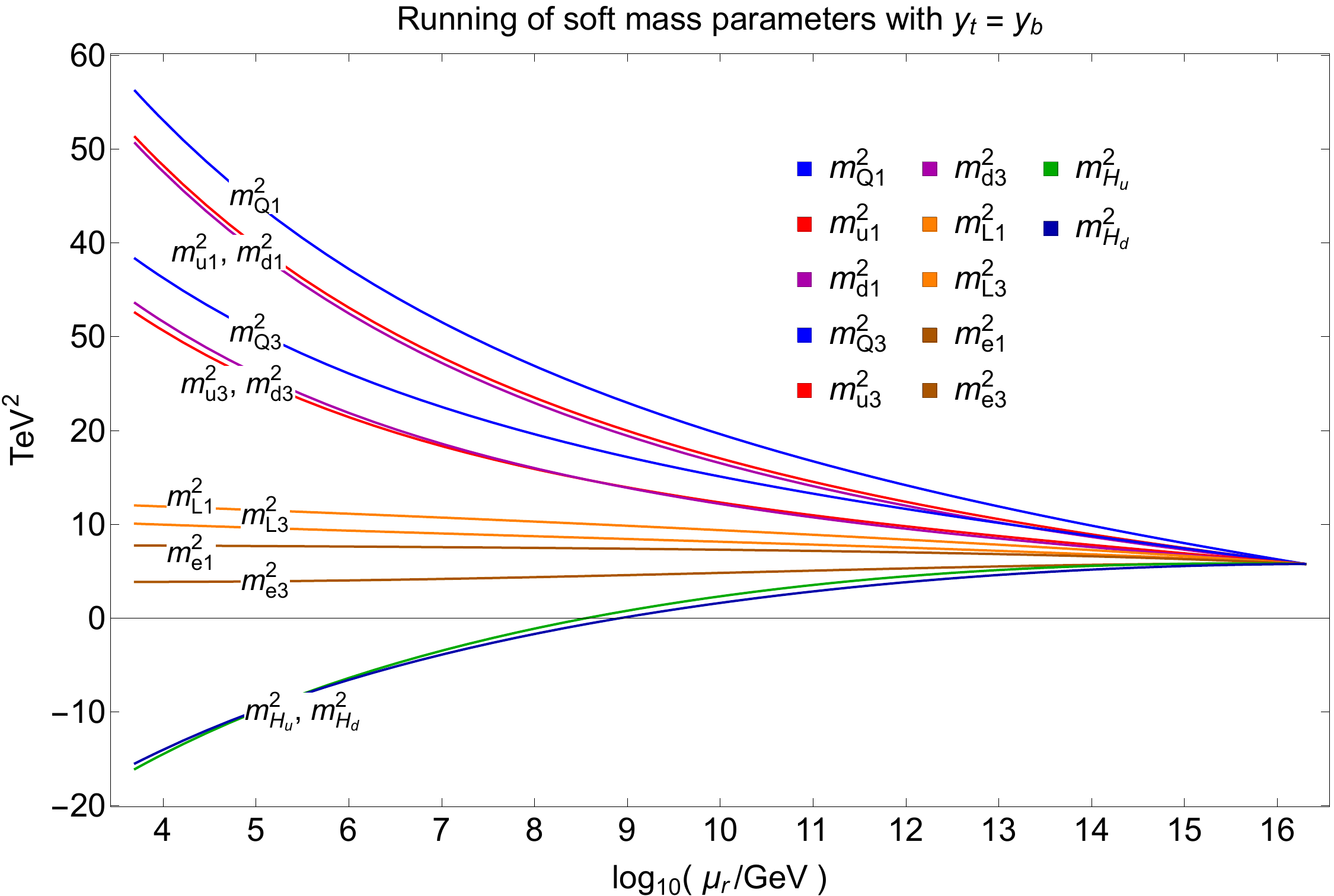}
	\vskip 0.5cm
	\includegraphics[width=12cm]{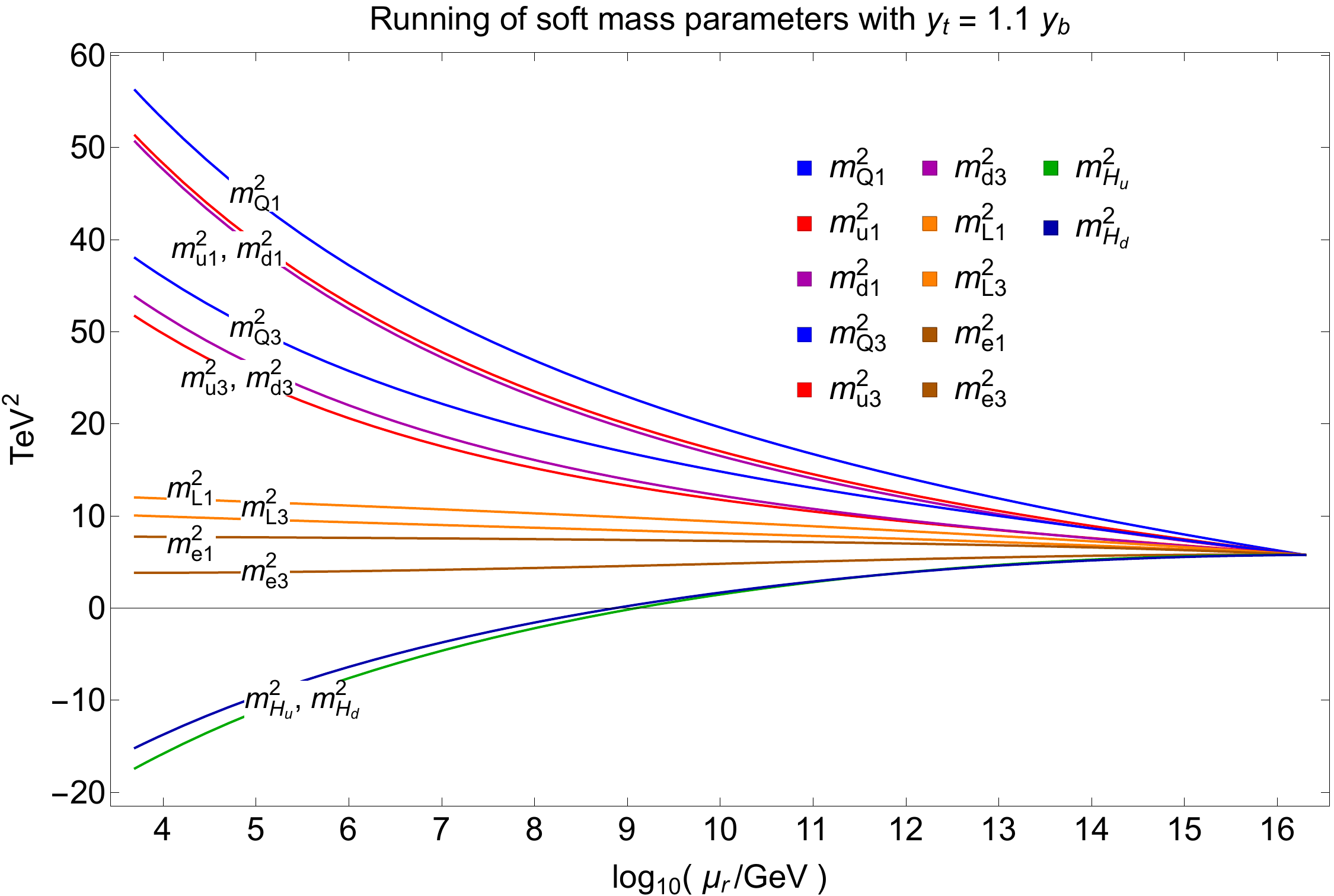}
	\end{center}
	\caption{The RGE running of all soft mass parameters. The two cases are for $t$-$b$ unification $y_t=y_b$ of the example point (1st panel) and for the modified point $y_t=1.1\,y_0$ (2nd panel).\label{Figure:RGE-soft-masses}}
	\end{figure}
	
\item \textbf{Effect of $t$--$b$ unification on $m^2_{H_d}-m^2_{H_u}$}\par
	We have seen from the RGE of the soft masses in the previous step that $t$-$b$ unification\footnote{We took $y_t=y_0$, which could at this point be deemed $t$-$\tau$ unification just as well as $t$-$b$ unification. It is only a later analysis of deformations in $y_\tau$ which confirms that it is the $t$-$b$ split, and not $t$-$\tau$ split, which is important.} has little qualitative effect on the running of these parameters taken in isolation, but has a crucial effect on $m^2_{H_d}-m^{2}_{H_u}$. Figure~\ref{Figure:RGE-tb-effect} shows RGE trajectories for $m^2_{H_d}-m^2_{H_u}$ 
	under different $y_t/y_b$ ratio boundary conditions at $\MGUT$, essentially demonstrating the sensitivity of this quantity to $t$-$b$ unification. We see that for our example point, the running expression $m^2_{H_d}-m^2_{H_u}$ increases essentially linearly with the $y_t-y_b$ difference (at least when relative differences are small), and with a substantial increase already when $y_t$ and $y_b$ differ at the percent level. The impact is even more dramatic when considered in terms of relative increases of $m^{2}_{H_d}-m^2_{H_u}$: a deviation of a mere $10\,\%$ from $t$-$b$ unification raises the value by a factor $4$, and consequently the masses of the extra MSSM Higgs particles by a factor of $2$. Looking at this from a reverse perspective, when approaching $t$-$b$-$\tau$ unification from a $t$-$b$ deformation direction, the predicted masses of the extra Higgses drop very quickly, typically below $\mathrm{1}\,\mathrm{TeV}$.
\par		
At  $\MSUSY$, the condition $m^{2}_{H_d}-m^{2}_{H_u}>0$ is necessary for (tree-level) EWSB. We can see in Figure~\ref{Figure:RGE-tb-effect} that this condition is fulfilled even for exact Yukawa unification (the $y_t=y_b$ curve), at least for this particular example point. This shows that there exist parameter points with exact Yukawa unification and successful EWSB. It is important to note that a successful EWSB with the $m^{2}_{H_d}=m^{2}_{H_u}$ GUT boundary condition (and universal gaugino masses) was not found in some of the prior literature \cite{Carena:1994bv,Matalliotakis:1994ft,Olechowski:1994gm,Murayama:1995fn,Ajaib:2013zha} due to extensive use of semi-analytic approximate formulas from 
e.g.~\cite{Hempfling:1994sa}, as was discussed in Section~\ref{section:introduction}. Part of the pessimism also stemmed from observing that the slope at $\MGUT$ is positive, thus driving the value $m^{2}_{H_d}-m^{2}_{H_u}$
in the wrong direction towards negative values; it is only later at low $\mu_r$ that the slope becomes negative and eventually manages to run the expression back to positive values, an indirect effect due to the running of other couplings. 
\par	 
Note that we plot the RGE solutions for all curves down to a fixed scale $\mu_r=\MSUSY$, which was computed for the $y_t=y_b$ case. This scale is defined as the geometric mean of the stop masses. Strictly speaking, the scale $\MSUSY$ shifts slightly with different ratios $y_t/y_b$, so that comparing the running quantity $m^{2}_{H_d}-m^{2}_{H_u}$ at a fixed scale is not exactly the same as comparing the mass scales of the extra Higgses. This shift, however, is negligible, since 
the quantities determining the stop masses run logarithmically with $\mu_r$ and change only slightly with the ratio $y_t/y_b$, as argued in the previous analysis step. It is thus justified to compare the running expression for different curves at a fixed scale $\MSUSY$ for qualitative considerations.
	
	\begin{figure}[htb]
	\begin{center}
	\includegraphics[width=10cm]{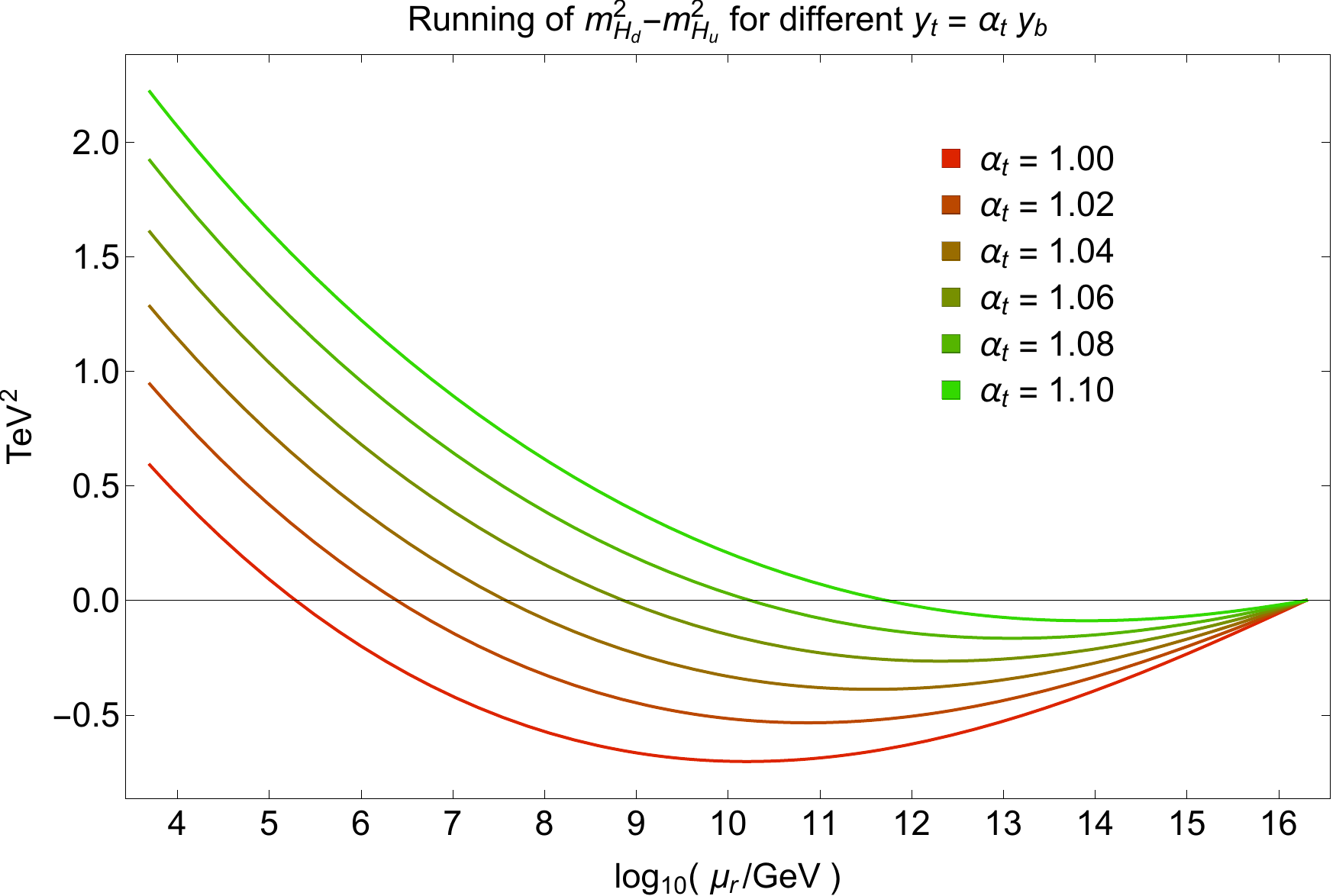}
	\end{center}
	\caption{The RGE running of $m^2_{H_d}-m^2_{H_u}$ for different values of the $y_t$ coupling. We can see the sensitivity of the value at $\MSUSY$ to the $y_t$ coupling: $t$-$b$ unification significantly lowers the values, but already a $\%$ level mismatch in $y_t$ and $y_b$ noticably changes the mass parameter difference. \label{Figure:RGE-tb-effect}}
	\end{figure}

\item \textbf{Contributions to $\beta(m^2_{H_d}-m^2_{H_u})$}\par
	To understand the effect that $t$-$b$-$\tau$ unification has on the RGE running, we consider the various contributions to $\beta(m^2_{H_d}-m^2_{H_u})$. One can combine the separate RGE in Eq.~\eqref{eq:RGE-simple-mHu} and \eqref{eq:RGE-simple-mHd} into the $\beta$ function of Eq.~\eqref{eq:RGE-UD}. For our example point, where the right-handed neutrinos are already integrated out at $\MGUT$, there are $4$ terms: terms proportional to $|y_t|^2$, $|y_b|^2$ and $|y_\tau|^2$, as well as a term proportional to $S$, which is a linear combination of scalar soft masses, see Eq.~\eqref{eq:RGE-definition-S}. We plot these contributions for the cases $y_t=y_b$ and $y_t=1.1\,y_b$ in Figure~\ref{Figure:RGE-ud-contributions}.
	\par
	The results show that in absolute terms the $|y_t|^2$ and $|y_b|^2$ contributions dominate over the $|y_\tau|^2$ one at $\MGUT$, an effect which only increases when running to lower $\mu_r$, while the contribution from the $S$ term stays numerically negligible throughout and will thus be ignored in the following discussion.  The larger contributions of the $t$ and $b$ terms start out due to  larger numeric prefactors (due to color) compared to the $\tau$ term. Furthermore, at lower energies the Yukawa couplings $y_t$ and $y_b$ rise with smaller scale, while $y_\tau$ falls, see Figure~\ref{Figure:RGE-gMYA}.  In addition, also the soft masses show the same trend, see Figure~\ref{Figure:RGE-soft-masses}. Note, however, that these terms in $\beta(m^{2}_{H_d}-m^{2}_{H_u})$ come with different signs; in particular, the $t$ and $b$ contributions have opposite signs. 
\par
It is thus convenient to compare the difference of the $t$ and $b$ terms (red curve) with the $\tau$ contribution (green curve), see right panels of Figure~\ref{Figure:RGE-ud-contributions}. We shall refer to these two contributions as the $t$-$b$ and $\tau$ contributions, respectively. The $t$-$b$ contribution comes into the $\beta$ function with a negative sign, so whenever the red curve dominates over the green curve, the beta function value becomes negative, i.e.~the RGE running of $m^{2}_{H_d}-m^{2}_{H_u}$ has a negative slope. Conversely, when the $\tau$ contribution dominates and the green curve is above the red, the slope is positive. 
As the figure shows, the slope is positive at large $\mu_r$ and negative at small $\mu_r$, which is consistent with Figure~\ref{Figure:RGE-tb-effect}. 
\par 
At low enough $\mu_r$ the $t$-$b$ contribution is expected to dominate over the $\tau$ contribution regardless of the starting $y_t/y_b$ ratio simply due to Yukawa coupling values at those energies, and that typically the squark soft masses are larger than the corresponding lepton ones. The ratio $y_t/y_b$ is crucial, however, for the $t$-$b$ contribution at energies near the GUT scale: when $y_t=y_b$ the $t$-$b$ contribution starts at zero, while $y_t/y_b>1$ implies a non-vanishing starting value for the RGE.\footnote{With a large enough $y_t/y_b$ ratio, the $t$-$b$ contribution may in fact already start out larger than the $\tau$ contribution, implying that the slope is always negative.} This crucially impacts the scale at which the $t$-$b$ contribution becomes bigger than the $\tau$ one, i.e.~when the red and green curves on the right panels of Figure~\ref{Figure:RGE-ud-contributions} cross. We see that for $y_t=1.1y_b$ the $t$-$b$ contributions already starts out almost as big as the $\tau$ contribution at $\MGUT$, so the curves intersect above $10^{14}\,\mathrm{GeV}$, while $t$-$b$ unification delays this until below $10^{11}\,\mathrm{GeV}$. Consequently, with $t$-$b$ unification the value of $m^{2}_{H_d}-m^2_{H_u}$ will be much lower, since the rise in its running value is delayed by several orders of magnitude in the energy scale $\mu_r$. 
\par
This completes our understanding of the effect of $t$-$b$ unification on $m^{2}_{H_d}-m^{2}_{H_u}$. Yukawa unification delays when the $t$-$b$ contribution in the beta function rises enough to dominate over the $\tau$ contribution, allowing for the running value of $m^{2}_{H_d}-m^{2}_{H_u}$ to rise much less by the scale $\mu_r=\MSUSY$. We emphasize that this effect is an indirect consequence of RG running of all parameters, and can thus be seen only when solving for the entire system of RGE numerically and evolving it over multiple orders of magnitude of $\mu_r$. In simplified analyses, such as studying the local RG behavior at $\MGUT$ by Taylor expansion or taking some running quantities in the beta function as constant to derive a linear-log semi-analytic approximation~\cite{Hempfling:1994sa}, not even the $m^{2}_{H_d}>m^{2}_{H_u}$ property at low $\mu_r$ is reproduced, let alone the more subtle effect of the $t$-$b$ deformation. 

	\begin{figure}[htb]
	\begin{center}
	\includegraphics[width=7.8cm]{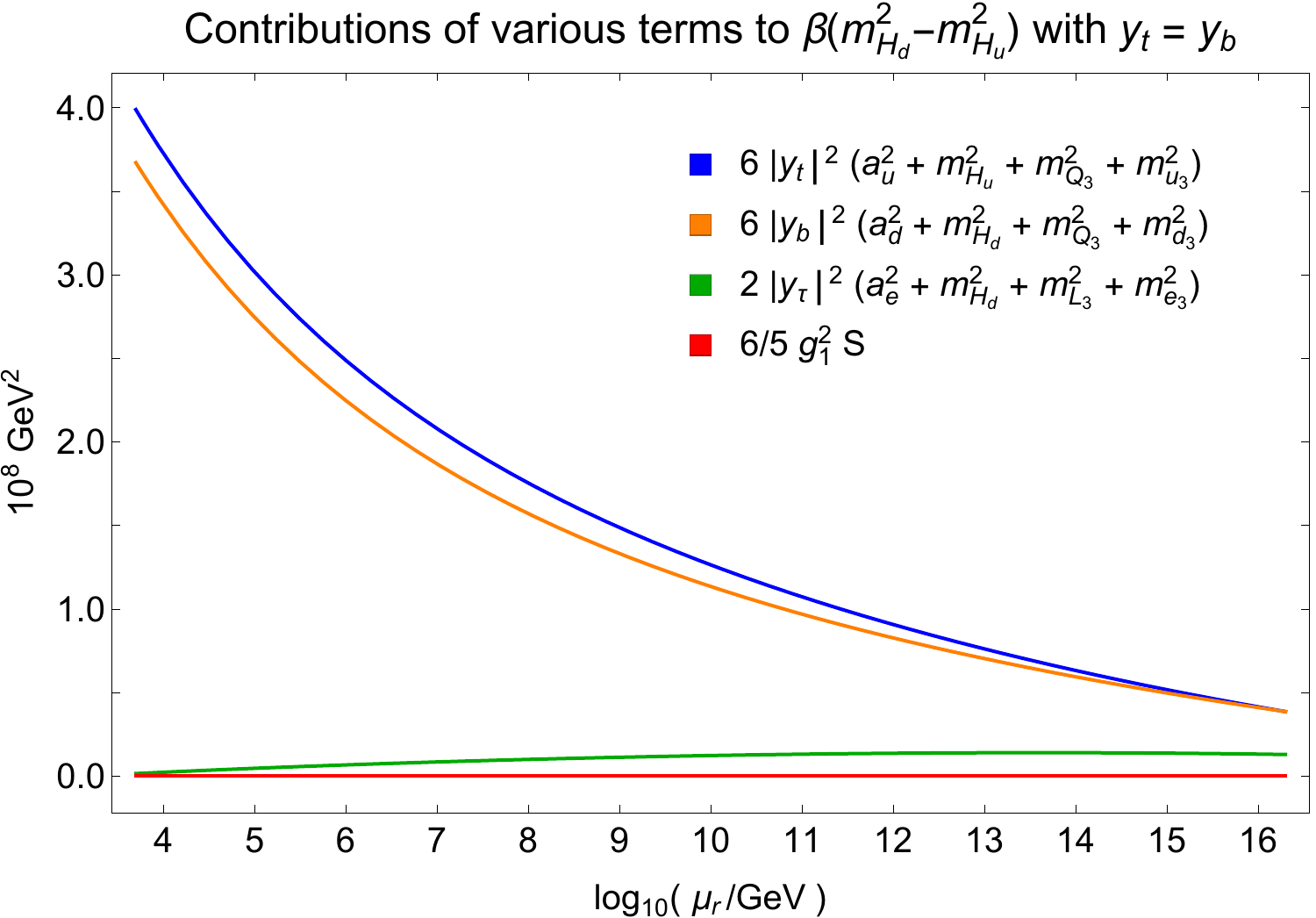}
	\hspace{0.5cm}
	\includegraphics[width=7.8cm]{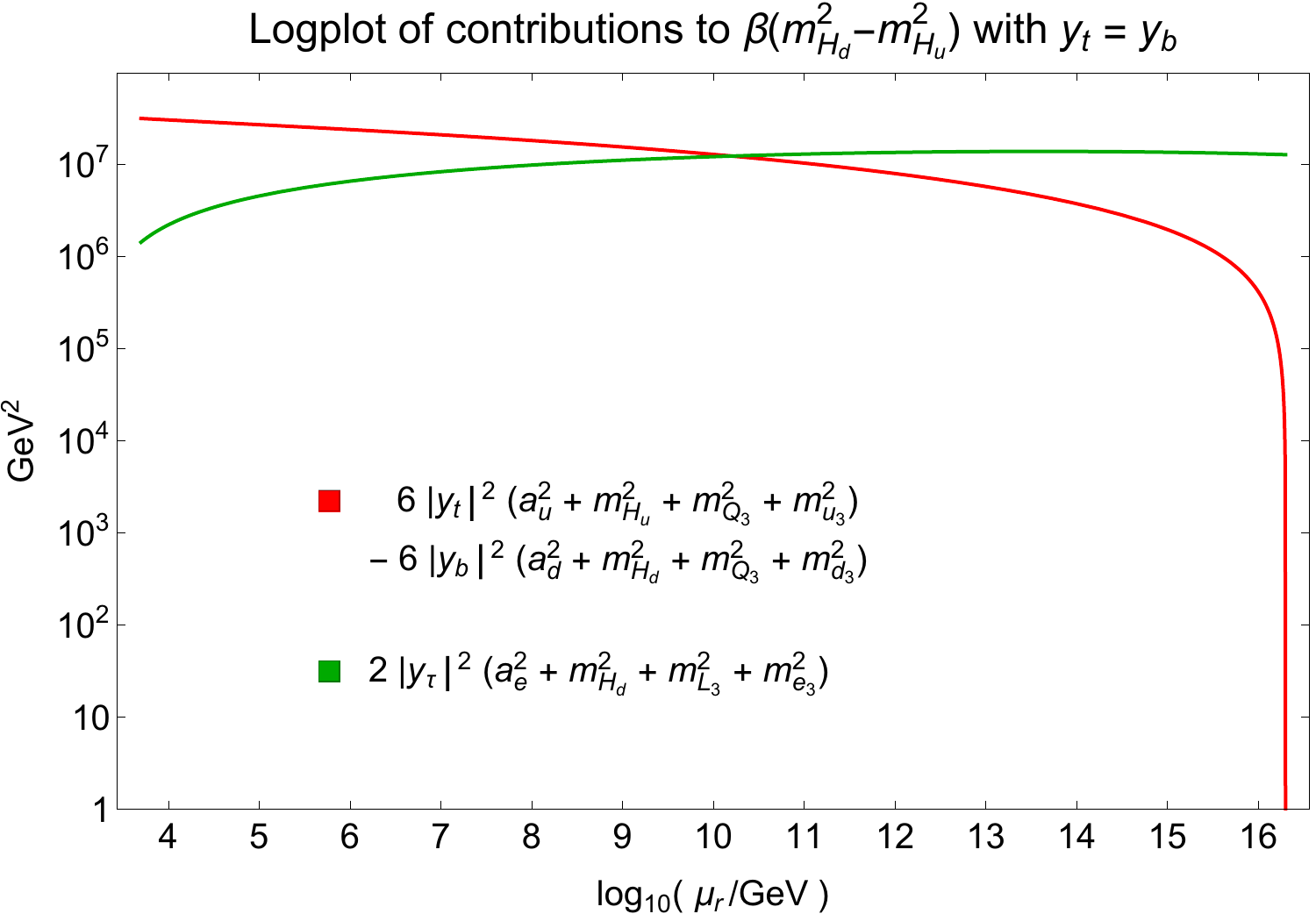}
	\vskip 0.5cm
	\includegraphics[width=7.8cm]{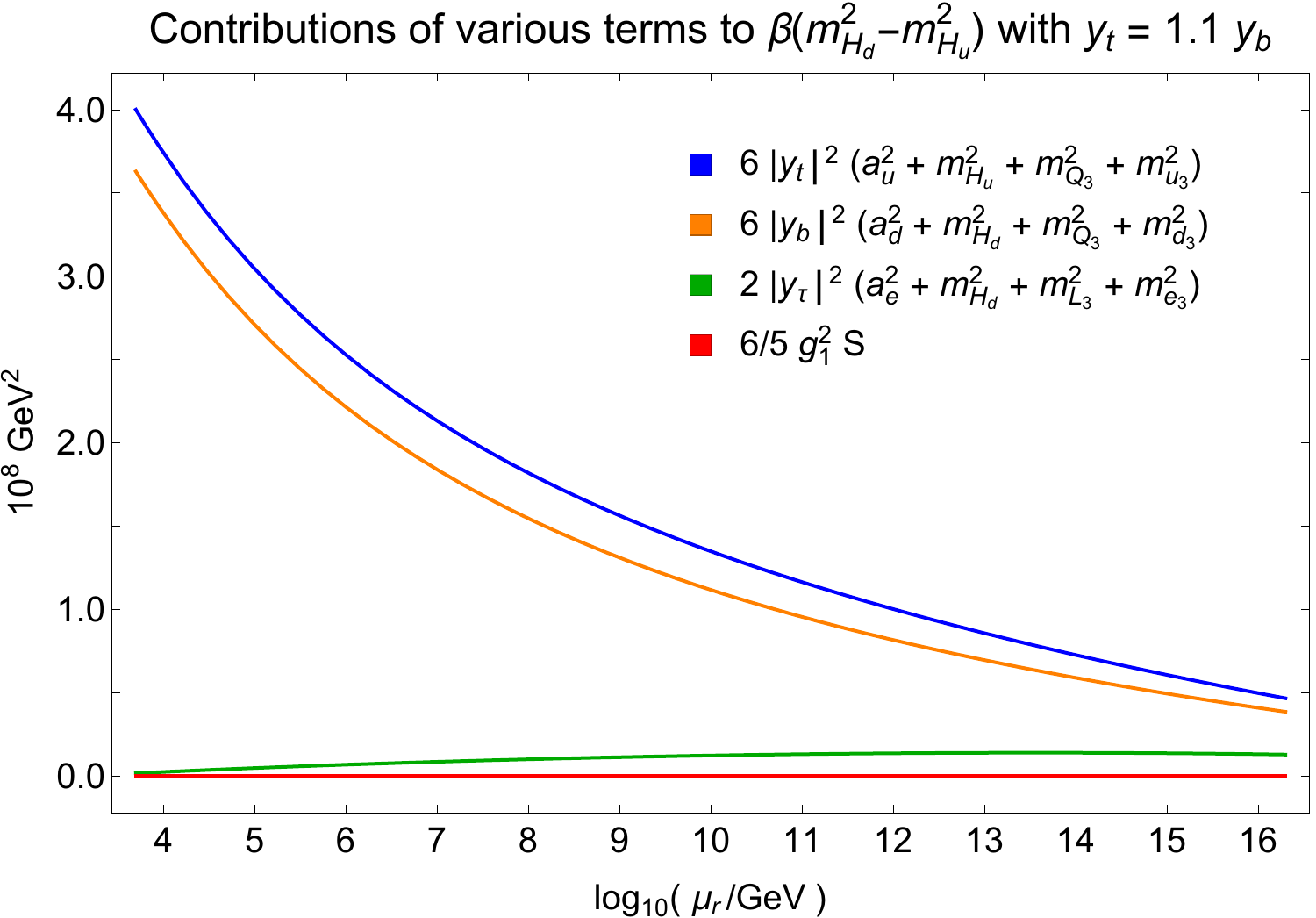}\hspace{0.5cm}
	\includegraphics[width=7.8cm]{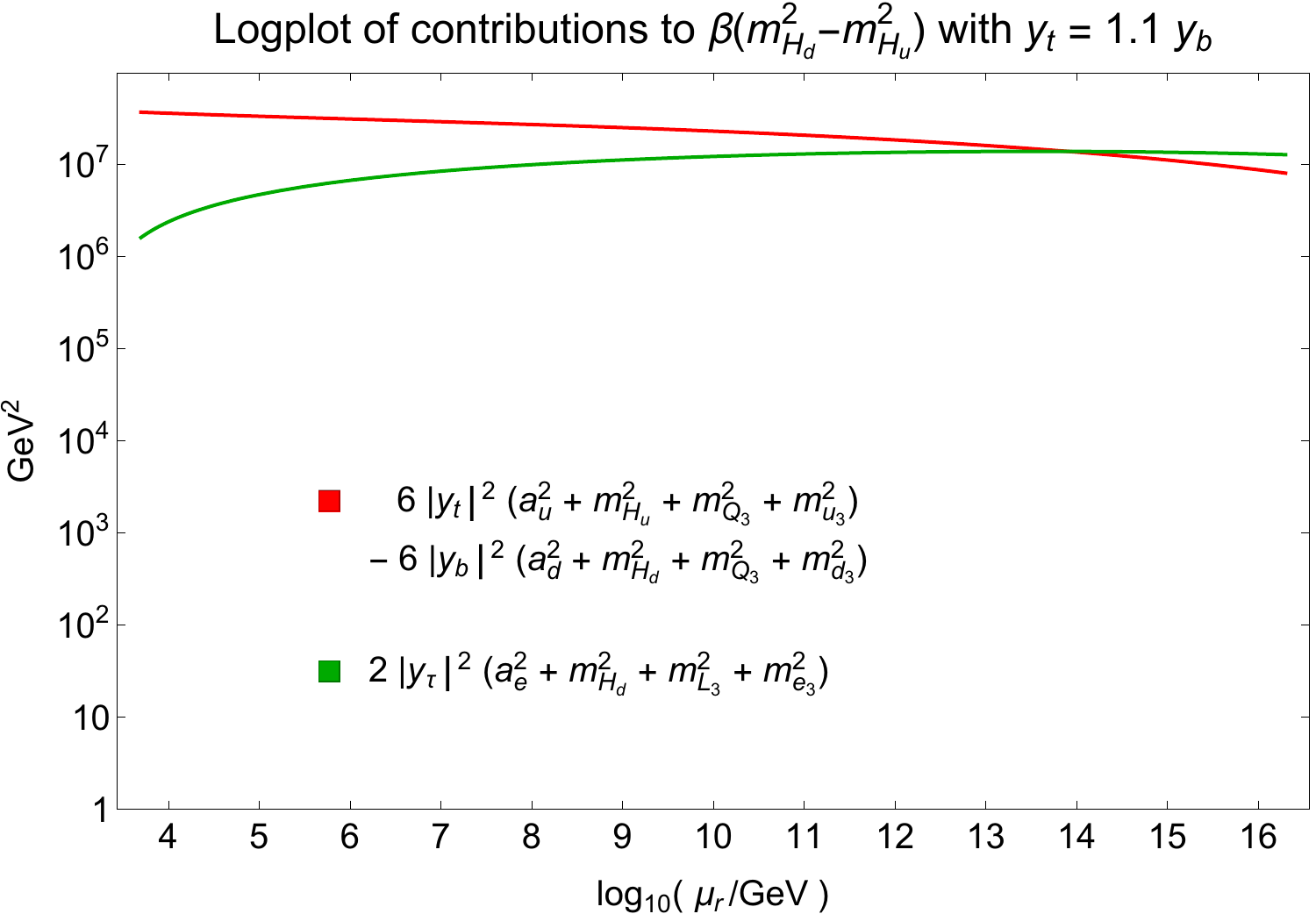}
	\end{center}
	\caption{The contributions of different terms to $\beta(m^2_{H_d}-m^2_{H_u})$ (left) and the comparison of the same contributions in a logplot (right). The plots are drawn for the case $y_t=y_b$ (above), and a deviation from that with $y_t=1.1\,y_b$ (below).\label{Figure:RGE-ud-contributions}}
\end{figure}

\item \textbf{Effect of $b$--$\tau$ unification on $m^2_{H_d}-m^2_{H_u}$}\\
	An interesting question now is what impact $b$-$\tau$ unification of couplings has on lowering the value $m^2_{H_d}-m^2_{H_u}$. It turns out that while $t$-$b$ unification is crucial for this effect, $b$-$\tau$ unification is not.
	\par
	We plot the RGE flow of  $m^2_{H_d}-m^2_{H_u}$ for different $\Delta_\tau:=y_b-y_\tau=y_0-y_\tau$ in Figure~\ref{Figure:RGE-neutrinos}. The results clearly show that $b$-$\tau$ unification has minimal effect on that quantity at the SUSY scale. The two sets of trajectories on the plot correspond to the $y_t=y_0$ case (red-blue) and the $y_t=1.1y_0$ case (green-cyan); 	trajectories in the same set differ in $\Delta_\tau$ from $0$ to $0.2$, which presents a relative drop in $y_\tau$ compared to $b$-$\tau$ unification of more than $40\,\%$, but trajectories in the same set nevertheless cluster together at $\MSUSY$, despite diverging at first at intermediate energies. 
	\begin{figure}[htb]
	\begin{center}
	\includegraphics[width=10cm]{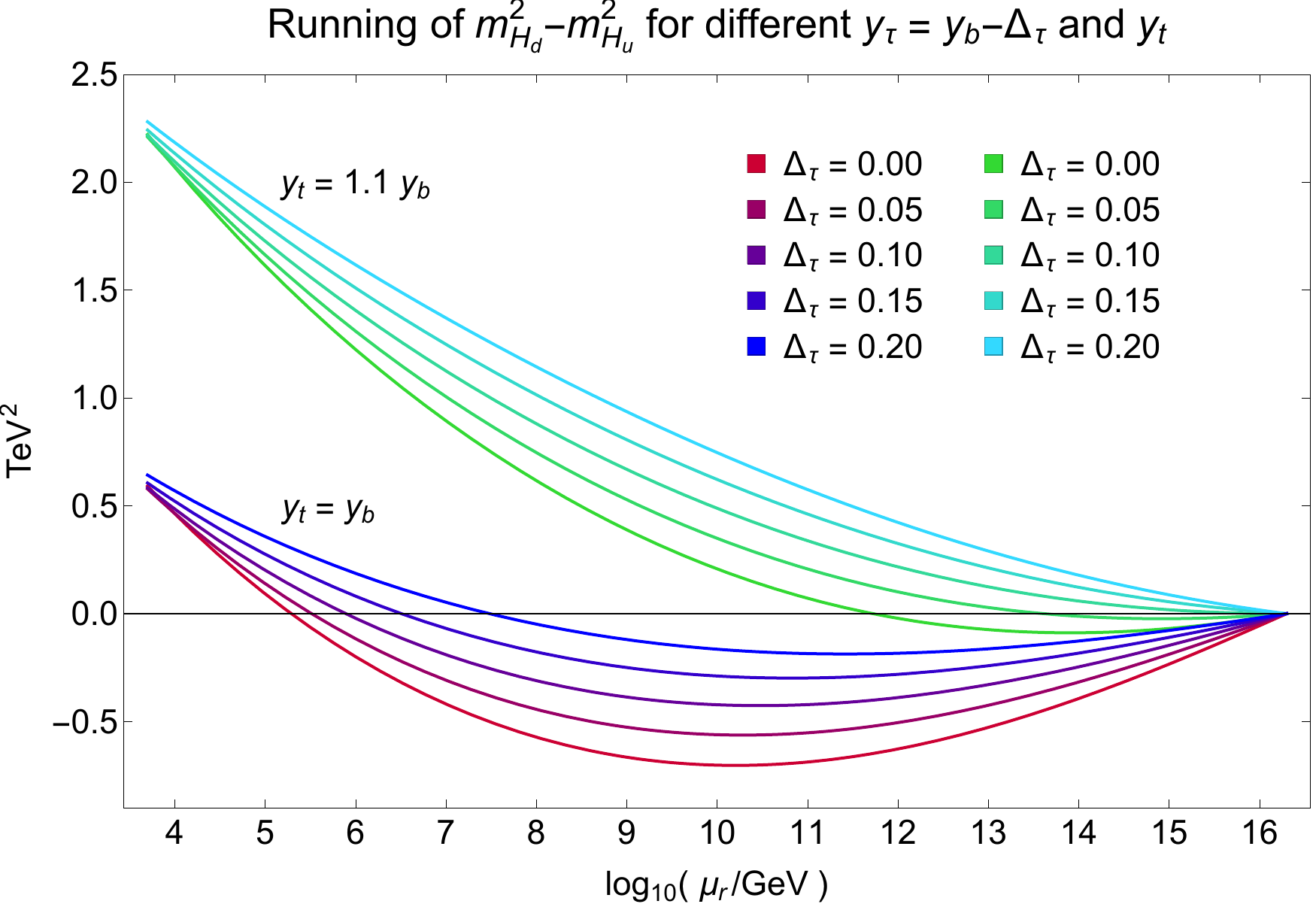}
	\end{center}
	\caption{The RGE running of $m^2_{H_d}-m^2_{H_u}$ for different deformations of $b$-$\tau$ unifiaction, i.e.~values of $y_\tau=y_b-\Delta_\tau$. There are two sets of trajectories: the red-to-blue trajectories are for $y_t=y_b$ ($t$-$b$ unification), while the green-to-cyan trajectories are for the case $y_t=1.1 y_b$. All other quantities are the same as in the example point~\eqref{eq:example-point-begin}-\eqref{eq:example-point-end}. \label{Figure:RGE-tau}}
\end{figure}

\item \textbf{Effect of $M_R$ on $m^2_{H_d}-m^2_{H_u}$}\\
	We see from Figure~\ref{Figure:RGE-tau} that the scale of right-handed neutrino $M_R$, associated with the large 3rd family neutrino Yukawa coupling $y_\nu$, has a comparatively small effect on the value of $m^2_{H_d}-m^2_{H_u}$ at $\MSUSY$, relative to effect of the $t$-$b$ deformation. The discontinuous changes in the slope happen at scales when the right-handed neutrino is integrated out, i.e.~at the scale $M_{R}$. We conclude that the right-handed neutrinos do not have a large direct effect on the mass scale of the extra Higgs particles, and we therefore do not include them in the analyses of Sections~\ref{section:RGE-analysis} and \ref{section:mass-scale}. It should be noted though that an indirect effect turns out to be possible, since their presence shifts the region of parameter space where good fits to low energy data are obtained, 
see Section~\ref{section:experimental-constraints}.
	 
	\begin{figure}[htb]
	\begin{center}
	\includegraphics[width=10cm]{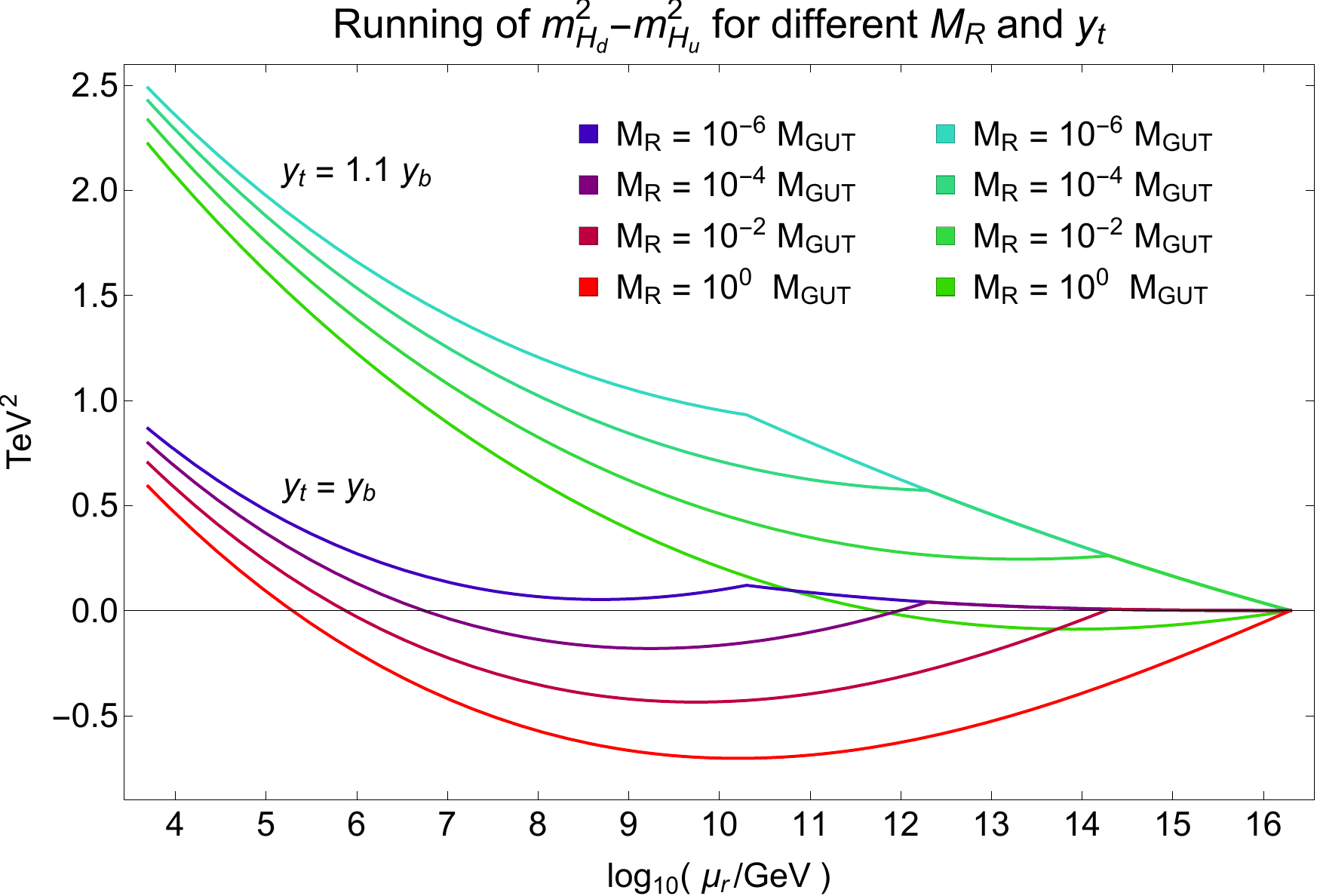}
	\end{center}
	\caption{The RGE running of $m^2_{H_d}-m^2_{H_u}$ for different values of the right-handed neutrino mass $M_R$. The set of red-to-blue trajectories are for $y_t=y_b$ ($t$-$b$ unification), while the green-to-cyan trajectories show the case $y_t=1.1 y_b$. All other quantities are the same as in the example point~\eqref{eq:example-point-begin}-\eqref{eq:example-point-end}.   \label{Figure:RGE-neutrinos}}
\end{figure}
\item \textbf{$\SO(10)$ boundary conditions: replace $m_0$ with $m_{16}$ and $m_{10}$}\\
We investigate whether having a simplified set of CMSSM parameters for the soft term boundary conditions is crucial for having light extra Higges. A more realistic, yet still minimalist, set of soft parameters for an $\SO(10)$ GUT theory is one where partial universality comes due to GUT symmetry. The universal gaugino mass parameter $M_{1/2}$ at the GUT scale can in this context be understood as arising from $\SO(10)$ symmetry of the gaugino masses. Similarly, since all SM fermions and right handed neutrinos come from the representations $\mathbf{16}$ of $\SO(10)$, a universal $a_0$ for different fermions can be understood in that way. On the other hand, there is no symmetry reason why the soft mass parameters of the sfermions should be equal to the soft mass parameters of the two MSSM Higgs doublets.
\par
We therefore consider a slightly more general case of parametrization for the soft terms, which we refer to as ``$\SO(10)$ boundary conditions''. We keep the $M_{1/2}$ and $a_{0}$ parameters, but have two different soft mass parameters $m_{16}$ and $m_{10}$ for the sfermions and Higgses, respectively:
\begin{align}
m_{16}^{2}\,\mathbf{1}&:=\mathbf{m}^2_Q\atMGUT=\mathbf{m}^2_L\atMGUT=\mathbf{m}^2_u\atMGUT=\mathbf{m}^2_d\atMGUT=\mathbf{m}^2_e\atMGUT=\mathbf{m}^2_\nu\atMGUT,\label{eq:m0-SO10-split-m16}\\
m_{10}^{2}&:=m^{2}_{H_u}\atMGUT = m^{2}_{H_d}\atMGUT.\label{eq:m0-SO10-split-m10}
\end{align}
The notation for $m_{16}$ and $m_{10}$ signifies which $\SO(10)$ 
representation the scalars of the soft term are part of. It is presumed here that $H_u$ and $H_d$ come from a $\mathbf{10}$ of $\SO(10)$, which allows for $t$-$b$-$\tau$ unification with the simple renormalizable 3rd family Yukawa operator $\mathbf{16}_3\cdot \mathbf{16}_3\cdot\mathbf{10}$. 
\par
We investigate the effect of such an $\SO(10)$ motivated split in the soft mass parameters in Figure~\ref{Figure:RGE-m10-m16-split}. We always take $m_{16}\equiv m_0$, while the deviation $\Delta_{m}\equiv m_{10}-m_0$ from the example parameter point occurs for the $H_u$ and $H_d$ soft masses. The figure shows a relative decrease or increase of $m_{10}$ by $1000\,\mathrm{GeV}$ (a relative difference of over $40\,\%$) from $m_0$. We see that the choice of $t$-$b$ unification or its deformation of $10\,\%$ again dominates over the soft mass split. The soft mass split thus does not qualitatively change the feature of the spectrum that $t$-$b$ unification leads to light extra MSSM Higgses, at least for similar scales of $m_{10}$ and $m_{16}$. Quantitatively, however, it can be seen from the figure that a $m_{10}>m_{16}$ split somewhat lowers the $m^2_{H_d}-m^2_{H_u}$ value, while $m_{10}<m_{16}$ raises it. For a large enough $m_{10}$, the value of $m^2_{H_d}-m^2_{H_u}$ may become negative, a problematic regime for EWSB.
\par
We have thus seen that the low masses of the extra Higgses persist even with $\SO(10)$ boundary conditions replacing CMSSM.
\begin{figure}[htb]
	\begin{center}
	\includegraphics[width=10cm]{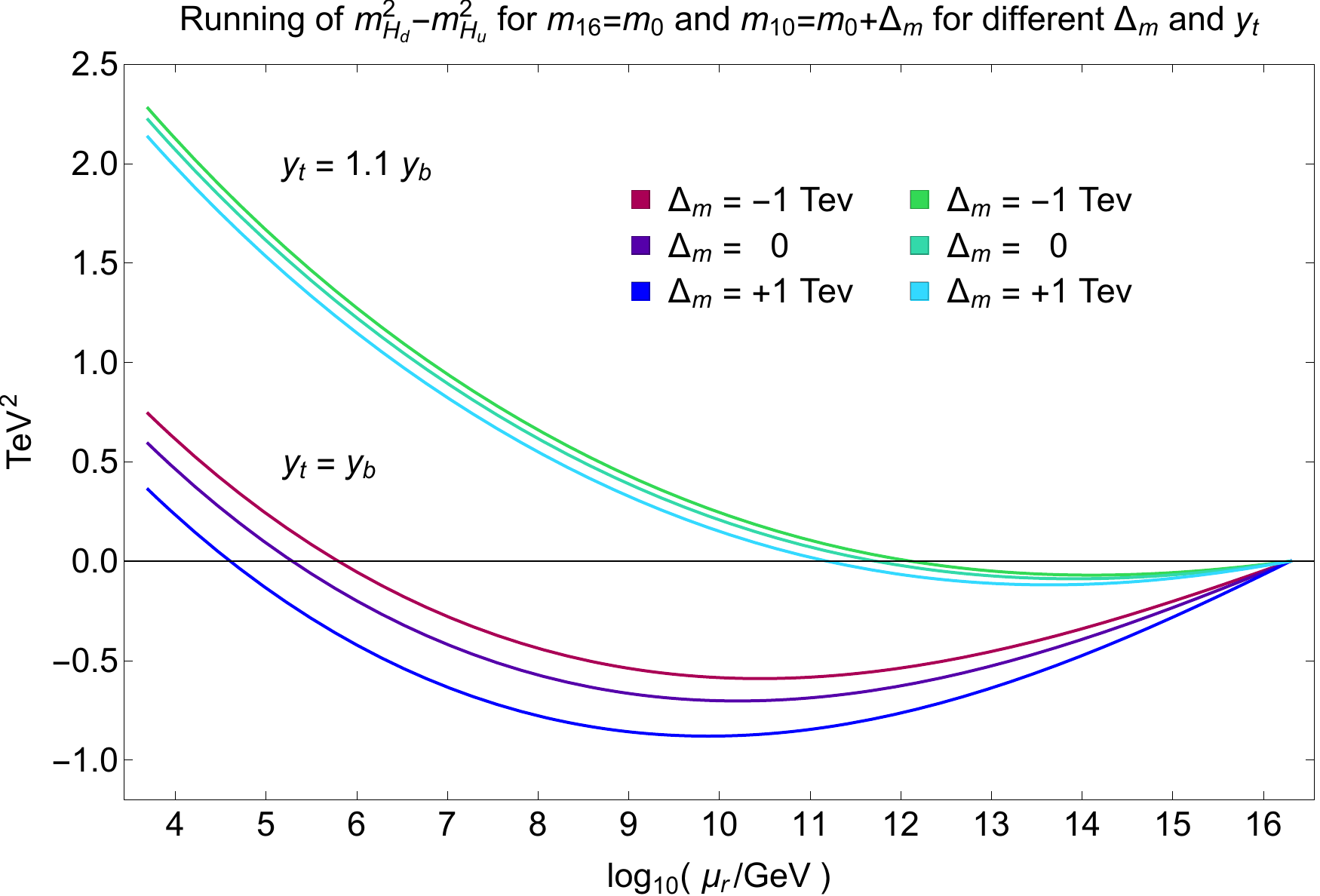}
	\end{center}
	\caption{The RGE running of $m^2_{H_d}-m^2_{H_u}$ for different splits 
	$\Delta_m$ in the soft mass parameters consistent with $\SO(10)$ unification: we take $m_{16}=m_0$ and $m_{10}=m_{0}+\Delta_m$, with $m_0$ of the 	
	example point~\eqref{eq:example-point-begin}-\eqref{eq:example-point-end}. We plot two sets of trajectories: the red-to-blue trajectories are for $y_t=y_b$ ($t$-$b$ unification), while the green-to-cyan trajectories are for the case $y_t=1.1 y_b$.  \label{Figure:RGE-m10-m16-split}}
\end{figure}
\item \textbf{Split in $m^2_{H_d}$ and $m^{2}_{H_u}$ at the GUT scale}\\
	As a final consideration, we consider how opening up a split in the soft mass parameters $m^{2}_{H_d}$ and $m^{2}_{H_u}$ at the GUT scale influences the value of the running quantity $m^2_{H_d}-m^{2}_{H_u}$ at the SUSY scale.
	\par
	The results are shown in Figure~\ref{Figure:RGE-md-mu-split}, for splits $\alpha$ in the GUT boundary conditions specified by $m_{H_d}=\alpha m_0$ and $m_{H_u}=m_0$, with $\alpha$ changing from no increase (red trajectory) to $20\,\%$ (blue trajectory). Note that separations at the GUT scale more or less carry over to the SUSY scale, at least if considered at orders of magnitude level. For example, a $(1\,\mathrm{TeV})^2$ gap between $m^2_{H_d}$ and $m^2_{H_u}$ in the boundary conditions at the GUT scale results in a similar gap of a bit less than $(1\,\mathrm{TeV})^2$ at the SUSY scale. This implies that opening up a gap of e.g.~$m^{2}_{H_d}\approx 1.13\,m^{2}_{H_u}$ at the GUT scale, as is common in the literature~\cite{
Baer:1999mc,
Baer:2000jj,
Baer:2001yy,
Blazek:2001sb,
Blazek:2002ta,
Baer:2009ie,
Gogoladze:2010fu,
Gogoladze:2011ce,
Anandakrishnan:2012tj,
Joshipura:2012sr,
Anandakrishnan:2013cwa,
Ajaib:2013zha,
Anandakrishnan:2014nea,
Altin:2017sxx}, 
can erase the effect of low extra Higgs masses.  
	 
	\begin{figure}[htb]
	\begin{center}
	\includegraphics[width=10cm]{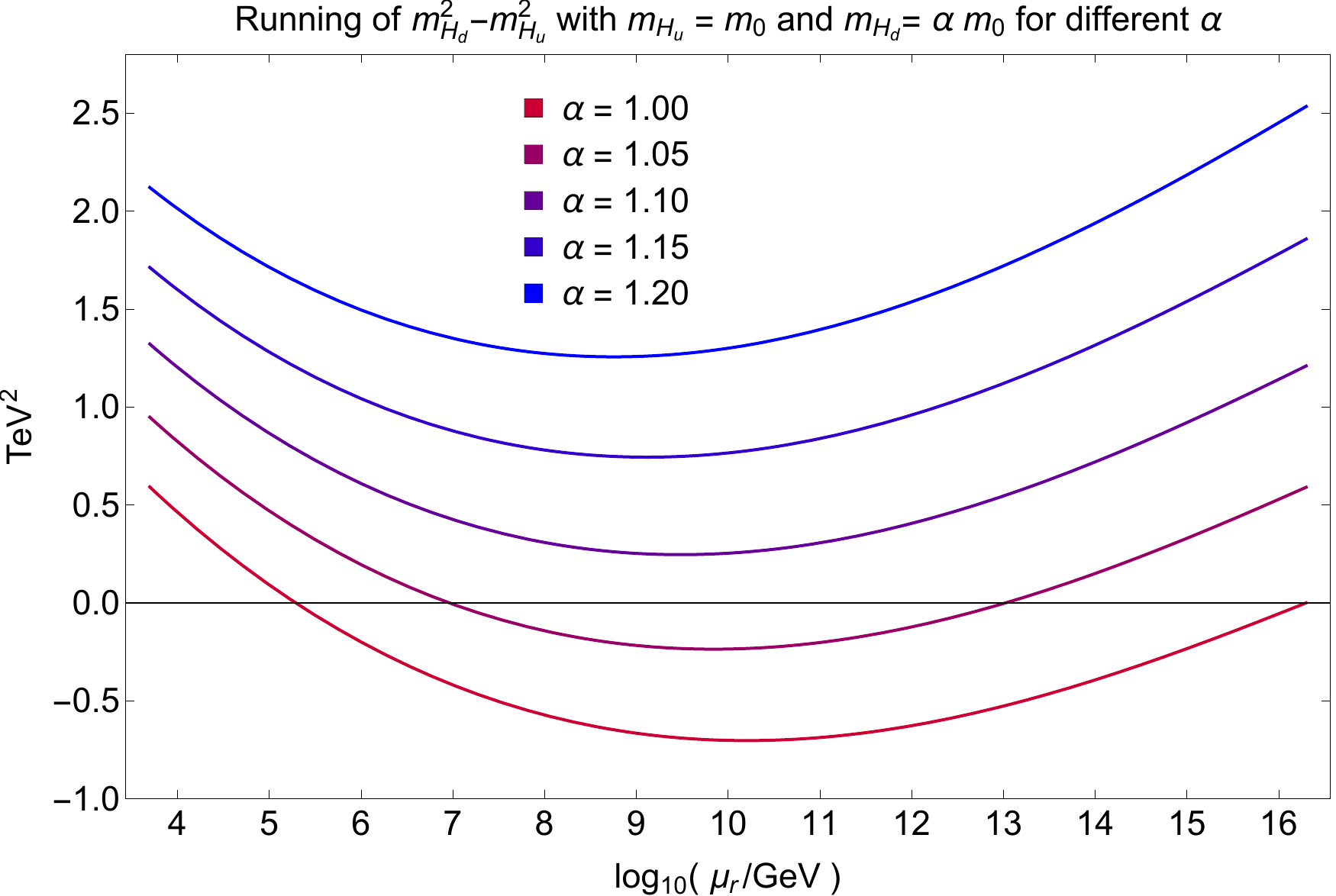}
	\end{center}
	\caption{The RGE running of $m^2_{H_d}-m^2_{H_u}$ for different splits 
	$\alpha$ between the initial values of $m^{2}_{H_d}$ and $m^{2}_{H_u}$ at the GUT scale, where $m_{H_u}=\alpha\,m_0$ and $m_{H_d}=m_0$ and $m_0$ has the value from the example point~\eqref{eq:example-point-begin}-\eqref{eq:example-point-end}. We see that RGE running to the SUSY scale preserves the initial split in the mass parameters.\label{Figure:RGE-md-mu-split}}
\end{figure}
\end{enumerate}

Our results show that the effect of low masses of the extra MSSM Higgses crucially depends on $t$-$b$ unification, while $b$-$\tau$ unification and the right-handed neutrino mass scale $M_R$ do not have as large an impact on this effect. The effect persists even if we deform CMSSM to introduce a split between sfermion and Higgs soft masses, i.e.~$\SO(10)$ boundary conditions in Eq.~\eqref{eq:m0-SO10-split-m16} and \eqref{eq:m0-SO10-split-m10}, but can be erased by opening up a further gap between the Higgs soft masses $m^{2}_{H_d}$ and $m^2_{H_u}$. On a side note, the results also show that REWSB can be performed successfully for suitable parameter points.

The above discussion also makes it clear that unless one studies scenarios of $t$-$b$ unification, usually in the context of $t$-$b$-$\tau$ unification, this effect will be missed. In particular, this effect will not be present in any kind of $\mathrm{SU}(5)$ SUSY GUT model attempting merely $b$-$\tau$ unification.


\section{The typical mass scales of the extra Higgs particles \label{section:mass-scale}}
In this section, we turn to the broader question of the predicted mass range of the extra MSSM Higgses when considering the entire region of parameter space that yields good fits to low energy data. 

We established in Section~\ref{section:MSSM-Higgs-masses} that the running difference of soft masses $m^{2}_{H_d}-m^{2}_{H_u}$ crucially determines the scale of the extra MSSM Higgs particles. A $1$-loop RGE analysis of this quantity was performed in Section~\ref{section:RGE-analysis}; results showed that with $t$-$b$-$\tau$ unification and CMSSM boundary conditions, the difference $m^{2}_{H_d}-m^{2}_{H_u}$ was indeed smaller than expected based on the mass scales of the soft parameters. Sensitivity analysis showed that this effect crucially depends on $t$-$b$ unification, while considerations such as $b$-$\tau$ unification, the right-handed neutrino scale $M_R$, and a split in the soft masses of sfermions $m_{16}$ and Higgses $m_{10}$ are of secondary concern.

The next step is a more precise calculation going beyond the proxy quantity $m^{2}_{H_d}-m^{2}_{H_u}$, instead considering the masses of the extra Higgs particles directly. We make the following improvements in the analysis for estimating the Higgs masses as accurately as possible:
\begin{enumerate}
\item The RGE running of the softly broken MSSM is performed at 2-loop level.
\item The masses of the extra Higgses are computed at 1-loop instead of tree level.
\end{enumerate}  

\noindent
To perform such improved calculations, we make use of the following tools:
\begin{itemize}
\item For 2-loop running, we make use of SusyTC \cite{Antusch:2015nwi} (version 1.2), an extension of the Mathematica based package REAP~\cite{Antusch:2005gp}. First, boundary conditions are input at the GUT scale. Then the RG running is performed by use of 2-loop RGEs for the softly broken 
MSSM\footnote{SusyTC also includes the superpotential and soft terms for right-handed neutrinos, which are automatically integrated out at the appropriate scale} from the GUT scale $\MGUT=2\cdot 10^{16}\,\mathrm{GeV}$ to the SUSY scale. The latter is computed dynamically as the geometric mean of the two lightest stop masses. At $\MSUSY$, matching of the MSSM and the ordinary SM is performed with the SusyTC option $\mathrm{sign}(\mu)=-1$, and the SUSY spectrum is computed. The sparticle masses are computed at tree-level, which we deem sufficient for all superpartners except for the masses of the Higgs sector, the details of which have an important impact on EW symmetry breaking and the scale of the extra Higgs particles. We also check for the existence of the EW symmetry-breaking vacuum at 1-loop level. 
The SM 2-loop running is then performed between the SUSY scale and the $Z$-boson scale $M_{Z}=91.2\,\mathrm{GeV}$.
\item The MSSM Higgs sector is computed to higher loop order by the program FeynHiggs~\cite{Bahl:2017aev,Bahl:2016brp,Hahn:2013ria,Frank:2006yh,Degrassi:2002fi,Heinemeyer:1998np,Heinemeyer:1998yj}, version 2.13.0. The output of SusyTC gives the Higgs masses at tree level, with the exception of $m^{2}_{\HPM}$ given at $1$-loop by using Eq.~\eqref{eq:extra-Higgs-1loop}. Using the output values of SusyTC as input for FeynHiggs, the SM Higgs mass is computed to $2$-loop and the extra Higgs particles' masses are computed to $1$-loop. 
\item For the computation of EW vacuum stability we make use of Vevacious \cite{Camargo-Molina:2013qva}. We use SusyTC to produce an SLHA file, amended with values of the MSSM $\mu$ and $b$ terms at tree and loop level, computed from the VIN file of the tree and $1$-loop potential for EW breaking produced by SARAH 4.14.1 \cite{Staub:2008uz,Staub:2013tta}. We use the SARAH predefined model with possible charge breaking via stau VEVs. 
\end{itemize}

We use these tools for improved computations of the $t$-$b$-$\tau$ unification model, where we still consider only the 3rd family Yukawa couplings to be non-vanishing as in Section~\ref{section:RGE-analysis}, and assume the right-handed neutrinos are integrated out at the GUT scale. The GUT scale values of the gauge couplings are taken to be those from Eq.~\eqref{eq:example-point-begin}-\eqref{eq:example-point-gauge-end}. We shall consider two scenarios of boundary conditions: the CMSSM scenario (5 parameters) and the $\SO(10)$ boundary condition scenario (6 parameters). The input parameters at the GUT scale are the following:
\begin{align}
\text{CMSSM scenario parameters:}&\qquad \tan\beta,\quad y_0,\quad M_{1/2},\quad a_0,\quad m_0.\label{eq:input-parameters-CMSSM}\\
\text{SO(10) scenario parameters:}&\qquad \tan\beta,\quad y_0,\quad M_{1/2},\quad a_0,\quad m_{16},\quad m_{10}.\label{eq:input-paramters-SO10}
\end{align}
We take $\mu<0$ in all cases. The standard notation of CMSSM parameters applies, the parameter $y_0$ is the $t$-$b$-$\tau$ unified Yukawa coupling, while $m_{16}$ and $m_{10}$ are defined according to Eq.~\eqref{eq:m0-SO10-split-m16} and \eqref{eq:m0-SO10-split-m10}.

Each parameter point in a scenario allows the computation of the Yukawa couplings at $M_{Z}$, the Higgs mass, as well as the SUSY spectrum. The part of the SUSY spectrum that is of greatest interest to us is the one of the masses of the extra MSSM Higgs particles; we would like to confirm that due to $t$-$b$-$\tau$ unification they should indeed be comparatively low.

As a first check, we recompute the example point from Eq.~\eqref{eq:example-point-begin}-\eqref{eq:example-point-end} with improvements of higher loop order. The results for the mass prediction of the CP-odd Higgs $\AZ$ are the following:
\begin{align}
m_{\AZ}&=\quad\underbrace{747\,\mathrm{GeV}}_{I}\quad\to\quad \underbrace{727\,\mathrm{GeV}}_{II}\quad\to\quad \underbrace{514\,\mathrm{GeV}}_{III}\;.
\end{align}
The result I corresponds to the tree level mass from Eq.~\eqref{eq:mass-A0} and 1-loop RGE, the result II corresponds to tree level mass and 2-loop RGE, while result III is the most accurate with the 2-loop RGE and 1-loop mass from FeynHiggs. We see that the predicted mass reduced after every improvement, which we find happens generically. This confirms that the low MSSM Higgs mass phenomenon persists (and may be further enhanced) even with the improved loop order in the calculation.

We now turn to a more general study of the parameter space beyond just the example point. In the subsequent analysis, the 3rd family Yukawa couplings and the SM Higgs mass are considered to be observables:
\begin{align}
\text{Observables:}&\qquad y_t,\quad y_b,\quad y_\tau,\quad m_{\hZ}.\label{eq:observables}
\end{align}
As a measure of goodness of fit we make use of the $\chi^2$ function: 
\begin{align}
\chi^2(\vec{x})=\sum_{i}\frac{\big(f_i(\vec{x})-y_i\big)^{2}}{\sigma_{i}^2},
\end{align}
where the vector $\vec{x}$ represents the input parameters of the model from either Eq.~\eqref{eq:input-parameters-CMSSM} or \eqref{eq:input-paramters-SO10}, 
while the index $i$ goes over all observables in Eq.~\eqref{eq:observables}. The $y_i$ denote the central values from the (experimental) data and $\sigma_i$ are their corresponding standard deviation errors, while $f_i(\vec{x})$ are the predictions for the $i$-th observable given the parameter point $\vec{x}$. Some observables may be equipped with asymmetric errors $\sigma_{i+}$ and $\sigma_{i-}$ when $f_{i}(\vec{x})>y_i$ or $f_{i}(\vec{x})<y_i$, respectively. 

The experimental values for the Yukawa couplings are considered in the $\overline{\text{MS}}$ scheme. The central values $y_i$ for the $3$ Yukawa couplings at the scale $M_{Z}$ are taken from Table~1 in \cite{Antusch:2013jca}, with relative errors adjusted upwards to $1\,\%$ due to limited  precision of our RGE procedure from $M_{GUT}$ to $M_{Z}$. The SM Higgs mass central value was taken to be $m_h=125.09\,\mathrm{GeV}$ \cite{Aad:2015zhl}, with a $\mathrm{3}\,\mathrm{GeV}$ error due to theoretical uncertainties in the computation.

\begin{figure}[htb]
	\begin{center}
	\includegraphics[width=10cm]{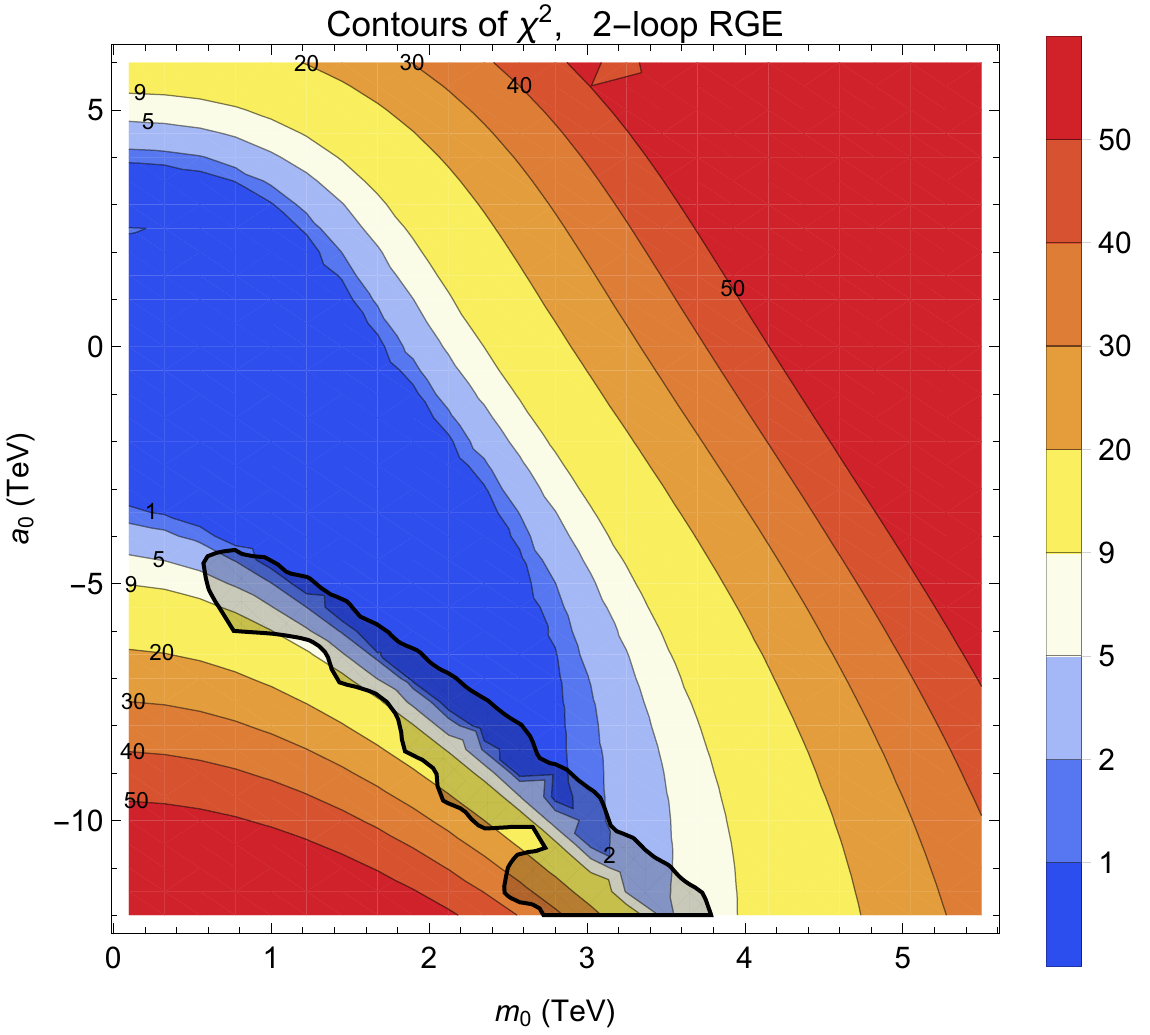}
	\end{center}
	\caption{A contour plot of the minimum $\chi^2$ for a point with fixed $m_0$ and $a_0$, while $\tan\beta$, $y_0$ and $M_{1/2}$ are varied. The darker region inside the black curve represents minimized points with an EW vacuum lifetime smaller than $10\times$ the age of the universe. In the lighter region outside of the black curve the vacuum is stable or sufficiently long-lived.
	} \label{Figure:CMSSM-fit-chi2}
\end{figure}

\begin{figure}[htb]
	\begin{center}
	\includegraphics[width=10cm]{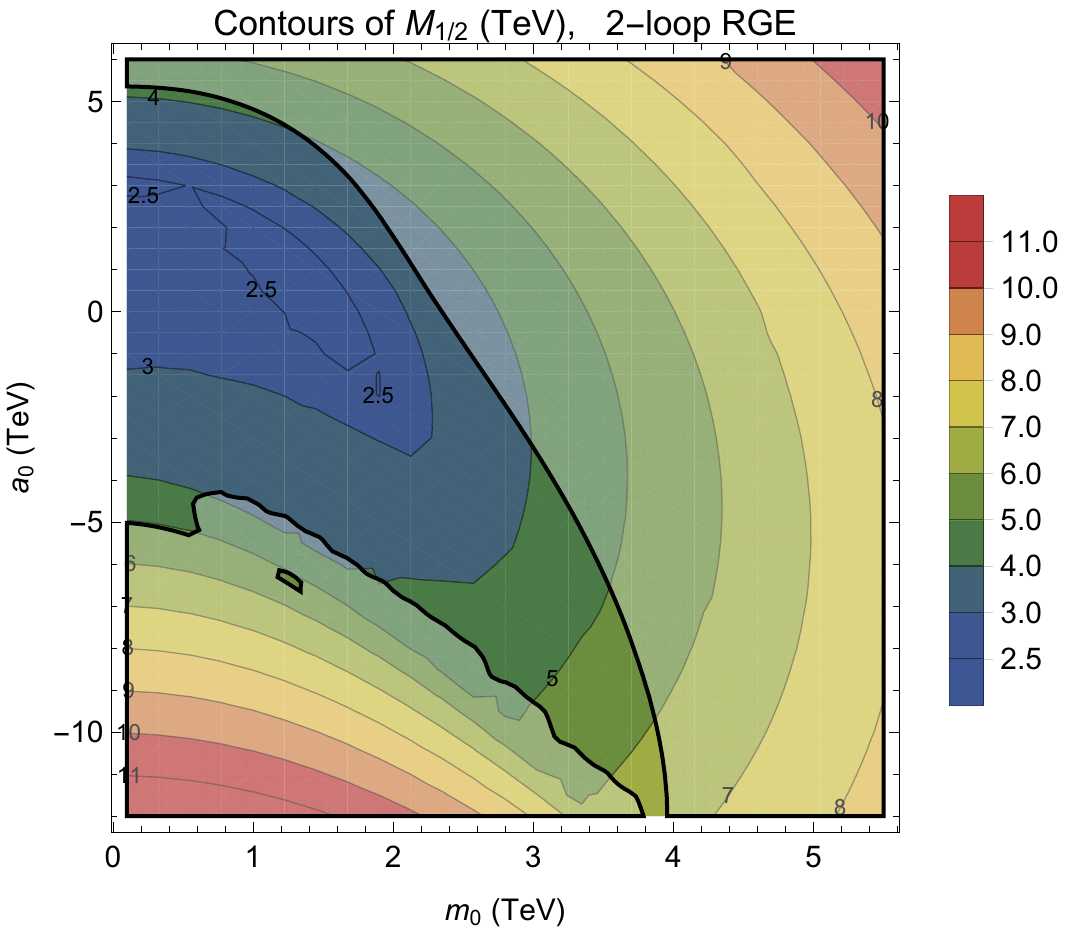}
	\end{center}
	\caption{A contour plot with values of the soft mass parameter $M_{1/2}$ for a $\chi^2$ best-fit point given a fixed $m_0$ and $a_0$. The fully-colored triangular region represents the one allowed by $\chi^2<9$ and vacuum stability from Figure~\ref{Figure:CMSSM-fit-chi2}. \label{Figure:CMSSM-fit-M12}}
\end{figure}

\begin{figure}[htb]
	\begin{center}
	\includegraphics[width=10cm]{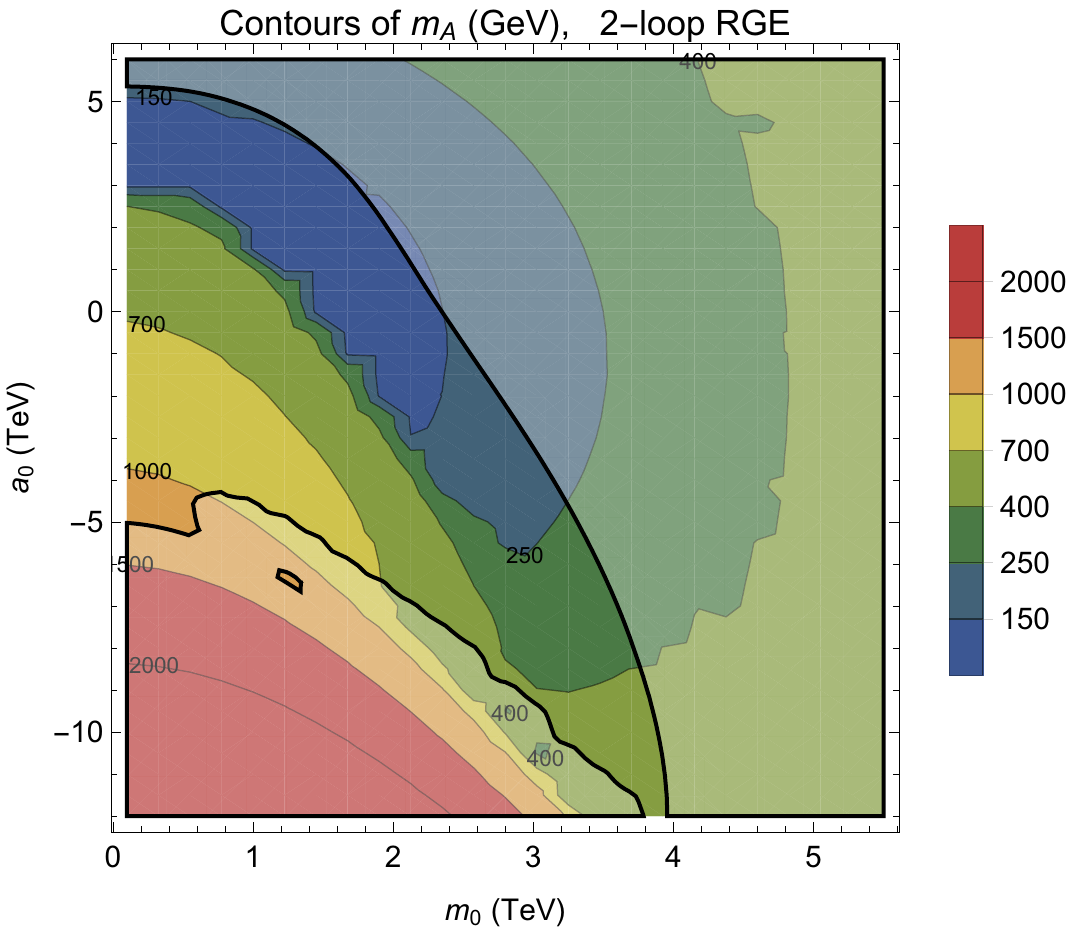}
	\end{center}
	\caption{A contour plot of the MSSM neutral CP-odd Higgs mass $m_{\AZ}$ for $\chi^2$ best-fit points given a fixed $m_0$ and $a_0$. This shows that lower than expected, (i.e.~sub-$\mathrm{TeV}$) masses of the extra MSSM Higgs particles are a general feature of $t$-$b$-$\tau$ unification. The fully-colored triangular region represents the one allowed by $\chi^2<9$ and vacuum stability from Figure~\ref{Figure:CMSSM-fit-chi2}.\label{Figure:CMSSM-fit-mHP}}
\end{figure}

We show that the prediction of a low extra Higgs mass is a generic feature of $t$-$b$-$\tau$ unification rather than of just the example point from the previous section. For this reason we search for a number of other points in the parameter space of CMSSM, which provide a good fit of the observables. We do this by a systematic search in the $m_0$-$a_0$ plane of parameters. For a fixed $m_0$ and $a_0$, we perform a minimization of the $\chi^2$ for the other $3$ input parameters $M_{1/2}$, $y_0$ and $\tan\beta$ in Eq.~\eqref{eq:input-parameters-CMSSM}. Remember that these $3$ free parameters are used to fit $4$ observables of Eq.~\eqref{eq:observables}, which may not necessarily be possible for an arbitrary point in the $m_0$-$a_0$ plane. The computation involves a minimization of $\chi^2$ for each point in a $25\times 37$ grid and subsequent interpolation between grid points; the points were taken equidistant and in the range
\begin{align}
&100\,\mathrm{GeV}\leq m_0 \leq 5500\,\mathrm{GeV},&&-12000\,\mathrm{GeV}\leq a_0 \leq 6000\,\mathrm{GeV},
\end{align} 
and include the edge points of these intervals. As we shall see, this range includes the entire region of admissibly low $\chi^2$, at least in the CMSSM context. The relevant results of this fit are summarized in Figures~\ref{Figure:CMSSM-fit-chi2}, \ref{Figure:CMSSM-fit-M12} and \ref{Figure:CMSSM-fit-mHP}. We analyze them below:
\begin{itemize}
\item Figure~\ref{Figure:CMSSM-fit-chi2} shows the contours of the minimal attainable $\chi^2$ for a point in the $m_0$-$a_0$ plane, with the shaded region excluding points due to vacuum stability, to be discussed below. Contour regions from blue to white represent points where a reasonable fit can be obtained: the darkest shade of blue represents almost perfect fits of $\chi^2<1$, while the white region represents the edge points where $\chi^2<9$, such that the deviation in any one observable cannot be more than $3\sigma$. We see that the allowed region in the $m_0$-$a_0$ plane is compact: the ranges are roughly
\begin{align}
&m_0 < 4\,\mathrm{TeV},&&-12\,\mathrm{TeV}<a_0<5\,\mathrm{TeV},
\end{align}
i.e.~the regions involve scales of a few $\mathrm{TeV}$. 
\item The darkly shaded region in Figure~\ref{Figure:CMSSM-fit-chi2} corresponds to points in the $m_0$-$a_0$ plane for which $\chi^2$ has been minimized, but the vacuum is not sufficiently stable. The threshold is taken to be at $10\times$ the current age of the universe, but the exponential sensitivity of the lifetime to the bounce action (see~\cite{Coleman:1977py,Callan:1977pt,Wainwright:2011kj}) means that one order of magnitude difference in the threshold does not appreciably change the excluded area. The unshaded region thus represents points with the EW vacuum either being metastable with a sufficiently long lifetime or stable. Note that the instability in the shaded region does not necessarily exclude all possible points with a given $m_0$ and $a_0$, but only the one minimizing $\chi^2$. Although an improved approach would be to include a sufficiently long vacuum lifetime as a necessary condition in the minimization of $\chi^2$, this would be much more demanding computationally. Ultimately, the vacuum computation performed here is sufficient to show that most of the low $\chi^2$ region consists of allowed points.
\item The minimization of $\chi^2$ gives the following ranges for $\tan\beta$ and $y_0$ for all best-fit points:
\begin{align}  
&48<\tan\beta<55,&&0.44<y_0<0.50.
\end{align}
These two parameters thus have small relative changes for best-fit points with different CMSSM soft parameters. The results are compatible with the well-known fact that $t$-$b$-$\tau$ unification requires $\tan\beta\approx 50$, while the unified coupling is approximately $y_0\approx 0.5$. A more interesting input parameter to track for different best-fit points in the $m_0$-$a_0$ plane, however, is the gaugino mass parameter $M_{1/2}$, since this provides the information for all CMSSM soft parameters of the well-fit points. A contour plot of the $M_{1/2}$ values is presented in Figure~\ref{Figure:CMSSM-fit-M12}; this data represents
a $2D$ surface of best ($3$rd family) Yukawa fits in the CMSSM soft-parameter space of $m_0$, $M_{1/2}$ and $a_0$. Any good fit of $t$-$b$-$\tau$ unification in the CMSSM would thus be expected to always lie in a compact region around the hypersurface: the $m_0$ and $a_0$ values would need to lie in the region of low $\chi^2$, while the $M_{1/2}$ value would need to lie near the one for the best-fit point. Results show that $M_{1/2}$ values of most best-fit points with $\chi^2<9$ lie in the range between $2.5\,\mathrm{TeV}$ and $6\,\mathrm{TeV}$, with the value increasing with increasing $m_0$ and $|a_0|$.
\item Figure~\ref{Figure:CMSSM-fit-mHP} shows the predicted mass $m_{\AZ}$ (at 1-loop) of the neutral CP-odd MSSM Higgs $\AZ$, which is the main result of interest. Note that CP is not broken at 1-loop, because our parameters do not have complex phases. We see that all best-fit points in the allowed region of the $m_0$-$a_0$ plane give a relatively low mass $m_{\AZ}$, roughly in the range between $150\,\mathrm{GeV}$ 
and $1200\,\mathrm{GeV}$. 
 Important note: the $m_{\AZ}$ values are given only for the best-fit points, so one should be careful not to interpret the figure as a precise prediction of the CP-odd Higgs mass as a function of only $a_0$ and $m_0$. 
 
 The results show our main premise: the low (or lower than expected scale of the extra Higgses, i.e. typically $<1\,\mathrm{TeV}$) is a relatively universal feature of $t$-$b$-$\tau$ unification, and does not depend on the precise values of the soft parameters. The extra Higgses are typically by far the lightest MSSM particles in such scenarios.
 This justifies our assertion that the example point chosen in Section~\ref{section:RGE-analysis} indeed exhibits generic features in regard to the low Higgs mass.
 
 Note the following important reservation about the results: they merely show the ``naive'' predicted mass of the extra Higgs particles in the CMSSM model. Potential experimental constraints have not been considered in this plot. In fact, as shall be discussed in the next section, practically the entire region predicted here (assuming exact $t$-$b$-$\tau$ unification) is under severe stress from ATLAS and CMS searches of $\HZ\to\tau\tau$. 
\end{itemize}

\section{Challenges to $t$-$b$-$\tau$ unification\label{section:experimental-constraints}}
We have seen in Section~\ref{section:mass-scale} that the scale of the extra MSSM Higgses is generically expected to be low in $t$-$b$-$\tau$ unification. The ultimate reason lies in the RG flow of the quantity $m^2_{H_d}-m^{2}_{H_u}$, which was analyzed in Section~\ref{section:RGE-analysis}, and found to have a relatively small yet positive value, the latter being important for consistent EWSB. In this section, we analyze the predictions of $t$-$b$-$\tau$ further and confront them with experimental data from the LHC.

As a first step, we extend the CMSSM scenario to the more general one with $\SO(10)$ boundary conditions, where the parameters consist of those in Eq.~\eqref{eq:input-paramters-SO10}, while the $\chi^2$ is again defined with the observables of Eq.~\eqref{eq:observables}. The standard deviations are taken as follows: the relative errors of the 3rd family Yukawa couplings are taken to be $1\,\%$, while the error of for the SM Higgs mass is taken to be $2\,\mathrm{GeV}$ due to theoretical uncertainties in the computation. 

This time we compute the overall expectations from this setup (with no fixed parameter values) by computing posterior probability densities of quantities of interest in a Bayesian approach by use of the Markov Chain Monte Carlo algorithm. 

This paragraph contains some technical details of the computation. The MCMC algorithm was performed with 12 parallel chains, each yielding $1.3\cdot 10^5$ points after discarding the initial bunch of $10^4$ in the burn-in period. The total number of used data points is thus $1.56$ million. Vacuum existence at $1$-loop was checked, but not vacuum stability under EM charge breaking.

\begin{figure}[htb]
	\begin{center}
	\includegraphics[width=14cm]{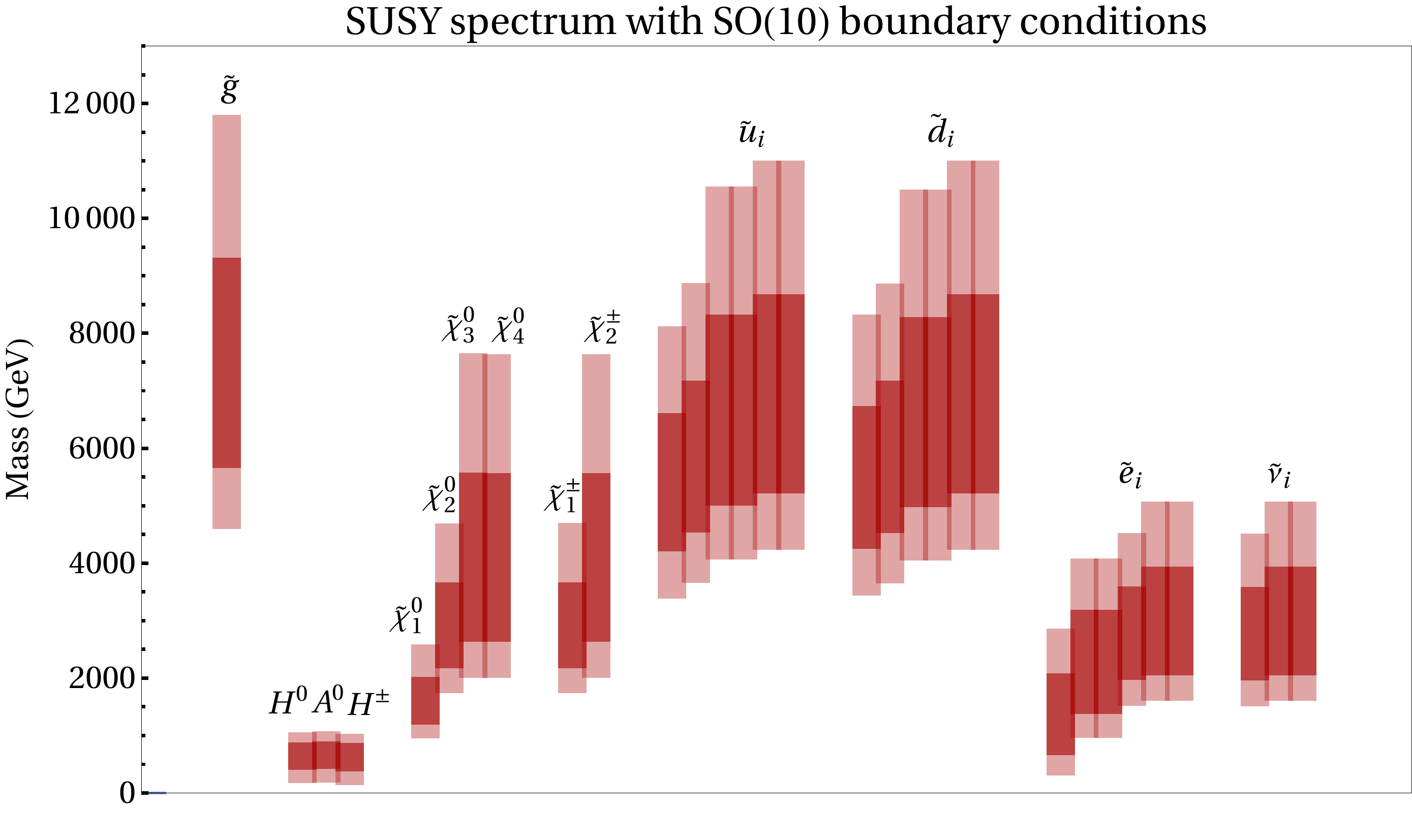}
	\end{center}
	\vskip -0.5cm
	\caption{The $1$-$\sigma$ (dark) and $2$-$\sigma$ (light) HPD intervals of the SUSY spectrum in the model with $\SO(10)$ boundary conditions. The lowest lying part of the spectrum are the extra MSSM Higgs states.
	\label{Figure:susy-spectrum}}
\end{figure}

The result of interest from the MCMC computation is the SUSY sparticle spectrum, which turns out to be quite predictive, due to good fits obtained only in a compact region of parameter space, analogously to Section~\ref{section:mass-scale}. The results are presented in Figure~\ref{Figure:susy-spectrum}, where we draw the $1$-$\sigma$ and $2$-$\sigma$ highest posterior density (HPD) intervals for the masses of the sparticles. We use the labels $\tilde{g}$ for gluinos, $\tilde{\chi}^{0}_i$ for neutralinos, $\tilde{\chi}^{\pm}_i$ for charginos, $\tilde{u}_{i}$ for up-type squarks, $\tilde{d}_{i}$ for down-type squarks, $\tilde{e}_{i}$ for charged sleptons and $\tilde{\nu}_{i}$ for sneutrinos, where the index $i$ goes over different ranges for different types of superpartners, but always corresponds to increasing mass (these are mass eigenstates, so the index $i$ is not directly related to flavor).

We make the following comments on the sparticle spectrum results:
\begin{itemize}
\item The lowest part of the SUSY spectrum are the extra Higgs particles $\HZ$, $\AZ$ and $\HP$. They are expected in the rough range between $\mathrm{500}\,\mathrm{GeV}$ and $\mathrm{1000}\,\mathrm{GeV}$. This reproduces the results for the case of CMSSM from Section~\ref{section:mass-scale}.
\item The next lightest states are the lightest neutralino $\tilde{\chi}^0_1$ and the lightest charged slepton $\tilde{e}_1$. We see from the expected ranges that the lightest supersymmetric particle (LSP) for some points must be the lightest charged slepton (i.e.~the stau) instead of the neutralino. Such points are experimentally problematic, since they would predict a charged LSP as a dark matter candidate. We performed a second MCMC analysis with the added constraint that the LSP must be the neutralino; this addition only minimally changes the quantitative predictions for HPD intervals of the other parts of the spectrum, so we choose not to include a separate plot. 
\item The rest of the spectrum is higher than $2\,\mathrm{TeV}$, with gluinos typically at $>5\,\mathrm{TeV}$. An interesting feature is that the sleptons are expected to have lower masses than squarks.
\end{itemize}

The predicted sparticle spectrum is mostly compatible with the LHC data and searches for these particles, with one notable exception: the extra MSSM Higgs particles. The most stringent constraint comes from the possible ditau decay of neutral Higgses $\HZ/\AZ\to \tau\tau$. The general scenario relevant in our case is  the so called hMSSM~\cite{Djouadi:2013uqa}, which assumes for all SUSY particles other than Higgses to be above $1\,\mathrm{TeV}$. It was shown that specifying only 
two parameters, $\tan\beta$ and $m_{\AZ}$, is sufficient to uniquely predict other tree-level quantities. The observed ditau rate is consistent with the SM background, so the non-observation of $\HZ$ or $\AZ$ is summarized by upper bounds on $\tan\beta$ for a given $m_{\AZ}$ in the $m_{\AZ}$-$\tan\beta$ plane. The latest ATLAS~\cite{Aaboud:2017sjh} and CMS~\cite{Sirunyan:2018zut} results on this, using the dataset with $36\,\mathrm{fb}^{-1}$ of integrated luminosity at $\sqrt{s}=14\,\mathrm{TeV}$, suggest a bound of $m_{\AZ}\gtrsim 1.5\,\mathrm{TeV}$ at $\tan\beta\approx 50$.

Based on Figure~\ref{Figure:susy-spectrum}, the $t$-$b$-$\tau$ model prediction for the mass of $\HZ$ and $\AZ$ is clearly in tension with the experimental bounds, at least for most of the otherwise available parameter space. In fact, a search among computed MCMC points showed that the extra Higgs masses in the scenario of $\SO(10)$ boundary conditions cannot go much higher than $1200\,\mathrm{GeV}$ (since that would incur a severe $\chi^2$ penalty). Comparing the various contributions to $\chi^2$ shows that the tension comes from the SM Higgs mass, which tends to be dragged too high for high values of the extra Higgses.

This result is consistent with the upper limit for the best fit points in the more constrained CMSSM scenario, see Figure~\ref{Figure:CMSSM-fit-mHP}; the additional parameter gained by the split of $m_{0}$ to $m_{16}$ and $m_{10}$ in the $\SO(10)$ boundary conditions thus does not appear to gain much maneuvering space over CMSSM for increasing the masses of the extra Higgs states. The CMSSM region in Figure~\ref{Figure:CMSSM-fit-mHP} with high extra Higgs masses is 
located at small $m_0$, i.e.~$m_0\lesssim 500\,\mathrm{GeV}$, while $a_0\sim -5\,\mathrm{TeV}$. 

This result indicates that exact $t$-$b$-$\tau$ unification, at least within the $\SO(10)$ boundary conditions scenario, is under strain exactly because of the low masses of the extra MSSM Higgses, the very feature pointed out and studied in this paper.

There are some possibilities, however, how to potentially relax the tension with experiment and allow for higher masses of extra Higges, while keeping the SM Higgs at the measured value:
\begin{enumerate}
\item We have seen that the low mass feature in extra Higgs states is especially sensitive to $t$-$b$ unification, cf.~Figure~\ref{Figure:CMSSM-fit-mHP}. Even just a few percent deformation in $t$-$b$ unification can substantially help with raising the masses of the extra Higgses. Such magnitudes for the deformation of $t$-$b$-$\tau$ unification could easily occur either due to GUT threshold corrections, which depend on the extra states in the $\SO(10)$ GUT breaking sector, or Planck scale suppressed operators, which could break the discrete symmetry responsible for the dominance of the $\mathbf{16}_i\cdot\mathbf{16}_j\cdot\mathbf{10}$ operator for the flavor entry $i=j=3$. Although all Yukawa couplings could obtain a threshold correction, we shall study only the case where $y_t$ splits from the others.
\item One expected extension of the MSSM at high energies, especially in the context of $\SO(10)$ GUT, is the extension by right-handed neutrinos. Although this does not influence the low masses of the extra Higgses directly, cf.~Figure~\ref{Figure:RGE-neutrinos}, it may have an indirect effect due to changing the running of Yukawa couplings at scales near the GUT scale. We shall investigate this possibility below.
\item Our analysis also assumed the GUT scale to be fixed at $2\cdot 10^{16}\,\mathrm{GeV}$. Changing the GUT scale could change the length of running of all the quantities, thus changing the value of the running quantity $m^2_{H_d}-m^{2}_{H_d}$. Trying this out numerically in our setup, we surprisingly found that the fit is helped by lowering and not raising the GUT scale, which is undesired from the point of view of proton decay. Nevertheless, this option remains a possibility, especially if one considers modifications of RGE due to other GUT particles, but we shall not pursue this possibility further in the paper.
\item The location of the MSSM Higgs doublets in $\SO(10)$ representations depends on the GUT breaking sector and details of doublet-triplet splitting. It may happen that the low mass MSSM doublets $H_u$ and $H_d$, which are mass eigenstates, are not aligned with the (flavor) doublet states in $\mathbf{10}$ of $\SO(10)$ due to the presence of other representations; the coefficients of $H_u$ and $H_d$ in that case may not be the same. In such a scenario the 3rd family Yukawa coefficients still come from an operator $\mathbf{16}\cdot\mathbf{16}\cdot\mathbf{10}$, but the different coefficients with which $H_u$ and $H_d$ are present in the doublet states of the $\mathbf{10}$ spoil $t$-$b$-$\tau$ unification in the effective MSSM theory below the GUT scale. Although possible, we do not consider this case further, since the spoiling of $t$-$b$-$\tau$ unification can essentially then be of any magnitude and pattern; what we are really interested in this analysis is keeping the $t$-$b$-$\tau$ unification pattern in the MSSM effective theory. 
\end{enumerate}

\begin{figure}[htb]
	\begin{center}
	\includegraphics[width=10cm]{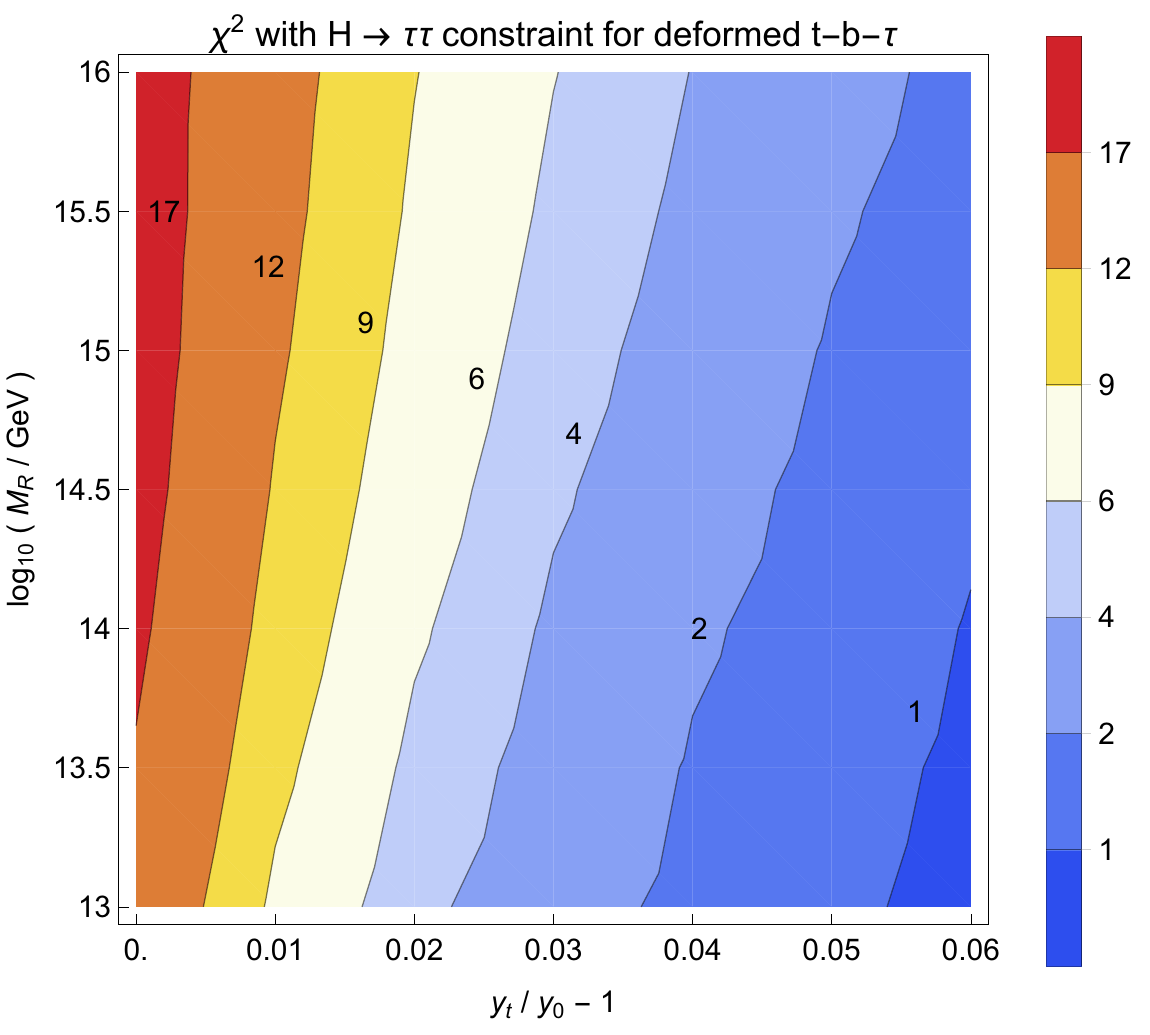}
	\skip -0.2cm
	\caption{The contour lines of minimal $\chi^2$ achieved by $\SO(10)$ boundary conditions for a fixed $t$-$b$ deformation $y_t/y_0-1$ and fixed scale $M_R$ of the right-handed neutrino of the 3rd family. The $\chi^2$ includes the constraint from ditau decays of $\HZ/\AZ$, thus requiring a large $m_{\AZ}$. Larger $t$-$b$ deformations substantially help with the fit, and to a smaller degree a lower $M_R$ does as well.\label{Figure:chi2-for-deformed}}
	\end{center}
\end{figure}

Out of the 4 caveats mentioned, we study now the effect of the first two, which we deem to be the most relevant for our analysis. The results are presented in  Figure~\ref{Figure:chi2-for-deformed}. We first provide some technical details regarding the computation of this plot and what was minimized, and then discuss the results.

Since we are now interested also in the masses of the extra Higgses, we perform the minimization with more observables in the $\chi^2$. For the input we have the $\SO(10)$ boundary condition parameters, now also assuming a possible split in $t$-$b$ and one right-handed neutrino (the one with the largest Yukawa coupling, i.e.~the unified coupling, in the Dirac mass term) at the scale $M_R$, which may now be below $\MGUT$. The other two Majorana type masses of the right-handed neutrinos are again set at the GUT scale. The input parameters are now
\begin{align}
\text{Deformation scenario parameters:}&\qquad \tan\beta,\quad y_0,\quad y_t,\quad M_{1/2},\quad a_0,\quad m_{16},\quad m_{10},\quad M_R,\label{eq:input-paramters-deformation}
\end{align}
where the unified Yukawa coupling now excludes the top coupling $y_t$: 
\begin{align}
y_0:=y_b\atMGUT=y_\tau\atMGUT=y_{\nu}\atMGUT.
\end{align}
As for the $\chi^2$, we consider the observables from Eq.~\eqref{eq:observables}, with two additional penalty terms. The first penalty term is associated to the non-observation of $\HZ/\AZ\to\tau\tau$ at the LHC, and is present only if $\tan\beta$ is too high given the value of $m_{\AZ}$. The expected values of the $\tan\beta$ upper bound and $1$-$\sigma$ upper error of the constraint (extended to bigger errors assuming a Gaussian profile) are taken from Figure~10b from the ATLAS analysis~\cite{Aaboud:2017sjh}. The other penalty 
basically enforces the 
neutralino to be the LSP, which turns out to be easily possible.

We now fix the $t$-$b$ deformation quantity $y_t/y_0-1$ and $M_R$, and perform a minimization in the other parameters. We do so for each point in a $7\times 7$ grid of equidistant points in the ``deformation plane'' of $y_t/y_0-1$ and $M_R$. The results of the minimized $\chi^2$ (using interpolation of the grid results to show contours) is shown in Figure~\ref{Figure:chi2-for-deformed}. The range of $t$-$b$ deformations is taken from $0$ to $6\,\%$, while the right-handed neutrino scale $M_R$ is considered on a logarithmic axis in the range between $10^{13}\,\mathrm{GeV}$ and $10^{16}\,\mathrm{GeV}$. Note: the points were checked for the existence of the EW vacuum at $1$-loop, but not explicitly for vacuum stability due to too excessive computation time. On the other hand, the points are close to points which have been checked with Vevacious, and overall in an unproblematic region with respect to vacuum stability. 
All points in the Figure have the extra MSSM Higgs particles as the lowest lying states at around $1.3$-$1.5\,\mathrm{TeV}$ in the sparticle spectrum, followed by the neutralino with a mass $>2\,\mathrm{TeV}$.  

As stated earlier, the main difficulty is the reconciliation of the SM Higgs mass with the $\HZ/\AZ\to\tau\tau$ constraint on extra Higgs masses. The best fit points all have small $m_{10}$, i.e.~$m_{10}<500\,\mathrm{GeV}$, as in the CMSSM case, but the $m_{16}$-$m_{10}$ split now allows for a bit bigger $a_0$ in magnitude without compromising $\chi^2$: $a_{0}\sim -10\,\mathrm{TeV}$.

The results clearly show that the $t$-$b$ deformation at a few percent level can indeed greatly reduce the tension (for example the blue region in the plot corresponding to $\chi^2<6$). This actually happens in two ways: first, it increases the masses of the extra Higgs particles and thus $m_{\AZ}$ (RGE effect), and second, it allows for a smaller $\tan\beta$ of around $46$, which also relaxes tension, since $\HZ/\AZ\to\tau\tau$ constraints are in the form of an upper bound on $\tan\beta$. In addition, Figure~\ref{Figure:chi2-for-deformed} also shows that the fit is improved by a lower right-handed neutrino scale, but the effect is sub-dominant compared to the $t$-$b$ deformation. 

Another important result of the minimization in the grid worth stating is also the following: the best fit points still tend to have the extra Higgs masses at the lower end of the allowed range. The non-deformed points under tension have the Higgs just above $1300\,\mathrm{GeV}$, while the deformed points not-under tension have those masses up to $1500\,\mathrm{GeV}$. Though the ditau constraint did not require them to be higher than around $1500\,\mathrm{GeV}$, this still shows that the deformed points have a preference for lower rather than higher masses of $m_{\AZ}$. A continuing non-observation of the ditau decay coming from $\HZ/\AZ$ neutral MSSM Higgses at the LHC would thus put the other points under increasing strain as well, requiring an ever larger $t$-$b$ deformation.

\section{Conclusions}

We considered in this paper $t$-$b$-$\tau$ Yukawa unification in the context of $\SO(10)$ SUSY GUTs with $\mu<0$. The $\mu<0$ is the preferred sign for Yukawa unification, since it provides the SUSY threshold corrections to the $b$ quark in the correct direction. Below the GUT scale, a good effective description is a softly broken MSSM possibly extended by right-handed neutrinos (if they are not yet integrated out). The boundary condition for the soft parameters at the GUT scale are assumed to be CMSSM-like, except for an additional split of the scalar soft mass parameter $m_0$ into sfermion masses $m_{16}$ and the mass parameter $m_{10}$ of the Higgs doublets $H_{u}$ and $H_{d}$, since these two soft mass parameters involve particles from different $\SO(10)$ representations. In particular, the features most important for comparison with the existing literature are exact Yukawa unification as opposed to quasi-unification, $m^{2}_{H_d}=m^{2}_{H_u}$ at the GUT scale, $\mu<0$, and universal gaugino masses.

We consider the above scenario to be the vanilla setup for Yukawa unification in $\SO(10)$, yet this has remained a largely unexplored possibility in the literature, where one or more of our stated assumptions are violated in an important way. The reason for that was a pessimistic outlook on the possibility of REWSB, based on approximate semi-analytic solutions of RGEs. In contrast, we show in this paper that REWSB is in fact possible to achieve by solving the full set of RGEs numerically.

The quantity of interest for successful EWSB is $m^{2}_{H_d}-m^{2}_{H_u}$, which must be positive at the SUSY scale. In the large $\tan\beta$ regime needed for Yukawa unification, this same quantity determines also the mass scale of the extra MSSM Higgs particles $\HZ$, $\AZ$ and $\HPM$ (cf.~Section~\ref{section:MSSM-Higgs-masses}). We find that the running quantity $m^{2}_{H_d}-m^{2}_{H_u}$ vanishes at the GUT scale due to the boundary conditions, first runs to negative values at lower scales, but the trend then reverses and it results in a positive value at $\MSUSY$. Crucially, this positive value is smaller than might be expected based on the scale of the soft parameters, typically below $\mathrm{TeV}$ (when assuming exact $t$-$b$-$\tau$ Yukawa unification at the GUT scale). This yields a SUSY mass spectrum with the characteristic feature that the extra Higgs states are the lowest lying sparticle states, a feature that we focused on in this paper.

We study in detail the 1-loop RGE running of the quantity $m^{2}_{H_d}-m^{2}_{H_u}$ in Section~\ref{section:RGE-analysis}; we analyze the various contributions to its beta function, as well as determine the sensitivity to various deformations of boundary conditions. We find that the low mass feature for the extra MSSM Higgs particles is very sensitive to the exactness of $t$-$b$ unification, with a $10\,\%$ percent deformation easily raising the scale by a factor of $2$. The $b$-$\tau$ unification, presence of right-handed neutrinos, or a split of a universal scalar soft mass $m_0$ into the sfermion and Higgs parameters $m_{16}$ and $m_{10}$, on the other hand, produce numerically a far more modest effect. Given the large sensitivity to $t$-$b$ deformations, we conclude that a top-down RGE calculation is more suitable to accurately model the extra Higgs masses in exact $t$-$b$-$\tau$ unification. 

This effect of low extra Higgs masses is ubiquitous in the entire parameter space, at least where $t$-$b$-$\tau$ unification leads to realistic Yukawa values at low energies. Most of the parameter space, both in the CMSSM and in the $\SO(10)$ boundary condition scenario, where a good fit to the 3rd family Yukawa couplings and the SM Higgs mass can be obtained, favors the extra Higgs masses at less than $1\,\mathrm{TeV}$ (for the case of exact $t$-$b$-$\tau$ unification), as presented in Sections~\ref{section:mass-scale} and \ref{section:experimental-constraints}. 
 
These model predictions, however, are in tension with ATLAS and CMS searches of ditau decays of neutral extra Higgses, i.e.~$\HZ/\AZ\to\tau\tau$. The experimental searches result in upper bounds on $\tan\beta$ as a function of $m_{\AZ}$. Since $t$-$b$-$\tau$ unification requires a large $\tan\beta\approx 50$, this suggests the extra Higgses to be above roughly $1.5\,\mathrm{TeV}$. In exact $t$-$b$-$\tau$ unification with correct Yukawa predictions at low scales, it is hard to achieve masses above $\sim 1.3\,\mathrm{TeV}$; the main obstacle turns out to simultaneously obtain heavy extra Higgses alongside a sufficiently low SM Higgs mass near $125\,\mathrm{GeV}$.

The tension with experiment can be reduced by relaxing exact $t$-$b$-$\tau$ unification. As shown in Section~\ref{section:experimental-constraints}, a deformation of $t$-$b$ unification at a level of a few percent can completely relieve the tension with experiment, both by raising the masses of the extra Higgs particles and lowering the required $\tan\beta$. Such a deformation of a few percent could come about from GUT threshold corrections, especially given the large numbers of particles in the $\SO(10)$ representations in the Higgs sector (which are of course model dependent), or Planck scale suppressed operators. It should be noted, however, that even deformed $t$-$b$-$\tau$ unification prefers lower rather than higher extra Higgs masses.

In summary, we have shown that $t$-$b$-$\tau$ (quasi-)unification in $\SO(10)$ SUSY GUTs with $\mu<0$ generically features comparably light extra MSSM Higgs particles. For exact $t$-$b$-$\tau$ unification we find a tension with LHC constraints from $\HZ/\AZ\to\tau\tau$, due to predicting too light masses of the extra MSSM Higgses. The tension can be successfully alleviated by relaxing the scenario to quasi-unification of Yukawa couplings: 
 a few percent split of the top Yukawa from the unified value (most importantly from the bottom Yukawa) can bring the extra Higgs states to sufficiently high values to avoid the present experimental constraints. Nevertheless, masses of these states close to the present bounds are still preferred. 
This implies that a continuing non-observation of the extra MSSM Higgses would require ever bigger deformation of $t$-$b$-$\tau$ unification, finally disfavoring the scenario. 
Conversely, an observation of an extra Higgs state in the ditau decay channel could be the first sparticle observation of the $t$-$b$-$\tau$ unified $\SO(10)$ SUSY GUT model, and measuring a sparticle spectrum with extra Higgses having the lowest masses could be a hint for the realization of this scenario in nature.


\section*{Acknowledgements}
The work of S.A., C.H.~and V.S.~has been supported by the Swiss National Science Foundation. The authors would like to thank Werner Porod for discussion, and  Ahmed Hammad for useful tips on the use of Vevacious.

\newpage
\appendix

\section{General RGE for softly broken MSSM with neutrinos\label{appendix:general_rge}}
In this Appendix, we present the $1$-loop RGE of a softly broken MSSM, which also contains right-handed neutrinos.
Below the mass thresholds of the right-handed neutrinos, they have to be integrated out of the theory, which essentially removes them from the RGEs \cite{Antusch:2002rr}.These equations are well known and are presented here merely for completeness; the MSSM equations can be found in \cite{Martin:1993zk,Antusch:2005gp,Antusch:2015nwi}. The equations assume the convention of REAP~\cite{Antusch:2005gp} and SusyTC~\cite{Antusch:2015nwi}, which in particular is an RL convention for the Yukawa matrices and the trilinear couplings, and which is used throughout this paper. In Table~\ref{tab:conventions} the convention for the quantities in $W_\text{MSSM}$ and $\mathcal{L}_{\text{soft}}$ (cf.~Eq.~\eqref{eq:MSSM-superpotential} and \eqref{eq:MSSM-soft-terms}) in this paper and in SusyTC is compared to the ones in Martin's Supersymmetry Primer~\cite{Martin:1997ns} and in SUSY Les Houches Accord (SLHA) 2~\cite{Allanach:2008qq}. We use $t=\log\mu_r$, where $\mu_r$ is the renormalization scale. Also, the hypercharge coupling $g_1$ of $\mathrm{U}(1)_Y$ is in the GUT normalization, related to the SM-normalized $\mathrm{U}(1)_Y$ coupling $g$ by $g_1^2=5/3\,g^2$.

\begin{table}[htb]
\begin{center}
\begin{tabular}{llll}
\toprule
\makebox[2.4cm][l]{paper} & \makebox[2.4cm][l]{SusyTC} & \makebox[2.4cm][l]{SUSY Primer} & \makebox[2.4cm][l]{SLHA~2}\\
\midrule
$\mathbf{Y}$ & $+\mathbf{Y}$ & $-\mathbf{y}$ & $+\mathbf{Y}^T$ \\
$\mu$ & $+\mu$ & $+\mu$ & $+\mu$ \\
$M$ & $+M$ & $-M$ & $+M$ \\
$\mathbf{m}^2$ & $+\mathbf{m}^2$ & $+\mathbf{m}^2$ & $+\mathbf{m}^2$ \\
$\mathbf{A}$ & $+\mathbf{T}$ & $-\mathbf{a}$ & $+\mathbf{T}^T$ \\
$m^2_{H_{u/d}}$ & $+m^2_{H_{u/d}}$ & $+m^2_{H_{u/d}}$ & $+m^2_{H_{u/d}}$ \\
$b$ & $+m_3^2$ & $+b$ & $+m_3^2$ \\
\bottomrule
\end{tabular}
\end{center}
\caption{Comparison of the labels and conventions chosen in this paper for the quantities present in $W_\text{MSSM}$ in Eq.~\eqref{eq:MSSM-superpotential} and $\mathcal{L}_{\text{soft}}$ in Eq.~\eqref{eq:MSSM-soft-terms}, with the conventions used by SusyTC, Martin's Supersymmetry Primer and SLHA~2. All labels and family indices are neglected and right-handed neutrinos are not considered. The convention in this paper corresponds to the one in SusyTC. SusyTC and the SUSY Primer adhere to the RL convention for the Yukawa matrices and the trilinear couplings, SLHA 2 employs the LR convention. \label{tab:conventions}}
\end{table}

The multiple family RGEs are the following:

\begingroup
\allowdisplaybreaks   
\begin{align}
    c_1\,\DT g_i&=\beta_ig_i^3\,,\\
    c_1\,\DT M_i&=2\beta_i g_i^2 M_i\,,\\
    c_1\,\DT \mu&=\mu\;\left(\TR(3\YUR+3\YDR+\YER +\YNR)-\tfrac{3}{5}g_1^2-3g_2^2\right),
\end{align}
\endgroup

\begingroup
\allowdisplaybreaks   
\begin{align}
    c_1\,\DT\YU&=\YU\,\left(3\TR(\YUR)\ONE+\TR(\YNR)+3\YUL+\YDL-\ONE (\tfrac{13}{15}g_1^2+3g_2^2+\tfrac{16}{3}g_3^2)\right),\\
    c_1\,\DT\YD&=\YD\,\left(3\TR(\YDR)\ONE+\TR(\YER)\ONE+3\YDL+\YUL-\ONE (\tfrac{7}{15}g_1^2+3g_2^2+\tfrac{16}{3}g_3^2)\right),\\
    c_1\,\DT\YE&=\YE\,\left(3\TR(\YDR)\ONE+\TR(\YER)\ONE+3\YEL+\YNL-\ONE (\tfrac{9}{5}g_1^2+3g_2^2)\right),\\
    c_1\,\DT\YN&=\YN\,\left(3\TR(\YUR)\ONE+\TR(\YNR)\ONE+3\YNL+\YEL-\ONE (\tfrac{3}{5}g_1^2+3g_2^2)\right),\\
    c_1\,\DT\MNU&=2\,(\YNR)\,\MNU+2\,\MNU\,(\YNR)^T,
\end{align}
\endgroup

\begingroup
\allowdisplaybreaks   
\begin{align}
    \begin{split}
        c_1\,\DT\AU&=\YU\,(4\YU^\dagger\AU+2\YD^\dagger\AD)+\AU\,(5\YUL+\YDL)+\\
         &\quad +\YU\left(6\TR(\AU\YU^\dagger)+2\TR(\AN\YN^\dagger)+\tfrac{26}{15}g_1^2 M_1+6g_2^2 M_2+\tfrac{32}{3}g_3^2M_3\right)+\\
        &\quad +\AU\left(3\TR(\YUR)+\TR(\YNR)-\tfrac{13}{15}g_1^2-3g_2^2-\tfrac{16}{3}g_3^2\right)\,,
    \end{split}\\
    \begin{split}
        c_1\,\DT\AD&=\YD\,(4\YD^\dagger\AD+2\YU^\dagger\AU)+\AD\,(5\YDL+\YUL)+\\
        &\quad +\YD\,\left(6\TR(\AD\YD^\dagger)+2\TR(\AE\YE^\dagger)+\tfrac{14}{15}g_1^2 M_1+6g_2^2 M_2+\tfrac{32}{3}g_3^2 M_3\right)+\\
        &\quad +\AD\,\left(3\TR(\YDR)+\TR(\YER)-\tfrac{7}{15}g_1^2-3g_2^2-\tfrac{16}{3}g_3^2\right)\,,
    \end{split}\\
    \begin{split}
        c_1\,\DT\AE&=\YE\,(4\YE^\dagger\AE+2\YN^\dagger\AN)+\AE\,(5\YEL+\YNL)+\\
        &\quad +\YE\,\left(6\TR(\AD\YD^\dagger)+2\TR(\AE\YE^\dagger)+\tfrac{18}{5}g_1^2 M_1+6g_2^2 M_2\right)+\\
        &\quad +\AE\,\left(3\TR(\YDR)+\TR(\YER)-\tfrac{9}{5}g_1^2-3g_2^2\right)\,,
    \end{split}\\
    \begin{split}
        c_1\,\DT\AN&=\YN\,(4\YN^\dagger\AN+2\YE^\dagger\AE)+\AN\,(5\YNL+\YEL)+\\
        &\quad +\YN\,\left(6\TR(\AU\YU^\dagger)+2\TR(\AN\YN^\dagger)+\tfrac{6}{5}g_1^2 M_1+6g_2^2 M_2\right)+\\
        &\quad +\AN\,\left(3\TR(\YUR)+\TR(\YNR)-\tfrac{3}{5}g_1^2-3g_2^2\right)\,,
    \end{split}
\end{align}
\endgroup

\begin{align}
    \begin{split}
        c_1\,\DT B&=B\left(\TR(3\YUR+3\YDR+\YER +\YNR)-\tfrac{3}{5}g_1^2-3g_2^2\right)\\
        &\phantom{=}+\mu \left(\TR(6\AU\YU^\dagger+6\AD\YD^\dagger+2\AE\YE^\dagger + 2\AN\YN^\dagger)+\tfrac{6}{5}g_1^2 M_1+6g_2^2 M_2\right),
    \end{split}
\end{align}

\begingroup
\allowdisplaybreaks   
\begin{align}
    \begin{split}
    c_1\,\DT m^2_{H_u}&=6\TR\big((m^2_{H_u}\ONE +\M^2_Q)\YUL+\YU^\dagger\M^2_u\YU+\AU^\dagger\AU\big)+\\
        &\quad + 2\TR\big((m^2_{H_u}\ONE +\M^2_L)\YNL+\YN^\dagger\M^2_\nu\YN+\AN^\dagger\AN\big)-\\
        &\quad -\tfrac{6}{5}g_1^2 |M_1|^2-6g_2^2 |M_2|^2+\tfrac{3}{5}g_1^2 S\,,
    \end{split}\\
    \begin{split}
        c_1\,\DT m^2_{H_d}&=6\TR\big((m^2_{H_d}\ONE +\M^2_Q)\YDL+\YD^\dagger\M^2_d\YD+\AD^\dagger\AD\big)+\\
        &\quad + 2\TR\big((m^2_{H_d}\ONE +\M^2_L)\YEL+\YE^\dagger \M^2_e\YE+\AE^\dagger\AE\big)-\\
        &\quad -\tfrac{6}{5}g_1^2 |M_1|^2-6g_2^2 |M_2|^2-\tfrac{3}{5}g_1^2 S\,,
    \end{split}
\end{align}
\endgroup

\begingroup
\allowdisplaybreaks    
\begin{align}
    \begin{split}
        c_1\,\DT \M^2_Q&=(\M^2_Q+2m^2_{H_u}\ONE )\YUL+(\M^2_Q+2m^2_{H_d}\ONE )\YDL+(\YUL+\YDL)\M^2_Q+\\
        &\quad + 2\YU^\dagger \M^2_u\YU+2\YD^\dagger \M^2_d\YD+2\AU^\dagger\AU+2\AD^\dagger\AD+\\
        &\quad +\ONE \left(-\tfrac{2}{15}g_1^2 |M_1|^2-6g^2_2|M_2|^2-\tfrac{32}{3}g_3^2|M_3|^2+\tfrac{1}{5}g_1^2 S\right),
    \end{split}\\
    \begin{split}
        c_1\,\DT \M^2_L&=(\M^2_L+2m^2_{H_u}\ONE )\YNL+(\M^2_L+2m^2_{H_d}\ONE )\YEL+(\YNL+\YEL)\M^2_L+\\
        &\quad + 2\YN^\dagger \M^2_\nu\YN+2\YE^\dagger \M^2_e\YE+2\AN^\dagger\AN+2\AE^\dagger\AE+\\
        &\quad +\ONE \left(-\tfrac{6}{5}g_1^2 |M_1|^2-6g^2_2|M_2|^2-\tfrac{3}{5}g_1^2 S\right),
    \end{split}\\
    \begin{split}
        c_1\,\DT \M^2_u&=(2\M^2_u+4m^2_{H_u}\ONE )\YUR+4\YU\M^2_Q\YU^\dagger+2\YUR\M^2_u+4\AU\AU^\dagger+\\
        &\phantom{=} +\ONE \left(-\tfrac{32}{15}g_1^2 |M_1|^2-\tfrac{32}{3}g_3^2|M_3|^2-\tfrac{4}{5}g_1^2 S\right),
    \end{split}\\
    \begin{split}
        c_1\,\DT \M^2_d&=(2\M^2_d+4m^2_{H_d}\ONE )\YDR+4\YD\M^2_Q\YD^\dagger+2\YDR\M^2_d+4\AD\AD^\dagger+\\
        &\phantom{=} +\ONE \left(-\tfrac{8}{15}g_1^2 |M_1|^2-\tfrac{32}{3}g_3^2|M_3|^2+\tfrac{2}{5}g_1^2 S\right),
    \end{split}\\
    \begin{split}
        c_1\,\DT \M^2_e&=(2\M^2_e+4m^2_{H_d}\ONE )\YER+4\YE\M^2_L\YE^\dagger+2\YER\M^2_e+4\AE\AE^\dagger+\\
        &\phantom{=} +\ONE \left(-\tfrac{24}{5}g_1^2 |M_1|^2+\tfrac{6}{5}g_1^2 S\right),
    \end{split}\\
    c_1\,\DT \M^2_\nu&=(2\M^2_\nu+4m^2_{H_u}\ONE )\YNR+4\YN\M^2_L\YN^\dagger+2\YNR\M^2_\nu+4\AN\AN^\dagger\,.
\end{align}
\endgroup

\noindent
The loop factor $c_1$ is defined as
	\begin{align}
	c_1&=16\,\pi^2,
	\end{align}
the values of the $\beta_i$ coefficients are 
	\begin{align}
	\beta_1=\tfrac{33}{5},\quad \beta_2&=1,\quad \beta_3=(-3)\,,\label{eq:RGE-beta-coefficients}
	\end{align}
and the quantity $S$ is defined as the following combination of soft scalar mass parameters:
\begin{align}
S&:=m^2_{H_{u}}-m^2_{H_d}+\TR(\M^2_Q-\M^2_L-2\M^2_u+\M^2_d+\M^2_e)\,.\label{eq:RGE-definition-S}
\end{align}

\section{Approximate RGE with 3rd family Yukawa couplings\label{appendix:rge-simplified}}
In this Appendix a simple approximation for the RGEs of the MSSM quantities (including right-handed neutrinos) is presented, which is self-consistent under RG running. Under the assumption that in each Yukawa matrix the $(3,3)$-entry dominates, all other entries are set to zero. Furthermore, the trilinear couplings are taken proportional to the Yukawa matrices. In order to have no extra flavor violation in the SUSY sector, the soft mass matrices are chosen diagonal, where soft masses of the first two families are identical. Taking these considerations together, the setup below presents the minimal set of self-consistent RGE equations, which simplifies the full set and captures the dominant effects of the running.

The ansatz is
\begin{align}
    \YU&=
    \begin{pmatrix}
    0&0&0\\
    0&0&0\\
    0&0&y_t\\
    \end{pmatrix},&
    \YD&=
    \begin{pmatrix}
    0&0&0\\
    0&0&0\\
    0&0&y_b\\
    \end{pmatrix},&
    \YE&=
    \begin{pmatrix}
    0&0&0\\
    0&0&0\\
    0&0&y_\tau\\
    \end{pmatrix},&
    \YN&=
    \begin{pmatrix}
    0&0&0\\
    0&0&0\\
    0&0&y_{\nu}\\
    \end{pmatrix},
    \label{eq:RGE-ansatz-begin}
\end{align}

\begin{align}
\MNU&=
    \begin{pmatrix}
    M_{\nu1}&0&0\\
    0&M_{\nu2}&0\\
    0&0&M_{\nu3}\\
    \end{pmatrix},
\end{align}

\begin{align}
\AU&=a_u\,\YU\,,&\AD&=a_d\,\YD\,,&\AE&=a_e\,\YE\,,&\AN&=a_\nu\,\YN\,.\label{eq:a-term-factor}
\end{align}

\begin{align}
\M^2_{Q}&=\begin{pmatrix} m^{2}_{Q_1}&&\\ &m^{2}_{Q_1}&\\ &&m^{2}_{Q_3}\\ \end{pmatrix},&
\M^2_{L}&=\begin{pmatrix} m^{2}_{L_1}&&\\ &m^{2}_{L_1}&\\ &&m^{2}_{L_3}\\ \end{pmatrix},&
\M^2_{u}&=\begin{pmatrix} m^{2}_{u_1}&&\\ &m^{2}_{u_1}&\\ &&m^{2}_{u_3}\\ \end{pmatrix},\\
\M^2_{d}&=\begin{pmatrix} m^{2}_{d_1}&&\\ &m^{2}_{d_1}&\\ &&m^{2}_{d_3}\\ \end{pmatrix},&
\M^2_{e}&=\begin{pmatrix} m^{2}_{e_1}&&\\ &m^{2}_{e_1}&\\ &&m^{2}_{e_3}\\ \end{pmatrix},&
\M^2_{\nu}&=\begin{pmatrix} m^{2}_{\nu_1}&&\\ &m^{2}_{\nu_1}&\\ &&m^{2}_{\nu_3}\\ \end{pmatrix}.\label{eq:RGE-ansatz-end}
\end{align}

Using this ansatz for the Yukawa couplings, the trilinear couplings and the soft masses, the RGEs from Appendix~\ref{appendix:general_rge} are simplified and now read as follows:

\begingroup
\allowdisplaybreaks   
\begin{align}
    c_1\,\DT g_i&=\beta_ig_i^3\,,\label{eq:RGE-simple-g}\\
    c_1\,\DT M_i&=2\beta_i g_i^2 M_i\,,\label{eq:RGE-simple-M}\\
    c_1\,\DT \mu&=\mu  \left(|y_\nu|^2+|y_{\tau}|^2+3 |y_{b}|^2 +3 |y_{t}|^2 -\frac{3 g_{1}^2}{5}-3 g_{2}^2\right),\label{eq:RGE-simpe-mu}
\end{align}
\endgroup

\begingroup
\allowdisplaybreaks   
\begin{align}
    c_1\,\DT y_t&=y_{t} \left(6 |y_{t}|^2+|y_{b}|^2 +|y_\nu|^2 -\frac{13}{15} g_{1}^2-3 g_{2}^2-\frac{16}{3} g_{3}^2\right),\label{eq:RGE-simple-t}\\
    c_1\,\DT y_b&=y_{b} \left(6 |y_{b}|^2+|y_{t}|^2+|y_{\tau}|^2-\frac{7}{15} g_{1}^2-3 g_{2}^2-\frac{16}{3} g_{3}^2\right),\label{eq:RGE-simple-b}\\
    c_1\,\DT y_\tau&=\,y_{\tau} \left(3 |y_{b}|^2+4 |y_{\tau}|^2 + |y_\nu|^2-\frac{9}{5} g_{1}^2-3 g_{2}^2\right),\label{eq:RGE-simple-tau}\\
    c_1\,\DT y_\nu&=\,y_{\tau} \left(3 |y_{t}|^2+4 |y_{\nu}|^2 + |y_\tau|^2-\frac{3}{5} g_{1}^2-3 g_{2}^2\right),\label{eq:RGE-simple-nu}
\end{align}
\endgroup

\begingroup
\allowdisplaybreaks   
\begin{align}
     c_1\,\DT M_{\nu1}&=0,\\
     c_1\,\DT M_{\nu2}&=0,\\
    c_1\,\DT M_{\nu3}&=4\,M_{\nu3}\,|y_\nu|^2\,,
\end{align}
\endgroup

\begingroup
\allowdisplaybreaks   
\begin{align}  
    c_1\,\DT a_u &= 2 a_{d} |y_{b}|^2 +12 a_{u} |y_{t}|^2 +2 a_\nu |y_\nu|^2 +\frac{26}{15} g_{1}^2 M_{1}+6 g_{2}^2 M_{2}+\frac{32}{3} g_{3}^2 M_{3}\,,\label{eq:RGE-simple-au}\\
    c_1\,\DT a_d &= 12 a_{d} |y_{b}|^2 +2 a_{e} |y_{\tau}|^2 +2 a_{u} |y_{t}|^2 +\frac{14}{15} g_{1}^2 M_{1}+6 g_{2}^2 M_{2}+\frac{32}{3} g_{3}^2 M_{3}\,,\label{eq:RGE-simple-ad}\\
    c_1\,\DT a_e &= 6 a_{d} |y_{b}|^2 +8 a_{e} |y_{\tau}|^2 +2 a_{\nu} |y_{\nu}|^2 +\frac{18}{5} g_{1}^2 M_{1}+6 g_{2}^2 M_{2}\,,\label{eq:RGE-simple-ae}\\
	c_1\,\DT a_\nu &= 6 a_{u} |y_{t}|^2 +2 a_{e} |y_{\tau}|^2 +8 a_{\nu} |y_{\nu}|^2 +\frac{6}{5} g_{1}^2 M_{1}+6 g_{2}^2 M_{2}\,,\label{eq:RGE-simple-anu}    
\end{align}
\endgroup

\begin{align}
    \begin{split}
    c_1\,\DT B&=3 |y_{b}|^2 (2 a_{d} \mu +B)+|y_{\tau}|^2(2 a_{e} \mu +B)+3 |y_{t}|^2 (2 a_{u} \mu +B) + |y_\nu|^2 (2 a_\nu \mu + B) \\
    &\phantom{=} -\frac{3}{5} B \left(g_{1}^2+5 g_{2}^2\right)+\frac{6}{5} \mu  \left(g_{1}^2 M_{1}+5 g_{2}^2 M_{2}\right),
    \end{split} \label{eq:RGE-simple-B}
\end{align}

\begingroup
\allowdisplaybreaks   
\begin{align}
    \begin{split}
    c_1\,\DT m^2_{H_u}&=6 |y_{t}|^2 \left(|a_{u}|^2+m^2_{H_u}+m^2_{Q_3}+m^2_{u_3}\right) + 2 |y_{\nu}|^2 \left( |a_\nu|^2 + m^2_{H_u} + m^2_{L_3} + m^2_{\nu_3} \right) \\
    &\phantom{=} -\frac{6}{5} g_{1}^2 |M_{1}|^2 -6 g_{2}^2 |M_{2}|^2+\frac{3}{5} g_{1}^2 S\,,
    \end{split}\label{eq:RGE-simple-mHu}\\
    \begin{split}
	c_1\,\DT m^2_{H_d}&=6 |y_{b}|^2 \left(|a_{d}|^2+m^2_{H_d}+m^2_{Q_3}+m^2_{d_3}\right)-\frac{6}{5} g_{1}^2 |M_{1}|^2 -6 g_{2}^2 |M_{2}|^2 -\frac{3}{5} g_{1}^2 S +\\
        &\phantom{=} +2 |y_{\tau}|^2 \left(|a_{e}|^2+m^2_{H_d}+m^2_{L_3}+m^2_{e_3}\right),
    \end{split}\label{eq:RGE-simple-mHd}
\end{align}
\endgroup

\begingroup
\allowdisplaybreaks
\begin{align}
    c_1\,\DT m^2_{Q_1}&=\frac{1}{15} \left(-2 g_{1}^2 |M_{1}|^2-90 g_{2}^2 |M_{2}|^2-160 g_{3}^2 |M_{3}|^2+3 g_{1}^2 S\right),\label{eq:RGE-simple-mQ1}\\
    c_1\,\DT m^2_{Q_3}&=c_1\,\DT m^2_{Q_1}+2 \left(|y_{b}|^2 \left(| a_{d}| ^2+m^2_{d_3}+m^2_{H_d}+m^2_{Q_3}\right)+|y_{t}|^2 \left(| a_{u}|^2+m^2_{H_u}+m^2_{Q_3}+m^2_{u_3}\right)\right),\label{eq:RGE-simple-mQ3}\\
    c_1\,\DT m^2_{L_1}&=-\frac{3}{5} g_{1}^2 \left(2 |M_{1}|^2+S\right)-6 g_{2}^2 |M_{2}|^2,\label{eq:RGE-simple-mL1}\\    
    c_1\,\DT m^2_{L_3}&=c_1\,\DT m^2_{L_1}+ 2 \left(|y_{\tau}|^2 \left(| a_{e}| ^2+m^2_{e_3}+m^2_{H_d}+m^2_{L_3}\right) + |y_\nu|^2 \left( |a_\nu|^2 + m^2_{H_u} + m^2_{L_3} + m^2_{\nu_3} \right)\right),\label{eq:RGE-simple-mL3}\\
    c_1\,\DT m^2_{u_1}&=-\frac{4}{15} \left(g_{1}^2 \left(8 |M_{1}|^2+3 S\right)+40 g_{3}^2 |M_{3}|^2 \right),\label{eq:RGE-simple-mu1}\\
    c_1\,\DT m^2_{u_3}&=c_1\,\DT m^2_{u_1}+ 4 |y_{t}|^2 \left(\left| a_{u}\right| ^2+m^2_{H_u}+m^2_{Q_3}+m^2_{u_3}\right),\label{eq:RGE-simple-mu3}\\
    c_1\,\DT m^2_{d_1}&=\frac{2}{15} \left(g_{1}^2 \left(3 S-4 |M_{1}|^2 \right)-80 g_{3}^2 |M_{3}|^2\right),\label{eq:RGE-simple-md1}\\
    c_1\,\DT m^2_{d_3}&=c_1\,\DT m^2_{d_1}+4 |y_{b}|^2 \left(\left| a_{d}\right| ^2+m^2_{d_3}+m^2_{H_d}+m^2_{Q_3}\right),\label{eq:RGE-simple-md3}\\
    c_1\,\DT m^2_{e_1}&=\frac{6}{5} g_{1}^2 \left(S-4 |M_{1}|^2\right),\label{eq:RGE-simple-me1}\\
    c_1\,\DT m^2_{e_3}&=c_1\,\DT m^2_{e_1}+4 |y_{\tau}|^2 \left(| a_{e}| ^2+m^2_{e_3}+m^2_{H_d}+m^2_{L_3}\right),\label{eq:RGE-simple-me3}\\
    c_1\,\DT m^2_{\nu_1}&=0\,,\label{eq:RGE-simple-mnu1}\\
    c_1\,\DT m^2_{\nu_3}&=4|y_\nu|^2 \left( |a_\nu|^2 + m^2_{H_u} + m^2_{L_3} + m^2_{\nu_3} \right).\label{eq:RGE-simple-mnu3}
\end{align}
\endgroup

\noindent
We also have
\begin{align}
    S&=m^2_{H_u}-m^2_{H_d}+2m^2_{Q_1}+m^2_{Q_3}-2m^2_{L_1}-m^2_{L_3}-4m^2_{u_1}-2m^2_{u_3}+2m^2_{d_1}+m^2_{d_3}+2m^2_{e_1}+m^2_{e_3}\,.
\end{align}
Note that the $a$-factors are defined via $\mathbf{A}_x=a_x \mathbf{Y}_x$, so their RGE have to be derived accordingly, e.g.~
\begin{align}
\DT(a_u y_t)&=\DT(a_u)y_t+a_u\DT(y_t)\,,
\end{align}
implying
\begin{align}
\DT a_u&=(1/y_t)\left(\DT(a_u y_t)-a_u \DT(y_t)\right).
\end{align}

For the Majorana neutrino mass associated to the large $3$rd family neutrino Yukawa coupling, we assume the value $M_{\nu 3}=M_R$ at the scale $M_R$, implying that this heavy neutrino is integrated out at the scale $M_R$. The $M_{\nu 3}$ does not appear in the RGE of any other quantity.

\end{document}